\documentclass[a4paper,11pt,dvipsnames]{article}

\pdfoutput=1 

\usepackage{jheppub} 

\usepackage{shuffle}
\usepackage{comment}
\usepackage[T1]{fontenc} 
\usepackage{tikz}
\usepackage{tikz-3dplot}
\usepackage{xcolor}
\usepackage{graphbox}
\usetikzlibrary{patterns, decorations.markings, arrows}
\usetikzlibrary{arrows.meta}
\usepackage{mathabx}
\usepackage[export]{adjustbox}
\allowdisplaybreaks
\usepackage[backgroundcolor=green!5, linecolor=blue!60]{todonotes}
\newcommand{\dd}{\text{d}}
\newcommand{\dlog}[1]{\frac{\dd #1}{#1}}

\newcommand{\la}{\langle}
\newcommand{\ra}{\rangle}
\newcommand{\ang}[1]{\langle #1\rangle}

\title{Poles At Infinity in On-shell Diagrams}

\author[a]{Taro V. Brown,}\emailAdd{tvbrown@ucdavis.edu}
\author[a]{Umut Oktem,}\emailAdd{ucoktem@ucdavis.edu}
\author[a,b]{Jaroslav Trnka}\emailAdd{trnka@ucdavis.edu}

\affiliation[a]{Center for Quantum Mathematics and Physics (QMAP) and Department of Physics,\\ University of California, Davis, California, USA}
\affiliation[b]{Institute of Particle and Nuclear Physics, Charles University, Prague, Czech Republic}

\abstract{In this paper we study on-shell diagrams in ${\cal N}{<}4$ supersymmetric Yang-Mills (SYM) theory. These are on-shell gauge invariant objects which appear as cuts of loop integrands in the context of generalized unitarity and serve as building blocks for amplitudes in recursion relations. In the dual formulation, they are associated with cells of the positive Grassmannian $G_+(k,n)$ and the on-shell functions can be reproduced as canonical differential forms. While for the case of the ${\cal N}{=}4$ maximally supersymmetric Yang-Mills theory all poles in on-shell diagrams correspond to IR poles when the momentum flows in edges are zero, for ${\cal N}{<}4$ SYM theories there are new UV poles when the loop momenta go to infinity. These poles originate from the prefactor of the canonical dlog form and do not correspond to erasing edges in on-shell diagrams. We show that they can be interpreted as a diagrammatic operation which involves pinching a loop and performing a ``non-planar twist'' on external legs, which gives rise to a non-planar on-shell diagram. Our result provides an important clue on the role of poles at infinite momenta in on-shell scattering amplitudes, and the relation to non-planar on-shell functions.}

\begin{document} 

\maketitle

\section{Introduction}

Modern on-shell methods for scattering amplitudes are based on a simple idea of building amplitudes from on-shell gauge invariant objects (see \cite{Elvang:2013cua,Dixon:2013uaa,Henn:2014yza,Cheung:2017pzi,Travaglini:2022uwo,Bern:2022jnl} for recent reviews). At tree-level, this idea is materialized in the form of on-shell recursion relations where the $n$-point tree-level amplitude is built recursively from the lower point (on-shell) amplitudes. The primary example is the Britto-Cachazo-Feng-Witten (BCFW) recursion relations \cite{Britto:2004ap,Britto:2005fq} where we shift external momenta of two external legs. There are also generalizations to multi-line shifts \cite{Cohen:2010mi,Cheung:2008dn,Cheung:2015cba}, as well as theories with special soft limits \cite{Cheung:2015ota,Luo:2015tat,Elvang:2018dco,Cheung:2018oki}. At loop level, unitarity methods allow us to write the amplitude as a linear combination of basis integrals with gauge-invariant on-shell prefactors \cite{Bern:1994zx,Bern:1994cg,Bern:2005iz,Bern:2009kd,Bern:2008ap,Bern:2012uc,Bern:2018jmv,Carrasco:2021otn,Bourjaily:2016evz,Bourjaily:2017wjl}. These prefactors are calculated using \emph{cuts of the loop integrand}, where we set propagators to be on-shell and the loop integrand factorizes into simpler objects. The basic cut is the \emph{unitarity cut} where two propagators from one loop are set to zero. This procedure can be iterated further and when the maximal number of propagators are cut, we talk about \emph{maximal cuts}.
\begin{equation}
\raisebox{-12mm}{\includegraphics[trim={0cm 0cm 0cm 0cm},clip,scale=0.77]{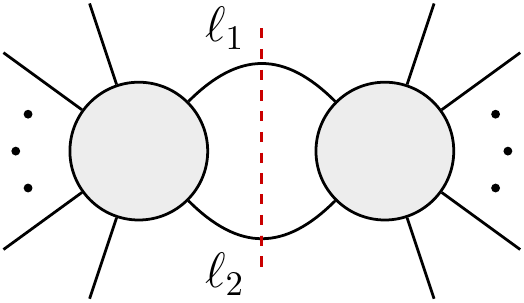}},\qquad \qquad 
 \raisebox{-17mm}{\includegraphics[trim={0cm 0cm 0cm 0cm},clip,scale=0.82]{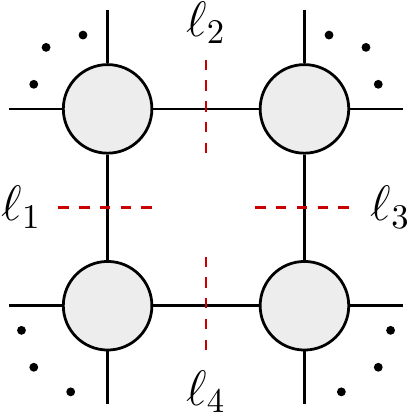}}.
\end{equation}
In these diagrams both internal and external lines are on-shell, and the result is given by the product of tree-level amplitudes in the corners. In theories with fundamental 3-point amplitudes we can proceed further and express tree-level amplitudes in the corners using BCFW recursion relations. In the end, we get \emph{on-shell diagrams} \cite{Arkani-Hamed:2012zlh,Franco:2013nwa,Arkani-Hamed:2014bca,Bai:2014cna,Franco:2015rma,Benincasa:2015zna,Benincasa:2016awv,Bourjaily:2016mnp,Heslop:2016plj,Herrmann:2016qea,Farrow:2017eol,Armstrong:2020ljm,Cachazo:2018wvl,Paranjape:2022ymg} which are on-shell gauge invariant objects. Each diagram is defined as a product of on-shell 3-point amplitudes (all lines, external and internal, are on-shell) integrated over momenta and Grassmann variables of internal legs,
\begin{equation} \label{eq:3-pointstitch}
\raisebox{-18.5mm}{\includegraphics[trim={0cm 0cm 0cm 0cm},clip,scale=1.1]{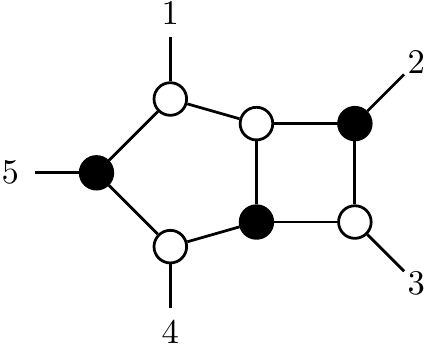}}
\begin{aligned}
     = \prod_k \int d^{\cal N}\widetilde{\eta}_k\int\frac{d^2\lambda_k\,d^2\widetilde{\lambda}_k}{{\rm GL}(1)}\left(\prod_j A_3^{(j)}\right)
\end{aligned}
\end{equation}
where ${\cal N}$ is the number of supersymmetries. On-shell diagrams play multiple roles, such as giving us the kinematical solutions for cuts of loop integrands and reproducing the associated on-shell functions on these cuts, but a subset of them also play a role as building blocks in the recursion relations for tree-level amplitudes or loop integrands. For example, the same diagram represents a maximal cut of 3-loop 6-point MHV amplitude or a term in the BCFW construction of the 6-point MHV tree-level amplitude,
\begin{equation}\raisebox{-19mm}{
\includegraphics[trim={0cm 0cm 0cm 0cm},clip,scale=1.1]{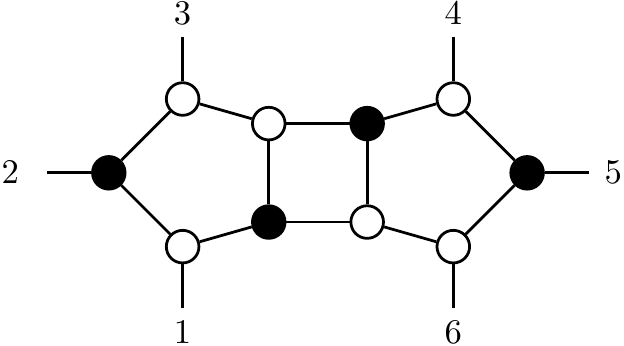}}.\label{onshell1}
\end{equation}
In ${\cal N}{=}4$ SYM theory this diagram evaluates to the Parke-Taylor factor,
\begin{equation}
\Omega = \frac{\delta^4(\lambda{\cdot}\widetilde{\lambda})\,\delta^4(\lambda{\cdot}\widetilde{\eta})}{\la12\ra\la23\ra\la34\ra\la45\ra\la56\ra\la61\ra}\label{PT6}.
\end{equation}
On-shell diagrams are fascinating objects and are tightly connected to the geometric formulations for on-shell scattering amplitudes using the positive Grassmannian and the Amplituhedron \cite{Arkani-Hamed:2013jha,Arkani-Hamed:2013kca,Arkani-Hamed:2017vfh,Damgaard:2019ztj,Ferro:2020ygk,Ferro:2022abq,Herrmann:2022nkh}, which have been successfully used to obtain various explicit results including certain all-loop order quantities \cite{Bai:2014cna,Franco:2014csa,Ferro:2016zmx,Ferro:2015grk,Ferro:2016ptt,Ferro:2018vpf,Ferro:2020lgp,Herrmann:2020oud,Herrmann:2020qlt,Kojima:2020tjf,Arkani-Hamed:2021iya,Rao:2019wyi,Kojima:2020gxs,Dian:2021idl,Dian:2022tpf}. In particular, each planar diagram, known in the math literature as a plabic graph \cite{Postnikov:2006kva}, is associated with a specific cell in the positive Grassmannian $G_+(k,n)$ represented by the matrix $C$. The matrix $C$ is parametrized by the variables $\alpha_i$ which are associated with edges (or faces) of the graph. In physics, the on-shell function in ${\cal N}{=}4$ SYM theory is reproduced by a canonical dlog form \cite{Arkani-Hamed:2012zlh} in all $\alpha_i$ and delta functions which solve for $\alpha_i$ in terms of kinematical variables,
\begin{equation} \label{eq:grassmannian}
    \Omega = \int \frac{d\alpha_1}{\alpha_1}\frac{d\alpha_2}{\alpha_2}\dots \frac{d\alpha_m}{\alpha_m}\,\delta (C{\cdot} \widetilde{\lambda})\,\delta(C^\perp{\cdot} \lambda)\,\delta(C{\cdot} \widetilde{\eta}).
\end{equation}
The kinematical poles in the on-shell function are related to the poles in the edge variables. One particular parametrization of the on-shell diagram (\ref{onshell1}) is
\begin{equation} \label{eq:MHV6point}
\raisebox{-18.5mm}{\includegraphics[trim={0cm 0cm 0cm 0cm},clip,scale=1.1]{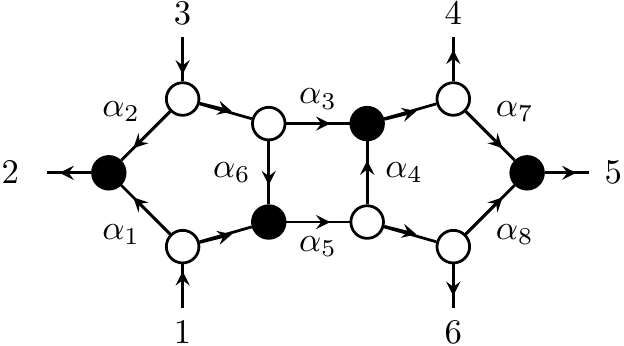}},
	\begin{aligned}
		~~~~&\alpha_1=\frac{\la 23 \ra}{\la 13 \ra},~~~~
		\alpha_2=\frac{\la 12 \ra }{\la 13 \ra},\\
		&\alpha_3=\frac{\la 46 \ra}{\la 36 \ra},~~~~\alpha_4=\frac{\la 34 \ra}{\la 36 \ra},\\
		&\alpha_5=\frac{\la 36 \ra}{\la 13 \ra},~~~~
		\alpha_6=\frac{\la 16 \ra}{\la 36 \ra},\\
		&\alpha_7=\frac{\la 56 \ra}{\la 46 \ra},~~~~
		\alpha_8=\frac{\la 45 \ra}{\la 46 \ra}.
	\end{aligned}
\end{equation}
and the formula (\ref{eq:grassmannian}) reproduces (\ref{PT6}). The kinematical poles in $\Omega$ correspond to poles in edge variables $\alpha_j$. In fact, taking a residue on the pole $\alpha_j{=}0$ erases a corresponding edge and produces a lower-loop on-shell diagram. Its on-shell function is equal to the residue of the original on-shell diagram on the kinematical pole. For example, in the diagram (\ref{eq:MHV6point}) the condition $\alpha_7{=}0$ sets the momentum flow in a corresponding edge to zero and forces $\la56\ra=0$. The result is a lower-loop on-shell diagram where external legs $5,6$ are attached to the same white vertex.
\begin{equation} \label{eq:MHV6point2}
\raisebox{-18.5mm}{\includegraphics[trim={0cm 0cm 0cm 0cm},clip,scale=1.1]{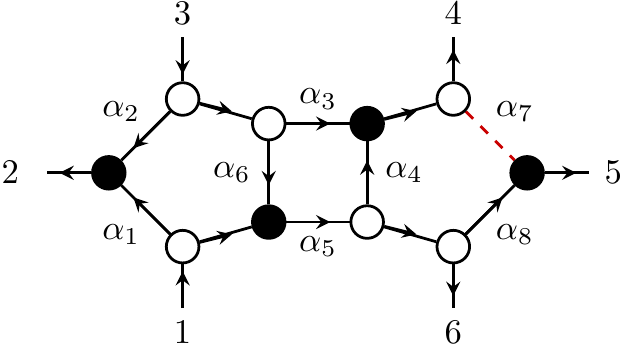}}\xrightarrow[\alpha_7=0]{}
\raisebox{-18.5mm}{\includegraphics[trim={0cm 0cm 0cm 0cm},clip,scale=1.1]{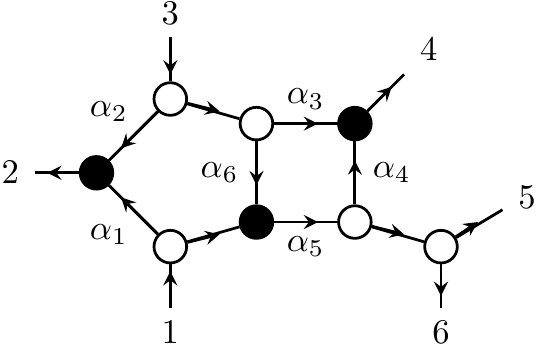}}.
\end{equation}
All poles in the on-shell function for ${\cal N}{=}4$ SYM are located deep in the IR region where the internal momenta are set to zero. For ${\cal N}{<}4$ there is an additional Jacobian factor in the Grassmannian form \eqref{eq:grassmannian} which introduces new types of poles other than $\alpha_j{=}0$ \cite{Arkani-Hamed:2012zlh}. These poles are in the UV where the loop momenta go to infinity. Nothing is known about these poles, i.e. what the precise locations in the loop momentum space  are (other than being at infinity), what type of kinematical conditions they impose on external momenta, and what the residues of on-shell diagrams on these poles are. Obviously, these poles also naturally arise when calculating the on-shell diagrams from the definition (\ref{eq:3-pointstitch}) but the Grassmannian picture nicely divides the well-understood IR poles (located at $\alpha_j{=}0$) from these new UV poles, which we also refer to as 
\emph{poles at infinity}.

The nature of these UV poles is a very important conceptual question, and we address it in the context of on-shell diagrams which properties are inherited from the properties of scattering amplitudes. In fact, the behavior on the IR poles, such as erasing edges in on-shell diagrams, taking residues on $\alpha_j{=}0$ etc., is just a consequence of perturbative unitarity for scattering amplitudes. However, the unitarity of the S-matrix does not predict what exactly happens with amplitudes (tree-level amplitudes and loop integrands) on UV poles when the loop momenta go to infinity (either external momenta or loop momenta) in a certain direction. The behavior on UV poles might be also tied to symmetries or special properties, for example the complete absence of any poles at infinity in planar ${\cal N}{=}4$ SYM is a consequence of the dual conformal symmetry \cite{Drummond:2006rz,Drummond:2008vq}. The same property has been also conjectured to hold for the non-planar sector of the same theory \cite{Arkani-Hamed:2014via,Bern:2014kca,Bern:2015ple,Bern:2017gdk,Bern:2018oao,Bourjaily:2018omh,Chicherin:2018wes,Bourjaily:2019gqu}, pointing to a possibility of a hidden symmetry in the full theory. The surprising behavior at infinity has also been observed in gravitational theories. This includes the improved $1/z^2$ scaling under BCFW shift \cite{Cachazo:2005ca,Bedford:2005yy,Arkani-Hamed:2008bsc}, enhanced cancelations on certain cuts \cite{Herrmann:2016qea,Herrmann:2018dja,Bourjaily:2018omh,Edison:2019ovj} or observed enhanced UV cancelations in higher loop amplitudes \cite{Bern:2012gh,Bern:2014sna,Bern:2017lpv,Herrmann:2018dja,Edison:2019ovj}. This is also likely related to surprisingly simple formulas for tree-level amplitudes \cite{Bern:1998sv,Elvang:2007sg,Drummond:2009ge,Mason:2009afn,Nguyen:2009jk,Hodges:2011wm,Hodges:2012ym,Trnka:2020dxl} or other special properties of gravity such as color-kinematics duality \cite{Bern:2008qj,Bern:2010ue,Bern:2019prr}. 

This naturally raises a question of a uniform framework to analyze amplitudes at infinite momenta, and search for implications of the \emph{unitarity at infinity} for the perturbative S-matrix. In this paper, we will make a modest step in this direction, and give a partial answer to this question in the context of planar on-shell diagrams in Yang-Mills theory. We will show where the poles at infinity are located in the kinematical space and what the kinematical constraints we have to impose to approach them are. The residue at infinity is then a diagrammatic operation on an on-shell diagram which we call a \emph{non-planar twist}. The residue on this pole of a planar diagram is a lower-loop \emph{non-planar diagram}. For ${\cal N}{=}3$ SYM all poles are simple while for ${\cal N}{<}3$ SYM we have multiple poles and the residue corresponds to a derivative operator acting on the lower-loop twisted on-shell diagram. 

The paper is organized as follows: In section 2, we review the basic facts about on-shell diagrams. In section 3, we study the poles at infinity in one-loop on-shell diagrams for ${\cal N}{=}3$ SYM and show that the imposed kinematical condition produces a simple tree-level factorization. In section 4, we show that in planar diagrams all poles at infinity are localized in one-loop subgraphs and we discuss their embedding into a higher-loop on-shell diagrams. We also show a number of examples and discuss Global Residue Theorems which relate UV poles to IR poles. In section 5, we generalize our procedure to ${\cal N}{<}3$ SYM and discuss the multiple poles at infinity. In section 6, we show that in non-planar diagrams the situation is more complicated as one kinematical condition can send multiple loop momenta to infinity. We end with conclusion and outlook in section 7.

\section{Fundamentals of on-shell diagrams}

On-shell diagrams are gauge invariant objects built from basic tree-level amplitudes. They appear in the context of tree-level recursion relations as building blocks for the amplitude, and in the context of generalized unitarity as coefficients of basis integrals. Most invariantly, they represent \emph{cuts of loop integrand} and provide important gauge-invariant data about the scattering amplitudes. In the theories with elementary three-point amplitudes they are built from two types of vertices, black and white, which represent two solutions of the massless on-shell 3-point kinematics
\begin{equation} \label{eq:momcon3}
    p_1^2=p_2^2=p_3^2=0,\quad p_1+p_2+p_3=0.
\end{equation}
Using the spinor helicity variables $p^\mu_{a\dot{a}} = \sigma^\mu_{a\dot{a}} \lambda^a\widetilde{\lambda}^{\dot{a}}$ the conditions on $p_1,p_2,p_3$ turn into collinearity of $\lambda$s or $\widetilde{\lambda}$s,
\begin{equation} \label{eq:collinear}
\text{MHV:}~~  \widetilde{\lambda}_1\, \propto \,\widetilde{\lambda}_2 \,\propto\, \widetilde{\lambda}_3 \qquad\text{and}\qquad \overline{\text{MHV}}:~~  \lambda_1\, \propto\, \lambda_2\,\propto\, \lambda_3,
\end{equation}
This highly restrictive kinematics fully fixes the form of the three-point amplitude for any external helicities. For the maximally ${\cal N}{=}4$ supersymmetric Yang-Mills theory we get two elementary 3-point MHV and $\overline{\rm MHV}$ amplitudes,
\begin{equation}
   \raisebox{-8mm}{\includegraphics[trim={0cm 0cm 0cm 0cm},clip,scale=1.1]{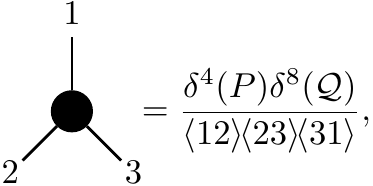}~~~~~~~~ \includegraphics[trim={0cm 0cm 0cm 0cm},clip,scale=1.1]{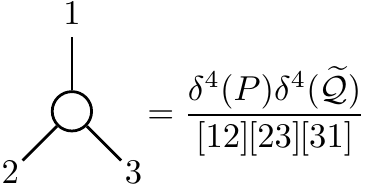}},
\end{equation}
where $\la ij\ra = \epsilon_{\alpha\beta}\lambda_i^\alpha\lambda_j^\beta$, $[ij]=\epsilon_{\dot{\alpha}\dot{\beta}}\widetilde{\lambda}_i^{\dot{\alpha}}\widetilde{\lambda}_j^{\dot{\beta}}$. Following the notation of \cite{Arkani-Hamed:2012zlh,Herrmann:2016qea} we have denoted
\begin{equation}
\begin{aligned}
        P \equiv \lambda\cdot\widetilde{\lambda} = \lambda_1\widetilde{\lambda}_1+\lambda_2&\widetilde{\lambda}_2+\lambda_3\widetilde{\lambda}_3,\quad {\cal Q} \equiv \lambda\cdot \widetilde{\eta} = \lambda_1\widetilde{\eta}_1 + \lambda_2\widetilde{\eta}_2 + \lambda_3\widetilde{\eta}_3,\\
        \widetilde{\cal Q}& = [12]\widetilde{\eta}_3 + [23]\widetilde{\eta}_1 + [31]\widetilde{\eta}_2,
\end{aligned}
\end{equation}
the momenta and supermomenta. The shorthand notation stands for $\lambda\cdot\widetilde{\lambda} = \sum_{a=1}^n\lambda_a\widetilde{\lambda}_a$, $\lambda\cdot\widetilde{\eta} = \sum_{a=1}^n\lambda_a\widetilde{\eta}_a$, where the sum goes over all external on-shell particles (we suppressed the SU(4) index $I$ in these formulas). Note that $\widetilde{\eta}$ are the fermionic variables in the expansion of the ${\cal N}=4$ superfield $\Phi$. 
\begin{align}
    \Phi(\widetilde \eta) = g^+ 
+ \widetilde \eta^I \ \widetilde {g}_I
+ \frac{1}{2!} \widetilde \eta^I\widetilde \eta^J\ \phi_{IJ}
+ \frac{1}{3!} \epsilon_{IJKL} \widetilde \eta^I\widetilde \eta^J \widetilde \eta^K\ \widetilde {g}^{L}
+ \frac{1}{4!} \epsilon_{IJKL} \widetilde \eta^I\widetilde \eta^J \widetilde \eta^K \widetilde \eta^L \ g^- .
\end{align}
If we have less than maximal supersymmetry, we need to indicate the helicities $h=\pm1$ of external particles. The elementary 3-point amplitudes in ${\cal N}{<}4$ SYM theory are then
$$
\raisebox{-8mm}{\includegraphics[trim={0cm 0cm 0cm 0cm},clip,scale=1.1]{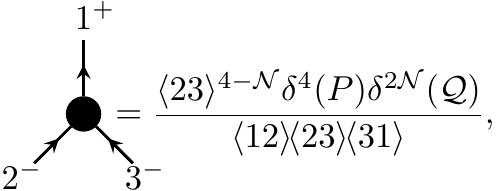}~~~~~~ \includegraphics[trim={0cm 0cm 0cm 0cm},clip,scale=1.1]{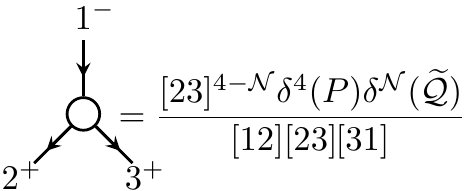}}.
$$
This includes the pure ${\cal N}{=}0$ Yang-Mills case. Note that the helicity flow (arrow) denotes the helicity assignment (negative helicity for incoming and positive helicity for outgoing), hence we can freely drop the $\pm$ labels on external legs.

\subsection{Building on-shell diagrams}

On-shell diagrams are given by the product of three-point amplitudes, multiplied by a Jacobian that stems from eliminating $\lambda$'s and $\eta$'s of internal on-shell legs in the momentum and super-momentum conservation delta functions. The final result depends on external kinematic data and possibly on free unfixed parameters.

Putting an internal edge on-shell gives one constraint equation, from which we can solve for the internal momenta in terms of the external kinematic data. Denoting the difference between the number of constraints and the internal degrees of freedom by $n_\delta$, we get three distinct cases. $n_\delta=0$ where we can express all internal momenta in terms of external data, $n_\delta>0$ which places additional conditions on the external data, and $n_\delta<0$ which leaves us with unsolved parameters that the amplitude will depend on. The simplest example of $n_\delta=0$, is the four-point one-loop box diagram,
\begin{equation} \label{eq:box}
\raisebox{-15mm}{\includegraphics[trim={0cm 0cm 0cm 0cm},clip,scale=1.2]{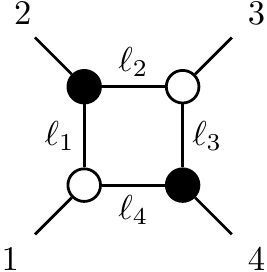}} 
\begin {aligned}\qquad\qquad
\ell_1&=\frac{\la 23 \ra}{\la 13 \ra}\lambda_1\widetilde{\lambda}_2,\\
\ell_2&=\frac{\la 12 \ra}{\la 13 \ra}\lambda_3\widetilde{\lambda}_2,
\end {aligned}
\qquad
\begin{aligned}
\ell_3&=\frac{\la 14 \ra }{\la 13 \ra}\lambda_3\widetilde{\lambda}_4,\\
\ell_4&=\frac{\la 34 \ra}{\la 13 \ra}\lambda_1\widetilde{\lambda}_4.
\end{aligned}
\end{equation}
You can see that each vertex imposes collinearity conditions on $\lambda$ or $\widetilde{\lambda}$. For example, the holomorphic part of momentum $\ell_2$ is given by the $\lambda_3$ spinor (from the upper right white vertex) and the anti-holomorphic part is given by the $\widetilde{\lambda}_2$ spinor (from the upper left black vertex). This itself solves for 3 of 4 on-shell conditions on internal on-shell propagators, $\ell_1^2=\ell_2^2=\ell_3^2=0$. The constraint from the final propagator $\ell_4^2=0$ fixes the overall constant in $\ell_2$ to be $\la12\ra/\la13\ra$. In ${\cal N}{=}4$ SYM theory we do not need to specify helicities of internal and external legs. For ${\cal N}{<}4$ SYM theory we have to specify them (or equivalently helicity flows denoted by arrows) to fully define an on-shell diagram. Solutions for internal momenta are the same regardless of the number of supersymmetries or theory (it can also be gravity on-shell diagrams or other). However, the actual on-shell function associated with the on-shell diagram is sensitive to these details. In particular, the on-shell function in ${\cal N}{=}4$ SYM for (\ref{eq:box}) is 
\begin{align}
    \Omega &= \int d^4\widetilde{\eta}_1\dots d^4\widetilde{\eta}_4 \int \frac{d^2\lambda_{\ell_1}d^2\widetilde{\lambda}_{\ell_1}}{{\rm GL}(1)} \dots  \frac{d^2\lambda_{\ell_4}d^2\widetilde{\lambda}_{\ell_4}}{{\rm GL}(1)}\label{four1}\\ &\hspace{2cm} \times \Bigg\{A_3(1,\ell_1,\ell_4)A_3(2,\ell_1,\ell_2)A_3(3,\ell_2,\ell_3)A_3(4,\ell_3,\ell_4)\Bigg\} = \frac{\delta^4(P)\,\delta^8(\mathcal{Q})}{\la 12 \ra \la 23 \ra \la 34 \ra \la 41 \ra}. \nonumber
\end{align}
All integrals are evaluated on the delta functions using
\begin{equation}
\int dx\,f(x)\,\delta(x-x_0) = f(x_0)
\end{equation}
(see \cite{Arkani-Hamed:2012zlh} for more details). All poles in the on-shell function correspond to sending one of the internal momenta of the diagram to zero as evident from (\ref{eq:box}). Taking the residue on this pole erases an edge and the result corresponds to a new on-shell diagram. For example, sending $\la12\ra=0$ implies $\ell_2=0$ and we get
\begin{equation}
\raisebox{-10mm}{\includegraphics[trim={0cm 0cm 0cm 0cm},clip,scale=1.2]{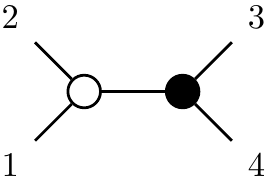}}
\begin {aligned}\quad
=\frac{\delta(\la 12 \ra )\delta^4(P)\delta^8(\mathcal{Q})}{\la 23 \ra \la 34 \ra \la 41 \ra} \label{four2}.
\end {aligned}
\end{equation}
This diagram is an example of $n_\delta>0$ where the extra constraint on the external kinematics $\la12\ra=0$ is evident from the delta function $\delta(\la12\ra)$ in the on-shell form (\ref{four2}). The condition on external kinematics can be also seen directly in the diagram where legs $1,2$ meet in the same white vertex. At the same time, legs $3,4$ meet in the same black vertex which imposes $[34]=0$ -- which is indeed a consequence of $\la12\ra=0$ for four-point kinematics. The simplest example of an on-shell diagram with $n_\delta<0$, is the four-point double-box diagram,
\begin{equation}
\raisebox{-15mm}{\includegraphics[trim={0cm 0cm 0cm 0cm},clip,scale=1.2]{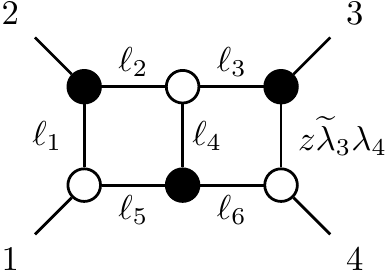}}
\begin {aligned}\qquad
\ell_1&=\frac{\la 23 \ra+z\la 24 \ra}{\la 13 \ra +z\la 14 \ra}\lambda_1\widetilde{\lambda}_2,\\
\ell_2&=\frac{\la 12 \ra}{\la 13 \ra +z\la 14 \ra}\widehat\lambda_3\widetilde{\lambda}_2,\\
\ell_3&=\widehat\lambda_3\widetilde{\lambda}_3,
    \end {aligned}\quad
\begin{aligned}
\ell_4&=\frac{\la 14 \ra}{\la 13 \ra+z\la 14 \ra}\widehat\lambda_3\widehat{\widetilde{\lambda}}_4,\\
\ell_5&=\frac{\la 34 \ra}{\la 13 \ra+z\la 14 \ra}\lambda_1\widehat{\widetilde{\lambda}}_4,\\
\ell_6&=\lambda_4\widehat{\widetilde{\lambda}}_4.
\end{aligned} \label{doublebox}
\end{equation}
where $\widehat{\widetilde{\lambda}}_4=\widetilde{\lambda}_4+z\widetilde{\lambda}_3$ and $\widehat{\lambda}_3=\lambda_3-z\lambda_4$. This diagram can be thought as a heptacut of the two-loop four-point amplitude in the ${\cal N}{=}4$ SYM theory, with $z$ being the last parameter in $\ell_1,\ell_2$ not localized by on-shell conditions from the cut. Alternatively, we can also interpret the diagram as a BCFW shift of the one-loop four-point amplitude (on legs $3,4$) with $z$ being the BCFW shift parameter. The on-shell function is given by
\begin{equation} \label{eq: doublebox}
    \Omega = \int\frac{dz\,\delta^4(P)\,\delta^8(\mathcal{Q})}{z\la 12 \ra (\la 23 \ra+z\la 24 \ra)\la 34 \ra \la 41 \ra}.
\end{equation}
Each pole corresponds to erasing an edge in the diagram but not all edges are erasable (hence we have five poles rather than seven). The residue on the pole $z=0$ corresponds the diagram \eqref{eq:box} already encountered.  

\subsection{Dual formulation}

In addition to calculating on-shell diagrams by treating them as the amalgamation of three-point amplitudes, there is a dual formulation where they can be viewed as differential forms on the positive Grassmannian space. To understand this picture better, let's consider the momentum conservation equation, namely:
\begin{equation}
    \delta^4(P) = \delta^4(\lambda \cdot \Tilde{\lambda}) = \delta^4(\lambda_1 \Tilde{\lambda}_1 + ... + \lambda_n \Tilde{\lambda}_n). \label{momcon}
\end{equation}
While this is a quadratic constraint, there is a way to re-write this condition in terms of two linear conditions for the $\lambda$'s and the $\Tilde{\lambda}$'s separately. To do so, we introduce a $k$-plane in $n$-dimensions represented by a $(k \times n)-$matrix (modded out by GL(k) since such row operations leave the $k$-plane invariant). The space of such planes is denoted $G(k,n)$ and is known as the Grassmannian. We will call the $(k \times n)$ matrix we use to represent a point in the Grassmannian space the $C$-matrix. Then we can re-write the momentum conservation equation geometrically as follows:
\begin{align}
    \delta(C\cdot Z)=\delta
    ^{((n-k)\times 2)}(C^\perp\cdot \lambda)\delta^{(k\times 2)}(C\cdot \widetilde\lambda)\delta^{(k\times \mathcal{N})}(C\cdot \widetilde\eta), \label{delta}
\end{align}
where $C^\perp$ is an orthogonal complement to $C$ satisfying $C^\perp\cdot C=0$.
We can think about this geometrically in the $n$-dimensional ``particle space" \cite{Arkani-Hamed:2009ljj}. In this space $\lambda$ and $\widetilde{\lambda}$ are 2-planes which are orthogonal to each other (\ref{momcon}). We define a $k$-dimensional plane $C$ which contains $\lambda$ and is orthogonal to $\widetilde{\lambda}$,
\begin{equation}
\raisebox{-15mm}{\includegraphics[trim={0cm 0cm 0cm 0cm},clip,scale=1.1]{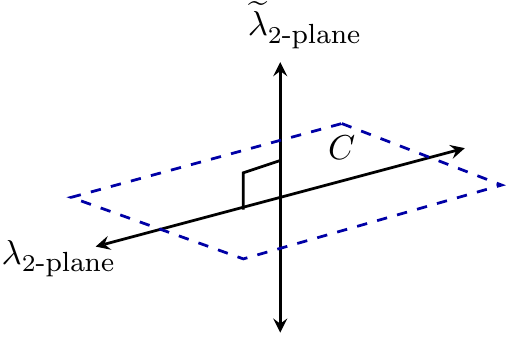}}.
\end{equation}
The $C$ matrix is then used to linearize the momentum conservation but conditions (\ref{delta}) are in fact stronger, $C\cdot\widetilde{\lambda} = C^\perp\cdot\lambda=0$ impose $2n$ conditions on the parameters of $C$. This reproduces four conditions from the momentum conservation and leaves us with additional $2n{-}4$ additional conditions. Depending on the exact dimensionality of the $C$-matrix, we can be left with some free parameters in $C$ (matrix is not completely fixed) or oppositely in order to satisfy (\ref{delta}) we need to impose additional conditions on external kinematics. Note that the last delta function in (\ref{delta}) involves fermionic variables $\widetilde{\eta}$ and it does not impose any conditions on $C$ -- we should just think about it as a polynomial in $\widetilde{\eta}$. 

Each on-shell diagram provides a particular parametrization of the $C$-matrix using edge variables $\alpha_i$, or face variables $f_i$. For planar diagrams, if the variables are real and take definite signs, all the main $(k\times k)$ minors are positive, and the $C$ matrix represents a cell in the {\it positive Grassmannian} $G_+(k,n)$. The connection between cells in positive Grassmannian and plabic graphs (name used in the math literature) was found by Postnikov in \cite{Postnikov:2006kva}, and later the connection to physics and scattering amplitudes was discovered in \cite{Arkani-Hamed:2012zlh}. Here, we just shortly review the key ingredients. 

For a given on-shell diagram, we choose an {\it orientation} by assigning arrows to all edges such that each black vertex has two incoming and one outgoing leg, while each white vertex has one incoming and two outgoing edges. Then we assign edge variables to all edges, fixing in each vertex one variable to one. It does not matter how we exactly choose the position of edge variables (fixing the GL(1) redundancy) but in each vertex there should be at least one edge with an edge variable. The diagram has $k$ sources, external legs with incoming arrows, and $n{-}k$ sinks, external legs with outgoing arrows. Then the element of the $(k\times n)$ matrix is given by the following prescription:
\begin{align}
    C_{\alpha a}=\sum_{\Gamma_{\alpha\to a}}\prod_j \alpha_j.
\end{align}
For each path from a source $\alpha$ to a sink $a$ we take a product of all edge variables along the path and sum over all paths. The element $C_{\alpha\alpha}=1$ while $C_{\alpha\alpha'}=0$ for the path between two different sources. For a simple, one-loop four-point box diagram we have
\begin{equation} \label{eq:4ptbox}
    \raisebox{-15mm}{\includegraphics[trim={0cm 0cm 0cm 0cm},clip,scale=1.2]{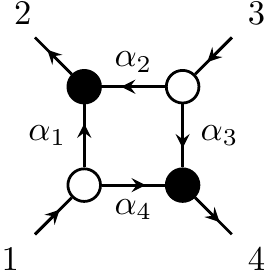}}
    \begin{aligned}
        \quad\Rightarrow \quad 
        C=\begin{pmatrix}
        1 & \alpha_1 & 0 & \alpha_4 \\
        0 & \alpha_2 & 1 & \alpha_3
        \end{pmatrix},\quad 
        C^{\perp}=\begin{pmatrix}
        -\alpha_1 & 1 & -\alpha_2 & 0 \\
        -\alpha_4 & 0 & -\alpha_3 & 1
        \end{pmatrix}.
    \end{aligned}
\end{equation}
Fixing the signs of edge variables to be
\begin{align}
\begin{array}{ccccccccc}
    \alpha_1>0\,, & \alpha_2>0\,, & \alpha_3>0\,, & \alpha_4 < 0\,,
\end{array}
\end{align}
guarantees that all $(2\times2)$ minors are positive and the $C$-matrix represents a top cell of $G_+(2,4)$. The on-shell function associated with the same on-shell diagram in ${\cal N}=4$ SYM theory is given by 
\begin{equation}
    \Omega = \int \prod_i \frac{d \alpha_i}{\alpha_i}\,\delta(C {\cdot} Z), \label{dual}
\end{equation}
where we are instructed to solve for the $\alpha_i$ variables from the delta functions which fix them as functions of external kinematics. For our example, the delta-function constraint on $C^\perp$ gives the following equations
\begin{equation}
		\delta^4(C^\perp{\cdot}\lambda)=
		\delta^2(-\alpha_1\lambda_1+\lambda_2-\alpha_3\lambda_3)
	    \delta^2(-\alpha_4\lambda_1-\alpha_2\lambda_3+\lambda_4),
\end{equation}
which after contracting with $\lambda_1$ and $\lambda_3$ turns into 4 equations $\prod_i^4\delta(E_i)$, that can then be rewritten as
\begin{equation}
J_\delta\times \langle 13 \rangle^2 \times  \delta\left(\alpha_2-\frac{\langle  1 2 \rangle}{\langle  1 3 \rangle} \right)
\delta\left(\alpha_1-\frac{\langle 2 3 \rangle}{\langle 1 3 \rangle}\right) \delta\left(\alpha_3-\frac{\langle 1 4 \rangle}{\langle 1 3  \rangle}\right) \delta\left(\alpha_4-\frac{\langle 4 3 \rangle}{\langle 1 3  \rangle}\right).
\end{equation}
Here $J_\delta\equiv \left| \frac{\partial E_{i}}{\partial \alpha_j}\right|^{-1}=\frac{1}{\langle13\rangle ^4}$ is the Jacobian that stems from solving the four $\delta$-function equations, and the factor $\langle 13 \rangle^2$ stems from the contraction with $\lambda_1$ and $\lambda_3$. Plugging the solutions for the edge-variables into the other $\delta$-function, one finds:
\begin{equation}
 \begin{aligned}
    \delta^4(C\cdot \widetilde\lambda)&
    =\delta^2\left(\widetilde\lambda_1+\frac{\ang{23}}{\ang{13}}\widetilde\lambda_2+\frac{\ang{43}}{\ang{13}}\widetilde\lambda_4\right)
    \delta^2\left(\frac{\ang{12}}{\langle 13 \rangle }\widetilde\lambda_2+ \widetilde\lambda_3+\frac{\ang{14}}{\langle 13 \rangle}\widetilde\lambda_4\right)\\
     &=\langle13 \rangle^4\,\delta^2\left(\langle 13 \rangle\widetilde\lambda_1+\ang{23}\widetilde\lambda_2+\langle 43 \rangle \widetilde\lambda_4\right)
    \delta^2\left(\langle 12 \rangle\widetilde\lambda_2+ \langle 13 \rangle\widetilde\lambda_3+\langle14 \rangle\widetilde\lambda_4\right).
 \end{aligned}
\end{equation}
The two equations can be obtained from a single momentum conservation equation by contracting with $\lambda_3$ and $\lambda_1$ respectively, i.e. we have,
\begin{equation}
	\begin{aligned}
		\delta^4(C\cdot \widetilde\lambda)=\langle13\rangle^2 \delta^{4}(\lambda_1\widetilde\lambda_1+\lambda_2\widetilde\lambda_2+\lambda_3\widetilde\lambda_3+\lambda_4\widetilde\lambda_4)\equiv\langle13\rangle^2\, \delta^{4}(P).
	\end{aligned}
\end{equation}
For the last delta-function we get the same thing except for replacing $\widetilde \lambda_i\to \widetilde \eta_i$,
\begin{equation}
	\begin{aligned}
			\delta^{8}(C\cdot \widetilde\eta)=\frac{1}{\langle 13 \rangle ^4}\delta^{8}(\lambda_1\widetilde\eta_1+\lambda_2\widetilde\eta_2+\lambda_3\widetilde\eta_3+\lambda_4\widetilde\eta_4)\equiv \frac{1}{\langle 13 \rangle ^4} \delta^{8}(\mathcal{Q}).
	\end{aligned}
\end{equation}
We are now in a position to calculate the on-shell function $\Omega$, 
\begin{align} \label{eq:4ptMHV}
\Omega &=\int \frac{d\alpha_1\dots d\alpha_4\,\delta^{8}(\mathcal{Q})\delta^{4}(P)}{\alpha_1\alpha_2\alpha_3\alpha_4\langle 13 \rangle^{4}}\delta\left(\alpha_2-\frac{\langle  1 2 \rangle}{\langle  1 3 \rangle} \right)
\delta\left(\alpha_1-\frac{\langle 2 3 \rangle}{\langle 1 3 \rangle}\right) \delta\left(\alpha_3-\frac{\langle 1 4 \rangle}{\langle 1 3  \rangle}\right) \delta\left(\alpha_4-\frac{\langle 4 3 \rangle}{\langle 1 3  \rangle}\right)\nonumber\\
		&\hspace{9cm} =
		\frac{\delta^{8}(\mathcal{Q})\,\delta^{4}(P)}{\ang{12}\ang{23}\ang{34}\ang{41}},
\end{align}
which reproduces the on-shell function (\ref{four1}) from the previous subsection. The only poles in the on-shell function $\Omega$ come from the edge variables being sent to $\alpha_i=0$ or $\alpha_i=\infty$, which removes an edge and imposes conditions on external kinematics, in complete agreement with our earlier discussion. Note that flipping the orientation of an edge variable also flips $\alpha \leftrightarrow 1/\alpha$ and thus flips the pole at $0$ and $\infty$. Let us now consider another on-shell diagram,
\begin{equation}
    \raisebox{-15mm}{\includegraphics[trim={0cm 0cm 0cm 0cm},clip,scale=1.2]{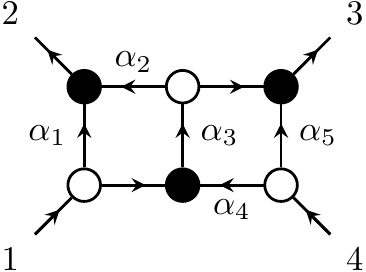}}
\begin{aligned}
\qquad
C=\begin{pmatrix}
1 & \alpha_1{+}\alpha_2\alpha_3 & \alpha_3 & 0\\
0 & \alpha_2 \alpha_3 \alpha_4 & \alpha_5 {+} \alpha_3 \alpha_4 & 1
\end{pmatrix}.
\end{aligned}
\end{equation}
The edge variables $\alpha_i$'s are given by,
\begin{equation}
    \alpha_1 = \frac{\la 23 \ra{-}\alpha_5\la 24 \ra}{\la 13 \ra{-}\alpha_5\la 14 \ra},~~~\alpha_2 = \frac{\la 12 \ra}{\la 13 \ra{-}\alpha_5\la 14 \ra},~~~\alpha_3 = \frac{\la 34 \ra}{\la 1 4 \ra },~~~\alpha_4 = \frac{\la 13\ra{-}\alpha_5\la 14 \ra}{\la 34 \ra}.
\end{equation}
The Jacobian from solving the $\delta$-functions is $J=\frac{1}{( \la 13 \ra-\alpha_5 \la 14 \ra)\la 34 \ra}$. From (\ref{dual}) we get
\begin{equation}
    \Omega = \int\frac{d\alpha_5\,\delta^4(P)\delta^8(\mathcal{Q})}{\alpha_5\la 12 \ra (\la 23\ra-\alpha_5\la 24 \ra)\la 3 4 \ra \la14\ra},
\end{equation}
which we recognize as \eqref{eq: doublebox} with $\alpha_5\to -z$. In some cases the perfect orientation can involve an infinite loop. For example, if we choose the external arrows of the box diagram in the following way:
\begin{equation} \label{eq:4ptbox-cycle}
    \raisebox{-15mm}{\includegraphics[trim={0cm 0cm 0cm 0cm},clip,scale=1.2]{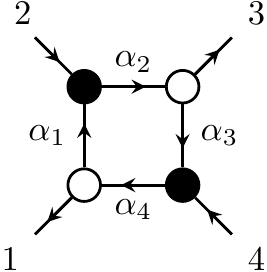}}
\begin{aligned}
\qquad
C=\begin{pmatrix}
\alpha_2 \alpha_3 \alpha_4 \delta & 1 & \alpha_2 \delta & 0\\
\alpha_4 \delta & 0 & \alpha_1 \alpha_2 \alpha_3 \delta & 1
\end{pmatrix}.
\end{aligned}
\end{equation}
The $\delta$ that appears here corresponds to a geometric series coming from an infinite loop, 
\begin{equation}
    \delta = \sum^\infty_{j = 0} (\alpha_1 \alpha_2 \alpha_3 \alpha_4)^j = \frac{1}{1-\alpha_1 \alpha_2 \alpha_3 \alpha_4}.
\end{equation}
Solving for edge variables we get

\begin{equation}
	\begin{aligned}
		&\alpha_1=\frac{\la 23 \ra}{\la 13 \ra},~~~
		\alpha_2=\frac{\la 13 \ra }{\la 12 \ra},~~~
		\alpha_3=\frac{\la 41 \ra}{\la 13 \ra},~~~
		\alpha_4=\frac{\la 13 \ra}{\la 34 \ra}\quad \mbox{and also}\quad \delta = \frac{\la12\ra\la34\ra}{\la13\ra\la24\ra}
	\end{aligned}
\end{equation}
and plugging in (\ref{dual}) we get the same formula as in \eqref{eq:4ptMHV}. Note that in all these cases we considered on-shell diagrams in ${\cal N}=4$ SYM theory which do not depend on the assignment of helicities. Hence, the formulas for diagrams (\ref{eq:4ptbox}) and (\ref{eq:4ptbox-cycle}) gave the same on-shell functions. For ${\cal N}<4$ SYM this is not the case and the helicity assignments are important.

\subsection{On-shell diagrams for \texorpdfstring{${\cal N}{<}4$}{N less than 4} SYM theory}

If the theory has less than the maximal supersymmetry, we do need to assign both external and internal helicities to fully specify the on-shell diagram. The same on-shell diagrams with different helicity assignments correspond to the same kinematics (solutions for momenta, edge variables etc) but they are given by different on-shell functions. The on-shell function for the ${\cal N}{<}4$ SYM on-shell diagram \cite{Arkani-Hamed:2012zlh} is 
\begin{equation}
	\begin{aligned}
		\Omega= \int \dlog{\alpha_1} \dlog{\alpha_2}\cdots  \dlog{\alpha_m} \,\mathcal{J}^{\mathcal{N}-4}\,\,\delta(C\cdot Z),\label{withJ}
	\end{aligned}
\end{equation}
where the Jacobian ${\cal J}$ is given by
\begin{equation}
	\begin{aligned}
		\mathcal{J}=1+\sum_if_i+\sum_{\substack{\text{disjoint}\\ \text{pairs } i,j}}f_i f_j+\sum_{\substack{\text{disjoint}\\ \text{pairs } i,j,k}}f_i f_j f_k+\cdots\,\,,
	\end{aligned}
\end{equation}
with $f_i$ denoting the face of a closed cycle, and the sums are over disjoint collections of these closed cycles. We mostly work with edge-variables, and each face $f_i$ can be defined as a clockwise-oriented product of edge-variables. Let us compare the results for two different orientations of the one-loop box diagram. For the diagram with no internal loop (\ref{eq:4ptbox}) we have ${\cal J}=1$ and
\begin{equation}
    \Omega = \frac{\la 13\ra^{4-{\cal N}}\,\delta^{4}(P)\,\delta^{2{\cal N}}({\cal Q})}{\la12\ra\la23\ra\la34\ra\la41\ra}.
\end{equation}
Note that this result is identical up to a trivial helicity factor, $\delta^{8}({\cal Q})\rightarrow \la13\ra^{4-{\cal N}}\,\delta^{2{\cal N}}({\cal Q})$, to the ${\cal N}=4$ on-shell function (\ref{eq:4ptMHV}). The second diagram (\ref{eq:4ptbox-cycle}) has an internal loop and the Jacobian reads 
\begin{equation}
    \mathcal{J} = 1 - \alpha_1 \alpha_2 \alpha_3 \alpha_4  = \frac{\la13\ra\la24\ra}{\la12\ra\la34\ra}.
\end{equation}
The on-shell function is then given by
\begin{equation}
    \Omega = \frac{\la 24 \ra^{4-\mathcal{N}}\delta^{4}(P)\delta^{2{\cal N}}({\cal Q})}{\la 12 \ra \la 23 \ra \la 34 \ra \la 41 \ra} \times \left( \frac{\la 12 \ra \la 34 \ra}{\la 13 \ra \la 24 \ra}\right)^{4-\mathcal{N}}\label{onshell4} = \frac{(\la 12\ra\la34\ra)^{4-{\cal N}}\,\delta^4(P)\delta^{2{\cal N}}({\cal Q})}{\la 12\ra\la23\ra\la34\ra\la41\ra\la13\ra^{4-{\cal N}}}.
\end{equation}
Note that the on-shell function (\ref{onshell4}) has a new pole for ${\cal N}{<}4$ located at $\la13\ra=0$. Looking back to (\ref{eq:box}) this sends $\ell_1\rightarrow\infty$, hence we refer to it as a \emph{pole at infinity} or a \emph{UV pole}. This pole is substantially different from all other poles which correspond to erasing edges in the on-shell diagram -- these poles correspond to sending the associated loop momenta to zero, we will refer to them as \emph{IR poles}. The presence of the Jacobian in (\ref{withJ}) dramatically changes the pole structure of the on-shell function $\Omega$. We will often divide the form into the bare function $\Omega^{\rm bare}$ and the appropriate power of the Jacobian,
\begin{equation}
    \Omega = \Omega^{\rm bare} \times {\cal J}^{{\cal N}-4}.
\end{equation}
The IR poles come from $\Omega^{\rm bare}$ while the UV poles always come from the Jacobian part. The role of UV poles is the main topic of this paper and we will elaborate on it much more in the next sections. Before diving into this discussion, it is worth noting one important point. For ${\cal N}{=}4$ SYM theory each on-shell diagram directly corresponds to a cut of the loop integrand. For ${\cal N}{<}4$ SYM this is a collection of on-shell diagrams in general: with fixed helicities of external legs we have to sum over all possible orientations of internal legs. For the one-loop example, if legs $1,3$ are incoming there is one orientation and a quadruple cut of the one-loop four-point ${\cal N}<4$ SYM integrand is given by a single diagram, 
\begin{equation} \label{eq:boxcut}
    \raisebox{-15mm}{\includegraphics[trim={0cm 0cm 0cm 0cm},clip,scale=1.1]{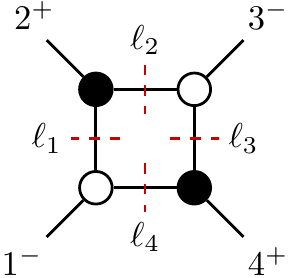}}\quad = \quad \raisebox{-15mm}{\includegraphics[trim={0cm 0cm 0cm 0cm},clip,scale=1.1]{Figures/Box2.pdf}}.
\end{equation}
For legs $2,4$ incoming there are two possible internal orientations,
\begin{equation} \label{boxcut2}
    \raisebox{-15mm}{\includegraphics[trim={0cm 0cm 0cm 0cm},clip,scale=1.1]{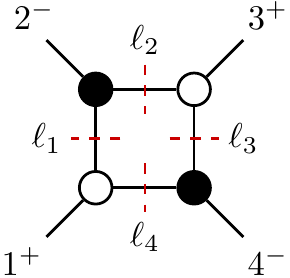}} ~~ = ~~ \raisebox{-15mm}{\includegraphics[trim={0cm 0cm 0cm 0cm},clip,scale=1.1]{Figures/Cyclebox.pdf}}~~ +~~ \raisebox{-15mm}{\includegraphics[trim={0cm 0cm 0cm 0cm},clip,scale=1.1]{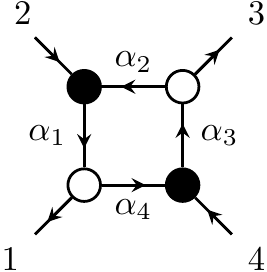}}.
\end{equation}
This is exactly the case of on-shell diagrams with internal loops as we always have two different orientations (clockwise and anti-clockwise). While in a cut they only appear in the sum, we can still treat these on-shell diagrams as independent well-defined objects and study their individual properties. For example, in the case of the amplitude $A_4(1^+2^-3^+4^-)$ (legs $2,4$ incoming) in ${\cal N}{=}3$ SYM each diagram has a single pole at infinity, located at $\la13\ra=0$ while the cut does not have this pole. Plugging (\ref{withJ}) for ${\cal N}=3$ we get,
\begin{equation*}
    {\rm Cut}\,A_4(1^+2^-3^+4^-) = \frac{\la12\ra\la23\ra\,\delta^4(P)\delta^{6}(Q)}{\la12\ra\la23\ra\la34\ra\la41\ra\la13\ra} + \frac{\la23\ra\la14\ra\,\delta^4(P)\delta^{6}({\cal Q})}{\la12\ra\la23\ra\la34\ra\la41\ra\la13\ra} = \frac{\la 24\ra\,\delta^4(P)\delta^{6}({\cal Q})}{\la12\ra\la23\ra\la34\ra\la41\ra}.
\end{equation*}
This is a general property: cuts of ${\cal N}{=}3$ SYM amplitudes are identical to their ${\cal N}{=}4$ SYM counterparts (up to trivial helicity factors) while ${\cal N}{=}3$ SYM on-shell diagrams are individually different and have poles at infinity. On a ${\cal N}{=}3$ cut we sum over contributing on-shell diagrams and any pole at infinity has to cancel.

\section{UV poles in one-loop graphs}

The bare on-shell functions describe the dynamics of scattering amplitudes in the IR sector when the (cut) loop momenta are small. All poles in $\Omega^{\rm bare}$ correspond to setting one of these momenta to zero and erasing an edge in the diagram, as a consequence of the perturbative unitarity. In the ${\cal N}{=}4$ SYM theory the on-shell function is given by $\Omega^{\rm bare}$, and hence we only have IR poles. This is true for both planar and non-planar sectors of the theory. In the planar sector, this is also a consequence of the dual conformal symmetry, while the symmetry explanation of the absence of poles at infinity in the non-planar sector is yet-to-be found. Note that the absence of these poles is related, but not identical, to the statement of the UV finiteness of final (integrated) scattering amplitudes in ${\cal N}{=}4$ SYM theory. In fact, the absence of poles at infinity in on-shell diagrams is a stronger property than the UV finiteness of the S-matrix, see \cite{Arkani-Hamed:2014via,Bern:2014kca,Bourjaily:2018omh} for more details.

For ${\cal N}{<}4$ SYM theory, the bare function $\Omega^{\rm bare}$ is decorated by a power of the Jacobian ${\cal J}$ as outlined in the previous section. The Jacobian changes the pole structure: it removes some of the existing IR poles, and adds new UV poles. Note that the UV poles are not fixed by the (standard) unitarity -- unlike IR poles where tree-level factorization or unitarity cuts fully fix corresponding residues. As a result, we can not naively expect that the residue on a UV pole, where the loop momentum goes to infinity, is an operation on an on-shell diagram. In the following, we will show that despite these assumptions the real situation is quite contrary: the UV pole can be understood as a (rather simple) diagrammatic operation, not erasing an edge, but shrinking a corresponding loop (which blows up on the pole) and rejoining internal legs via a ``non-planar twist''. As a result, the residue on the UV pole is generally a lower-loop \emph{non-planar} on-shell diagram. In our discussion, we first focus on the ${\cal N}{=}3$ case, where the on-shell function contains only one power of ${\cal J}$, and the pole at infinity in $\Omega$ is always a simple pole. For general ${\cal N}$ we get higher order poles at infinity and the residues correspond to derivatives of an on-shell function associated with the diagram.

\subsection{First \texorpdfstring{${\cal N}{=}3$}{N equal to 3} examples}

The simplest example is the one-loop box on-shell diagram with an internal loop discussed earlier. The on-shell function is given by,
\begin{equation}
        \raisebox{-12mm}{\includegraphics[trim={0cm 0cm 0cm 0cm},clip,scale=1.1]{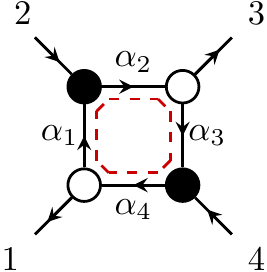}}
\begin{aligned}
    = \underbrace{\frac{\delta^4(P)\delta^6({\cal Q})\la 24 \ra }{\la 12 \ra \la 23 \ra \la 34 \ra \la 14 \ra}}_{\Omega^{\rm bare}} \times \underbrace{\left( \frac{\la 12 \ra \la 34 \ra }{\la 13 \ra \la 24 \ra}\right)}_{{\cal J}^{-1}} = \frac{\delta^4(P)\delta^6({\cal Q})}{\la 23 \ra \la 14 \ra \la 13\ra}.
\end{aligned} \label{four4N3}
\end{equation}
where the Grassmann variables are $\eta_i^I$, $I=1,2,3,$ and hence the supermomentum conservation is $\delta^6({\cal Q})$. We can see that the Jacobian ${\cal J}$ introduces two new poles, $\la13\ra$ and $\la24\ra$ and removes poles $\la12\ra$ and $\la34\ra$. The absence of poles $\la12\ra$ and $\la34\ra$ is caused by the fact that corresponding edges are no longer removable. For example, removing edge $\alpha_2$ sends $\la12\ra=0$ which leads to an illegal helicity flow
\begin{equation}
        \raisebox{-10mm}{\includegraphics[trim={0cm 0cm 0cm 0cm},clip,scale=1.1]{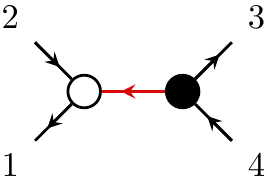}}
\end{equation}
as we have a wrong number of incoming/outgoing legs in black/white vertices. This pole was perfectly legal pole in the ${\cal N}{=}4$ SYM case but now the Jacobian removes it. Same is true for the pole $\la23\ra=0$ and the edge $\alpha_4$. Next, the pole $\la24\ra$ is canceled by the helicity factor in the numerator of $\Omega$ and only $\la13\ra=0$ stays as a new valid pole. As noted before, this is a pole at infinity as it sends the on-shell loop momentum $\ell_1\rightarrow\infty$. While this pole does not erase any edge, we can easily see what type of kinematical constraint it imposes: 
\begin{equation}
    \la13\ra=0 \,\leftrightarrow\, \lambda_1\sim\lambda_3,\,\,\mbox{and by momentum conservation}\,\,\,[24]=0\,\leftrightarrow\,\widetilde{\lambda}_2\sim\widetilde{\lambda}_4.
\end{equation}
But this is nothing else than the factorization diagram,
\begin{equation}
        \raisebox{-9mm}{\includegraphics[trim={0cm 0cm 0cm 0cm},clip,scale=1.1]{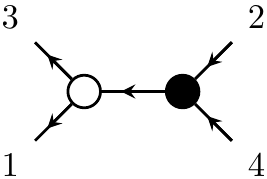}}
\begin{aligned}
    =\frac{\delta^4(P)\delta^6({\cal Q})\delta(\la 13 \ra)}{\la 23 \ra \la 14 \ra}. \label{resinf1}
\end{aligned}
\end{equation}
The internal momentum flow is here fixed by the number of incoming/outgoing legs in white/black vertex. Note that our procedure was based solely on the analysis of the kinematical conditions between external legs, i.e. we imposed $\la13\ra=0$ and searched for an on-shell diagram which materializes that. But very importantly, the on-shell function for this diagram (\ref{resinf1}) is equal to the residue of the original diagram (\ref{four4N3}) on the pole $\la13\ra=0$. In this case it might look a bit trivial as we only replaced $1/\la13\ra \rightarrow \delta(\la13\ra)$ but the same procedure is valid for the general case.

Checking the equality of the residue of the original on-shell diagram evaluated on the pole with the new on-shell diagram (with extra delta function imposed) might sometimes be tricky. For example, if we consider the same diagram with the internal loop and the opposite orientation, we get
\begin{equation}
        \raisebox{-12mm}{\includegraphics[trim={0cm 0cm 0cm 0cm},clip,scale=1.1]{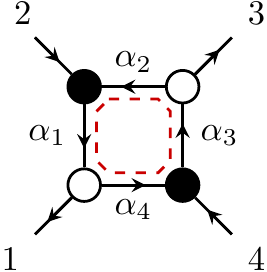}}
\begin{aligned}
    = \underbrace{\frac{\delta^4(P)\delta^6({\cal Q})\la 24 \ra}{\la 12 \ra \la 23 \ra \la 34 \ra \la 14 \ra}}_{\Omega^{\rm bare}} \times \underbrace{\left( \frac{\la 23 \ra \la 14 \ra }{\la 13 \ra \la 24 \ra}\right)}_{{\cal J}^{-1}} = \frac{\delta^4(P)\delta^6({\cal Q})}{\la 12 \ra \la 34 \ra \la13\ra}.
\end{aligned} \label{cyclebox5}
\end{equation}
Taking the residue on the pole $\la13\ra=0$ we arrive at the same diagram as before,
\begin{equation}
        \raisebox{-9mm}{\includegraphics[trim={0cm 0cm 0cm 0cm},clip,scale=1.1]{Figures/Factorization2.pdf}}
\begin{aligned}
    =\frac{\delta^4(P)\delta^6({\cal Q})\delta(\la 13 \ra)}{\la 12 \ra \la 34 \ra}. \label{resinf2}
\end{aligned}
\end{equation}
As you can see the on-shell functions look different in (\ref{resinf1}) and (\ref{resinf2}) but they are the same up to a sign, as $\la12\ra\la34\ra=\la23\ra\la14\ra$ for $\la13\ra=0$. In fact, if we sum (\ref{resinf1}) and (\ref{resinf2}) we get zero which is consistent with ${\cal N}=3$ SYM cuts not having any poles at infinity.

Next, let us consider the four-point double box on-shell diagram we discussed in the previous section. There are multiple different orientations of the same diagram which lead to different structures of poles at infinity. We can see from the parametrization of the loop momenta (\ref{doublebox}) that the left loop blows up for $z=-\la13\ra/\la14\ra$, while the right loop momentum goes to infinity for $z=\infty$. The accessibility of these poles depends on a particular orientation and the presence of closed loops. For the following orientation 
\begin{equation}
    \raisebox{-14mm}{\includegraphics[trim={0cm 0cm 0cm 0cm},clip,scale=1.1]{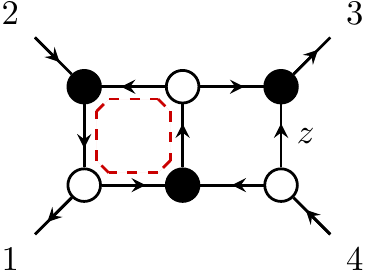}}
    \begin{aligned}
    &=\int dz \underbrace{\frac{\la 2 4 \ra \delta^4(P)\delta^6(Q)}{z(\la 2 3 \ra + z\la 24 \ra)\la 1 4 \ra \la 3 4 \ra \la 1 2 \ra}}_{\Omega^{\rm bare}} \times \underbrace{\frac{\la 1 4 \ra ( \la 2 3 \ra + z\la 2 4 \ra)}{\la 2 4 \ra (\la 1 3 \ra + z\la 1 4 \ra)}}_{{\cal J}^{-1}}\\
    &\hspace{3cm} = \int dz \frac{\delta^4(P)\delta^6({\cal Q})}{z\la34\ra\la12\ra(\la13\ra+z\la14\ra)},\label{twoloop1}
    \end{aligned}
\end{equation}
we have a closed cycle in the left loop and the on-shell function has a pole at $z=-\la13\ra/\la14\ra$. A different orientation has a closed cycle in the right loop and a pole at $z=\infty$,
\begin{equation}
     \raisebox{-15mm}{\includegraphics[trim={0cm 0cm 0cm 0cm},clip,scale=1.1]{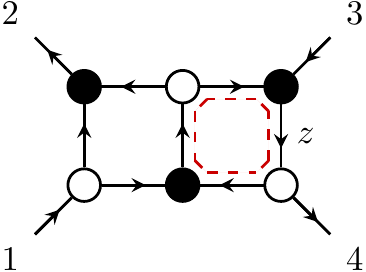}}
     \begin{aligned}
    &=\int dz \frac{\la 1 3 \ra \delta^4(P)\delta^6({\cal Q})}{z(\la 2 3 \ra + z\la 2 4 \ra)\la 1 4 \ra \la 3 4 \ra \la 1 2 \ra} \times \frac{z\la 1 4 \ra }{ \la 1 3 \ra}\\
    &\hspace{2.5cm} = \int dz\frac{\delta^4(P)\delta^6({\cal Q})}{\la12\ra\la34\ra(\la23\ra+z\la24\ra)}
     \end{aligned}\label{twoloop2}
\end{equation}
Finally, we can have an orientation with two internal cycles and both poles in $\Omega$,
\begin{equation} 
     \raisebox{-14mm}{\includegraphics[trim={0cm 0cm 0cm 0cm},clip,scale=1.1]{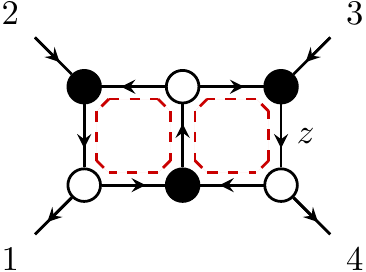}}
   \begin{aligned}
    &=\int dz \frac{\la 2 3 \ra \delta^4(P)\delta^6(Q)}{  z( \la 23 \ra + z\la 2 4 \ra)\la 1 2 \ra \la 1 4 \ra \la 3 4 \ra}
    \times \frac{z \la 1 4 \ra (\la 2 3 \ra + z\la 2 4 \ra)}{\la 2 3 \ra ( \la 1 3 \ra + z\la 1 4 \ra)}
    \\ 
    &\vspace*{-1cm}\hspace{3cm} = \int dz\frac{\delta^4(P)\delta^6({\cal Q})}{\la12\ra \la34\ra (\la 13\ra+z\la 14\ra)}
\label{twoloop3}
\end{aligned}
\end{equation}
Note that in this case there is no IR pole present as erasing any edge would lead to a contradiction with the helicity assignment. As in the previous case, we want to interpret the residue on the pole at infinity as another (now one-loop) on-shell diagram. As both cycles are just one-loop on-shell diagrams, we can treat them locally, see section \ref{sec:planar-proof} for proof of this. This means that if we blow up the left loop we need to replace the box on-shell diagram with external legs $1,2,\ell_3,\ell_6$ by the tree-level factorization diagram with legs $2,\ell_6$ in the black vertex and $1,\ell_3$ in the white vertex,
\begin{equation} 
     \raisebox{-15mm}{\includegraphics[trim={0cm 0cm 0cm 0cm},clip,scale=1.15]{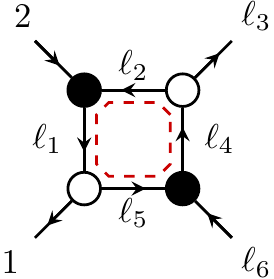}}\qquad
     \longrightarrow\qquad
    \raisebox{-15mm}{\includegraphics[trim={0cm 0cm 0cm 0cm},clip,scale=1.15]{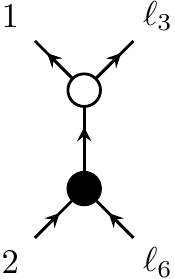}}
    \label{eq:Box6}
\end{equation}
Making this replacement in (\ref{twoloop2}) we get 
\begin{equation} \label{eq:doublebox-res}
         \raisebox{-15mm}{\includegraphics[trim={0cm 0cm 0cm 0cm},clip,scale=1.15]{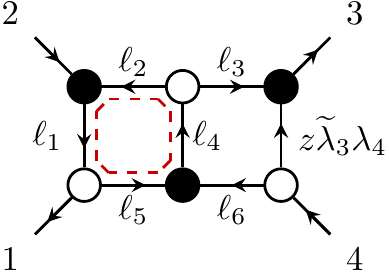}}~~
         \xrightarrow[z=-\frac{\la13\ra}{\la14\ra}]{}
     \raisebox{-15mm}{\includegraphics[trim={0cm 0cm 0cm 0cm},clip,scale=1.15]{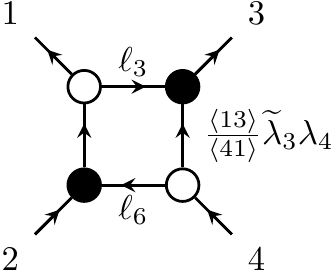}}
     \begin{aligned}
      ~  = ~ \frac{\delta^4(P)\delta^6({\cal Q})}{\ang{12}\ang{13}\ang{34}} 
    \end{aligned}
\end{equation}
The same expression for the on-shell function can be obtained by taking the residue of (\ref{twoloop1}) on the $z=-\la13\ra/\la14\ra$ pole. If we blow up the same loop in the third diagram (\ref{twoloop3}), we get a one-loop diagram with a closed cycle,
\begin{equation}
     \raisebox{-15mm}{\includegraphics[trim={0cm 0cm 0cm 0cm},clip,scale=1.15]{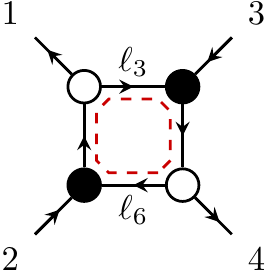}}
     \begin{aligned}
       = \frac{\delta^4(P)\delta^6({\cal Q})\la 23 \ra}{\la 12 \ra \la 13 \ra \la 24 \ra \la 34 \ra} \times \left( \frac{\la 12 \ra \la 34 \ra}{\la 23 \ra \la 14 \ra}\right) = \frac{\delta^4(P)\delta^6({\cal Q})}{\la23\ra\la14\ra\la13\ra}
    \end{aligned}
\end{equation}
which is precisely equal to the residue of (\ref{twoloop3}) on the $z=-\la13\ra/\la14\ra$ pole. We can do the same analysis for the $z=\infty$ pole, again finding an agreement between the new on-shell diagram and the residue of the original on-shell function (\ref{twoloop3}).

\subsection{Two-loop pentabox diagram}

Let us now look at a more complicated example: five-point MHV pentabox on-shell diagram. Solving for internal on-shell momenta gives
\begin{equation} \label{eq:pentabox}
\raisebox{-20mm}{\includegraphics[trim={0cm 0cm 0cm 0cm},clip,scale=1.15]{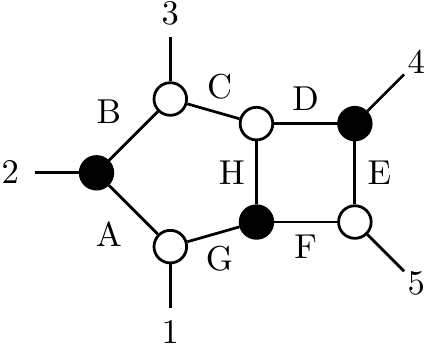}}
\begin {aligned}\qquad
\ell_A&=\frac{\la 23 \ra}{\la 13 \ra}\lambda_1 \Tilde{\lambda}_2, \,\,\, \ell_E =\frac{\la 34 \ra}{\la 35 \ra} \lambda_5 \Tilde{\lambda}_4,\\
\ell_B&=\frac{\la 12 \ra}{\la 13 \ra} \lambda_3 \Tilde{\lambda}_2, \,\,\, \ell_F =\frac{\lambda_5 \, (Q_{12}\cdot \lambda_3)}{\la 35 \ra},\\
\ell_C&=\frac{\lambda_3 \, (Q_{23} \cdot \lambda_1)}{\la 13 \ra}, \,\,\, \ell_G=\frac{\lambda_1 \, (Q_{12}\cdot \lambda_3)}{\la 13 \ra},\\
\ell_D&=\frac{\la 45 \ra}{\la 35 \ra} \lambda_3 \Tilde{\lambda}_4, \,\,\, \ell_H=\frac{\la 15 \ra}{\la 13 \ra \la 35 \ra} \lambda_3 \, (Q_{12} \cdot \lambda_3).
\end{aligned}
\end{equation}
We see that the left (pentagon) loop has a pole at infinity for $\la13\ra=0$, all momenta $\ell_A$, $\ell_B$, $\ell_C$, $\ell_G$, $\ell_H$ blow up, while the right (box) loop has a UV pole for $\la35\ra=0$ when $\ell_D$, $\ell_E$, $\ell_F$ and $\ell_H$ go to infinity. The loop momentum $\ell_H$ contains both poles as it participates in both internal cycles. Note that the assignment of cycles is unique and it critically depends on the \emph{planarity} of the diagram. In particular, we do not consider the external heptagon cycle -- there is no pole at infinity associated with it, see Section \ref{sec:non-planar} for elaboration,
\begin{equation} \label{eq:pentabox9}
\raisebox{-20mm}{\includegraphics[trim={0cm 0cm 0cm 0cm},clip,scale=1.2]{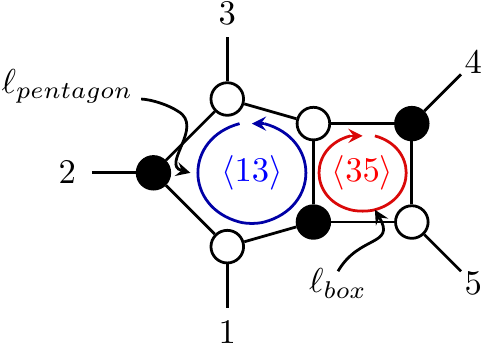}}.
\end{equation}
Now we choose a particular orientation where both poles at infinity are present due to helicity assignments and solve for the edge variables,
\begin{equation} \label{eq:pentabox5}
\raisebox{-20mm}{\includegraphics[trim={0cm 0cm 0cm 0cm},clip,scale=1.2]{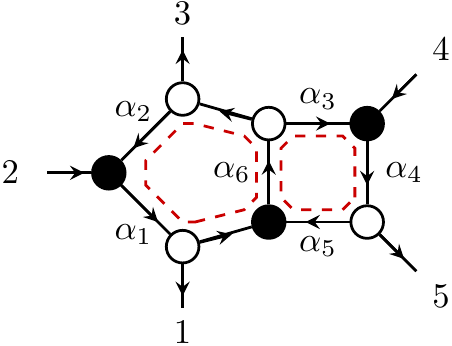}}
	\begin{aligned}
		~~~~&\alpha_1=\frac{\la 13 \ra}{\la 23 \ra},~~~~
		\alpha_2=\frac{\la 12 \ra }{\la 13 \ra},~~~~
		\alpha_3=\frac{\la 45 \ra}{\la 35 \ra},\\
		&\alpha_4=\frac{\la 35 \ra}{\la 34 \ra},~~~~
		\alpha_5=\frac{\la 13 \ra}{\la 35 \ra},~~~~
		\alpha_6=\frac{\la 35 \ra}{\la 15 \ra}.
	\end{aligned}
\end{equation}
This on-shell diagram is a part of the octacut of the 2-loop 5pt amplitude $A_5(1^+2^-3^+4^-5^+)$. In order to calculate the whole cut we would have to sum over all possible internal helicity assignments (for ${\cal N}{<}4$ SYM) as discussed earlier, but we can also study this one on-shell diagram individually. For this particular case (\ref{eq:pentabox5}) the bare on-shell function (which is identical for any assignment of internal helicities) is,
\begin{equation}
    \Omega^{\rm bare} = \int \frac{d \alpha_1\dots d\alpha_6\,\delta(C\cdot Z)}{\alpha_1\alpha_2\alpha_3\alpha_4\alpha_5\alpha_6} = \frac{\la24\ra\,\delta^4(P)\delta^6(Q)}{\la 12\ra\la 23\ra\la 34\ra\la 45\ra\la 51\ra}\label{bare5pt}
\end{equation}
and the Jacobian ${\cal J}$ consists of two cycles,
\begin{equation}
    {\cal J} = \frac{1}{1-\alpha_1 \alpha_2 \alpha_6 - \alpha_3 \alpha_4 \alpha_5 \alpha_6} = \frac{\la 13\ra\la 24\ra\la 35\ra}{\la 23\ra\la 15\ra\la 34\ra}\label{Jac5pt}
\end{equation}
The on-shell function $\Omega$ is then equal to
\begin{equation}
    \Omega = \underbrace{\frac{\la24\ra\,\delta^4(P)\delta^6({\cal Q})}{\la 12\ra\la 23\ra\la 34\ra\la 45\ra\la 51\ra}}_{\Omega^{\rm bare}} \times \underbrace{\left(\frac{\la 23\ra\la 15\ra\la 34\ra}{\la 13\ra\la 24\ra\la 35\ra}\right)}_{{\cal J}^{-1}} = \frac{\delta^4(P)\delta^6({\cal Q})}{\la 12\ra\la 45\ra\la 13\ra \la 35\ra} \label{pentabox2}
\end{equation}
We see that three of the IR poles $\la23\ra$, $\la34\ra$, $\la15\ra$ are removed as erasing $\alpha_1$, $\alpha_4$ and $\alpha_6$ edges would lead to inconsistent helicity flows. On the other hand, both UV poles $\la13\ra$, $\la35\ra$ are added. We first blow up the right loop and calculate the residue on the pole $\la35\ra=0$. The right loop can be treated as a one-loop box diagram with external legs $C,4,5,G$. The residue is a following on-shell diagram,
\begin{equation}
         \raisebox{-15mm}{\includegraphics[trim={0cm 0cm 0cm 0cm},clip,scale=1.1]{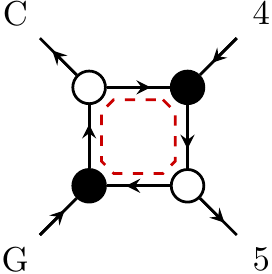}}
         \qquad
         \longrightarrow
         \qquad
         \raisebox{-15mm}{\includegraphics[trim={0cm 0cm 0cm 0cm},clip,scale=1.1]{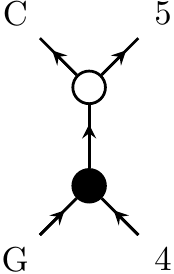}}
\end{equation}
Replacing the box by this tree graph in (\ref{eq:pentabox}) we get 
\begin{equation}
\raisebox{-18mm}{\includegraphics[trim={0cm 0cm 0cm 0cm},clip,scale=1.1]{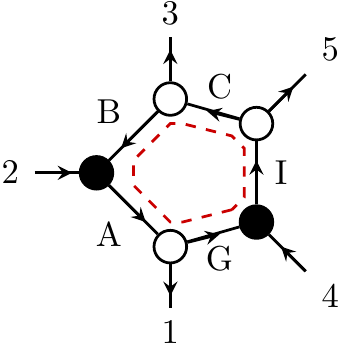}}
\end{equation}
Note that in this diagram the condition $\la35\ra$ is automatically imposed as evident from legs $3,5$ being connected by a chain of white vertices. We can now assign edge variables
\begin{equation}
 \raisebox{-18mm}{\includegraphics[trim={0cm 0cm 0cm 0cm},clip,scale=1.1]{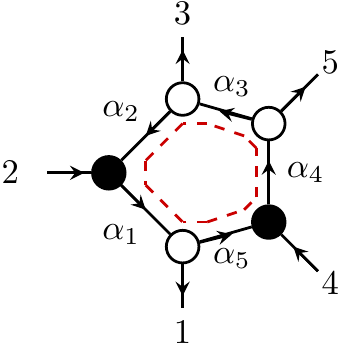}} \qquad 
 \begin{aligned}
\qquad&\alpha_1=\frac{\langle 15 \rangle
}{\langle 25 \rangle},~~
\alpha_2=\frac{\langle 12 \rangle
}{\langle 13\rangle},\\
\alpha_3&=\frac{\langle 23\rangle}{\langle 25 \rangle},~~
\alpha_4=\frac{\langle15 \rangle}{\langle 14\rangle},~~
\alpha_5=\frac{ \langle 45 \rangle}{\langle 51 \rangle}.
 \end{aligned}
\end{equation}
and calculate the on-shell function,
\begin{equation}
    \Omega_{\rm UV} = \int\frac{d \alpha_1\dots d\alpha_5\, \delta(C\cdot Z)}{\alpha_1\alpha_2\alpha_3\alpha_4\alpha_5} \times \frac{1}{1-\alpha_1\alpha_2\alpha_3\alpha_4\alpha_5} = \frac{\delta^4(P)\delta^6({\cal Q})\,\delta(\la35\ra)}{\la12\ra\la45\ra\la13\ra},
\end{equation}
which is equal to a residue of (\ref{pentabox2}) on $\la35\ra=0$ pole. Let us now consider the same pentabox on-shell diagram but calculate the residue on the other pole at infinity $\la13\ra=0$ which blows up the left pentagon. As we have already seen, the calculation of UV poles is a local operation and we need to figure out how to do it for the pentagon sub-diagram. As it turns out, the rule is very similar to what we saw for the box. The result is a tree-level on-shell diagram with legs $2,F$ in the black vertex and legs $1,3,D$ connected by a chain of white vertices, 
\begin{equation}
        \raisebox{-18mm}{\includegraphics[trim={0cm 0cm 0cm 0cm},clip,scale=1.1]{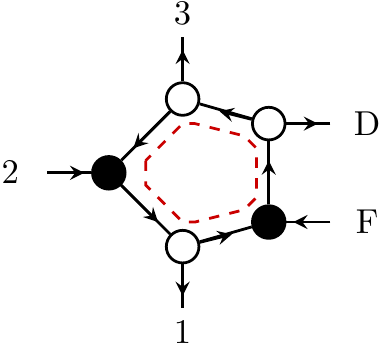}}~ =~ \raisebox{-12.5mm}{\includegraphics[trim={0cm 0cm 0cm 0cm},clip,scale=1.1]{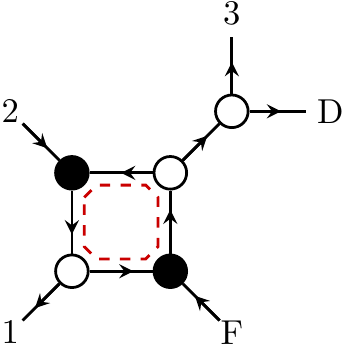}}~\longrightarrow
         ~~\raisebox{-14mm}{\includegraphics[trim={0cm 0cm 0cm 0cm},clip,scale=1.1]{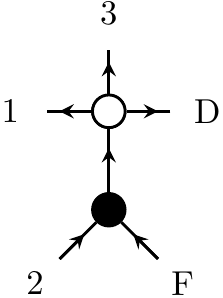}}
\end{equation}
where we first merge-expanded the pentagon diagram and then we used the same replacement rule as for the box. In the final tree-level on-shell diagram we merged all legs into one white vertex, which can be interpreted as a 4-point $k=1$ tree-level amplitude, or alternatively can be further expanded as a chain of regular white three-point vertices,
\begin{equation}
         \raisebox{-13.5mm}{\includegraphics[trim={0cm 0cm 0cm 0cm},clip,scale=1.05]{Figures/PentaFac.pdf}}\quad = \quad
         \raisebox{-18.5mm}{\includegraphics[trim={0cm 0cm 0cm 0cm},clip,scale=1.05]{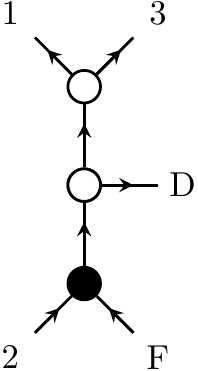}}
         \quad = \quad
         \raisebox{-18.5mm}{\includegraphics[trim={0cm 0cm 0cm 0cm},clip,scale=1.05]{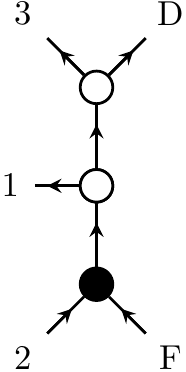}}
         \quad = \quad
         \raisebox{-18.5mm}{\includegraphics[trim={0cm 0cm 0cm 0cm},clip,scale=1.05]{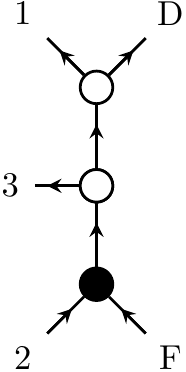}} \label{mergexp}
\end{equation}
The identity between all these diagrams is a consequence of a usual merge-expand rule \cite{Arkani-Hamed:2012zlh}. Using (\ref{mergexp}) the residue on the $\la13\ra=0$ pole is
\begin{equation}
 \raisebox{-18mm}{\includegraphics[trim={0cm 0cm 0cm 0cm},clip,scale=1.05]{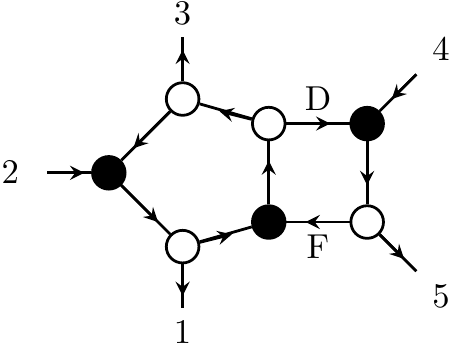}}
         \quad\longrightarrow \quad
         \raisebox{-18.5mm}{\includegraphics[trim={0cm 0cm 0cm 0cm},clip,scale=1.05]{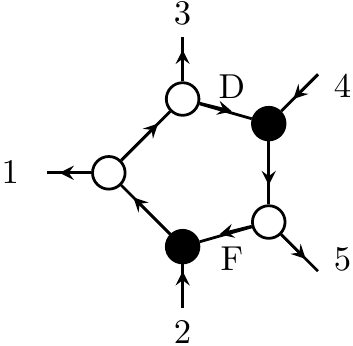}}\label{pent5}
\end{equation}
As expected this diagram automatically implements the condition $\la13\ra=0$. Note that using other representations of the tree graph (\ref{mergexp}) we would get other one-loop pentagons which differ from (\ref{pent5}) by the merge-expand move and do not change the on-shell function. It is easy to see that the on-shell function for (\ref{pent5}) is equal to
\begin{equation}
    \Omega_{\rm UV} = \frac{\delta^4(P)\delta^6({\cal Q})\delta(\la13\ra)}{\la12\ra\la45\ra\la35\ra}
\end{equation}
which is equal to the residue of (\ref{pentabox2}) on $\la13\ra=0$ pole. Naively, we could access the pole at infinity by solving for one of the edge variables from ${\cal J}^{-1}=0$. However, the meaning of the new edge variables is unclear as they do not correspond to a parametrization of the lower-loop on-shell diagram. Hence we have to work directly in the kinematical space when accessing the poles at infinity. 

\subsection{MHV one-loop diagrams}

As we have already seen, the calculation of the UV pole of an on-shell diagram reduced to the calculation of this pole in the one-loop sub-diagram. The result of this operation was glued into the rest of the graph. Our goal here is to calculate the UV pole for a general one-loop on-shell diagram. This is a direct generalization of what we have already seen for box and pentagon. Let us first look at the MHV hexagon,
\begin{equation}
 \raisebox{-20.5mm}{\includegraphics[trim={0cm 0cm 0cm 0cm},clip,scale=1.1]{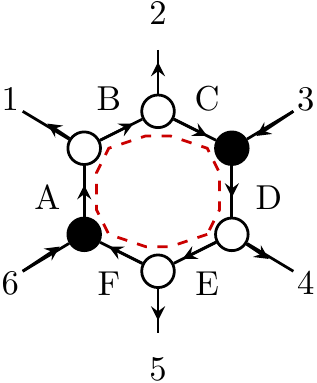}}
 \begin {aligned}\qquad\qquad
\ell_A&=\frac{[13]}{[36]}\lambda_1 \widetilde{\lambda}_6,~~~~
\ell_B&=\frac{[16]}{[36]}\lambda_1 \widetilde{\lambda}_3,\\
\ell_C&= \frac{[26]}{[36]}\lambda_2 \widetilde{\lambda}_3,~~~~
\ell_D&=  \frac{[46]}{[36]}\lambda_4 \widetilde{\lambda}_3,\\
\ell_E&=  \frac{[34]}{[36]}\lambda_4 \widetilde{\lambda}_6,~~~~
\ell_F&=  \frac{[35]}{[36]}\lambda_5 \widetilde{\lambda}_6.
\end{aligned}\label{hex1}
\end{equation}
As we can see from the chains of white vertices, the external kinematics is subject to two constraints $\la12\ra=\la45\ra=0$, i.e. $\lambda_2=\alpha \lambda_1$, $\lambda_5=\beta \lambda_4$. The pole at infinity is accessed by sending $[36]=0$ as evident from explicit solutions for internal loop momenta,
\begin{equation}
    0 = s_{36} = s_{1245} = (p_1+p_2)\cdot(p_4+p_5) = \la 14\ra ([14]+\alpha [24] + \beta [15] + \alpha\beta [25])
\end{equation}
Hence $[36]=0$ implies $\la14\ra=0$ on the support of $\la12\ra=\la45\ra=0$. In the end, all $\lambda$s of external legs connected to white vertices are collinear, $\lambda_1\sim\lambda_2\sim\lambda_4\sim\lambda_5$, i.e. $\la ij\ra=0$ for $i,j=1,2,4,5$. This merges legs $1,2,4,5$ in the white vertex and $3,6$ into black vertex
\begin{equation}
        \raisebox{-20.5mm}{\includegraphics[trim={0cm 0cm 0cm 0cm},clip,scale=1.1]{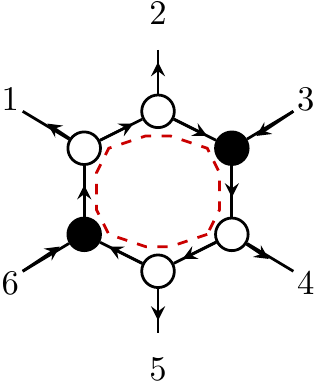}} \,\longrightarrow\,\,\,\,
         \raisebox{-10mm}{\includegraphics[trim={0cm 0cm 0cm 0cm},clip,scale=1.1]{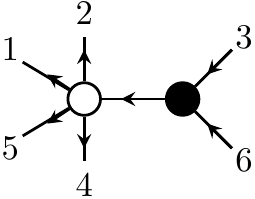}} \,\,=\,\,
         \raisebox{-10mm}{\includegraphics[trim={0cm 0cm 0cm 0cm},clip,scale=1.1]{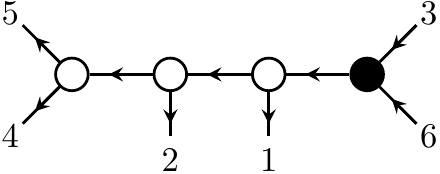}}
\end{equation}
where the five-point white vertex can be again expanded in any way we want. We could obtain the same conclusion if we first merge-expanded the original hexagon,
\begin{equation}
 \raisebox{-20.5mm}{\includegraphics[trim={0cm 0cm 0cm 0cm},clip,scale=1.1]{Figures/Hexagon.pdf}}\qquad = \quad\raisebox{-22mm}{\includegraphics[trim={0cm 0cm 0cm 0cm},clip,scale=1.1]{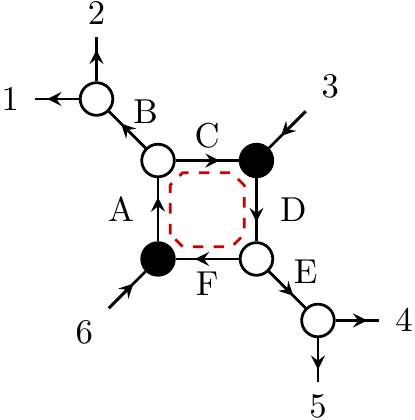}}
\end{equation}
By accessing the pole at infinity in the box diagram, we arrive at one particular representation of the resulting tree graph,
\begin{equation} \label{eq:factorization4}
 \raisebox{-22mm}{\includegraphics[trim={0cm 0cm 0cm 0cm},clip,scale=1.1]{Figures/Box-6pt.pdf}}\quad \longrightarrow \quad\raisebox{-18mm}{\includegraphics[trim={0cm 0cm 0cm 0cm},clip,scale=1.1]{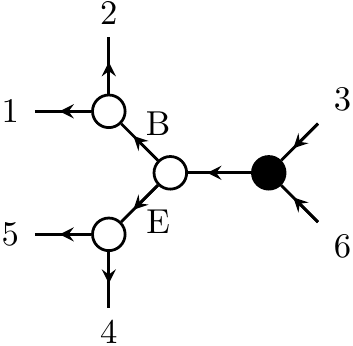}}
\end{equation}
This can be also verified by an explicit calculation. The on-shell function for the hexagon with closed internal loop (\ref{hex1}) is
\begin{equation}
    \Omega = \underbrace{\frac{\la36\ra\,\delta^4(P)\delta^6({\cal Q})\,\delta(\la12\ra)\delta(\la45\ra)}{\la 23\ra\la34\ra\la56\ra\la 16\ra}}_{\Omega^{\rm bare}} \times \underbrace{\frac{\la23\ra\la56\ra}{\la25\ra\la36\ra}}_{{\cal J}^{-1}} = \frac{\delta^4(P)\delta^6({\cal Q})\,\delta(\la12\ra)\delta(\la45\ra)}{\la34\ra\la16\ra\la25\ra}\label{res11}
\end{equation}
On the pole at infinity, $\la25\ra=0$, and replacing $1/\la25\ra\rightarrow \delta(\la25\ra)$ we do get the correct on-shell function for the on-shell diagram (\ref{eq:factorization4}). Note that this is all calculated on a very constrained kinematics so there are many equivalent ways how to write the formula (\ref{res11}). It is clear how the calculation of the UV pole generalizes for any one-loop MHV on-shell diagram,
\begin{equation} \label{eq:ngon}
 \raisebox{-28mm}{\includegraphics[trim={0cm 0cm 0cm 0cm},clip,scale=1.1]{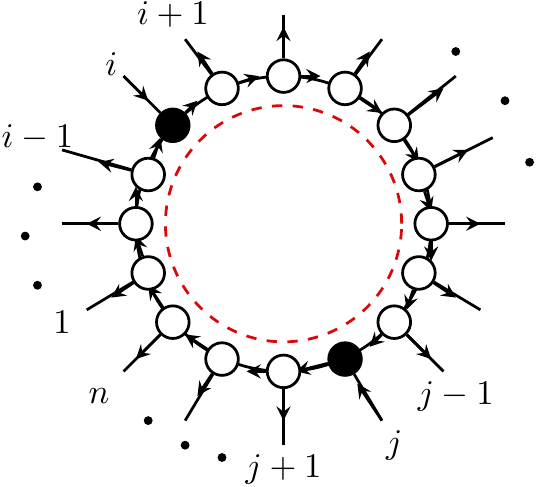}}~~ =~~
  \raisebox{-23mm}{\includegraphics[trim={0cm 0cm 0cm 0cm},clip,scale=1.1]{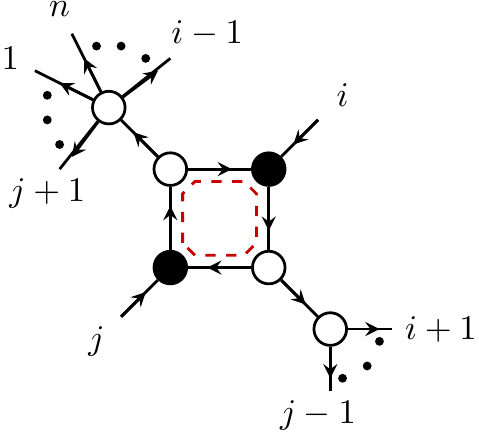}}
\end{equation}
where we conveniently reduced the $n$-gon into a ``core'' box diagram connected to white blobs of external legs. The $\lambda$-spinors of all external legs connected to the same white vertex are proportional, so the kinematics are highly constrained. As before, on the pole at infinity $[ij]=0$ we replace the box diagram by a simple tree-level factorization graph, and all but legs $i,j$ are now connected to the same white vertex,
\begin{equation}
         \raisebox{-20mm}{\includegraphics[trim={0cm 0cm 0cm 0cm},clip,scale=1.1]{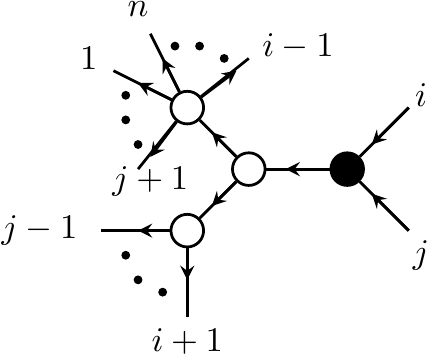}}\qquad = \qquad
         \raisebox{-14mm}{\includegraphics[trim={0cm 0cm 0cm 0cm},clip,scale=1.1]{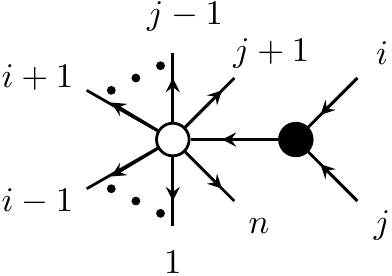}}
\end{equation}
We can expand the white vertex into three-point vertices in an arbitrary way without changing the kinematics or the on-shell function. In the original on-shell diagram many collinearity conditions are already imposed,
\begin{equation}
    \lambda_{i{+}1}\sim\lambda_{i{+}2}\sim\dots\sim \lambda_{j{-}1} \quad \mbox{and} \quad \lambda_{j{+}1}\sim\lambda_{j{+}2}\sim\dots\sim \lambda_{i{-}1}
\end{equation}
The pole at infinity is accessed by $\la pq\ra=0$ where $p,q$ are from two different sets of labels of external legs attached to white vertices, $p\in\{i{+}1,\dots,j{-}1\}$ and $q\in\{j{+}1,\dots,i{-}1\}$, or alternatively we can access the pole by sending $[ij]=0$, which are all equivalent conditions.

\subsection{Arbitrary \texorpdfstring{$n$}{n}-gons and helicities}

Our goal now is to generalize the rule for the UV pole for an $n$-gon with an arbitrary configuration of white and black vertices, and arbitrary $k$ degree. The first example is a very simple six-point NMHV hexagon
\begin{equation} \label{eq:nmhv-hexagon}
         \raisebox{-19mm}{\includegraphics[trim={0cm 0cm 0cm 0cm},clip,scale=1.1]{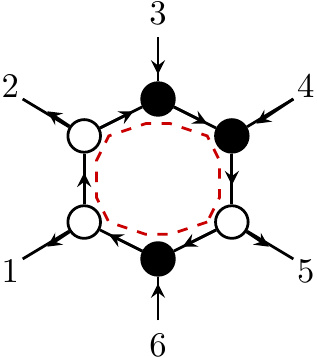}} 
\qquad = \qquad
         \raisebox{-17mm}{\includegraphics[trim={0cm 0cm 0cm 0cm},clip,scale=1.1]{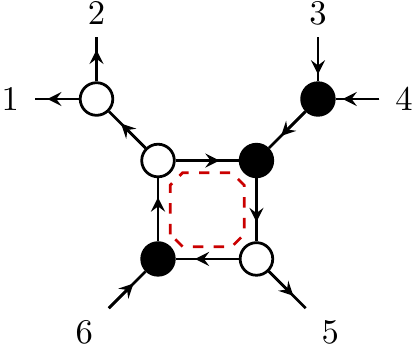}}
\end{equation}
where we used the merge-expand moves to redraw the diagram. We can see that the core diagram is again a box, hence the rule for the UV pole is the same as before and we have not learnt anything new. This makes it clear that the next qualitatively new case is the hexagon diagram where the white and black vertices are alternating, and the core diagram is a proper hexagon that can not be simplified,
\begin{equation} \label{eq:hexagon2}
\raisebox{-19mm}{\includegraphics[trim={0cm 0cm 0cm 0cm},clip,scale=1.1]{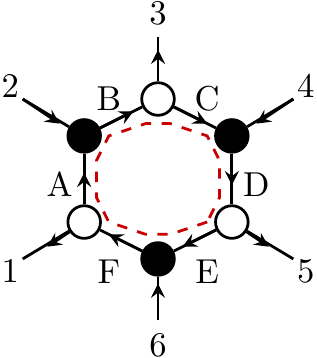}}
\begin {aligned}~~~
\ell_A&=\frac{\la 23 \ra}{\la 13 \ra}\lambda_1 \Tilde{\lambda}_2 =\frac{[16]}{[26]}\lambda_1 \Tilde{\lambda}_2,~
\ell_B= \frac{\la 12 \ra}{\la 13 \ra}\lambda_3 \Tilde{\lambda}_2 =\frac{[34]}{[24]}\lambda_3 \Tilde{\lambda}_2, \\
\ell_C&=  \frac{\la 45 \ra}{\la 35 \ra}\lambda_3 \Tilde{\lambda}_4= \frac{[23]}{[24]} \lambda_3 \Tilde{\lambda}_4,~
\ell_D=  \frac{\la 34 \ra}{\la 35 \ra} \lambda_5 \Tilde{\lambda}_4 = \frac{[56]}{[46]} \lambda_4 \Tilde{\lambda}_5,\\
\ell_E&=  \frac{\la 16 \ra}{\la 15 \ra} \lambda_5 \Tilde{\lambda}_6 = \frac{[45]}{[46]} \lambda_5 \Tilde{\lambda}_6, ~
\ell_F=  \frac{\la 56 \ra}{\la 15 \ra} \lambda_1 \Tilde{\lambda}_6 = \frac{[12]}{[26]}\lambda_1 \Tilde{\lambda}_6. \\
\end{aligned}
\end{equation}
The kinematics of this on-shell diagram is more involved. There is no chain of black/white vertices and no merge-expand operation can change it. There are two independent kinematical conditions imposed on external momenta but none of these conditions is a simple collinearity of $\lambda$ or $\widetilde{\lambda}$ spinors. Solving for internal momenta we learn that there are three cyclically related kinematical conditions imposed,
\begin{equation}
    \la 1|2{+}3|4] = \la 3|4{+}5|6] = \la 5|6{+}1|2] = 0 \label{NMHV6con}
\end{equation}
One of these conditions is redundant, i.e. any two of them imply the third one. Looking at the denominator of on-shell loop momenta we can see that the pole at infinity can be approached by setting $\la13\ra=0$. On the support of (\ref{NMHV6con}) this also sends 
\begin{equation}
    \la 13\ra = \la 15\ra = \la 35\ra = 0 = [24] = [26] = [46] \label{UV6pt}
\end{equation}
In other words, $\lambda_1\sim\lambda_3\sim\lambda_5$ and $\widetilde{\lambda}_2\sim\widetilde{\lambda}_4\sim\widetilde{\lambda}_6$ on the pole at infinity. The corresponding on-shell diagram is then 
\begin{equation} \label{eq:hexafac3}
\raisebox{-11mm}{\includegraphics[trim={0cm 0cm 0cm 0cm},clip,scale=1.15]{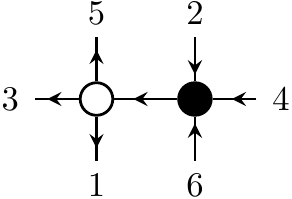}}
\end{equation}
and the on-shell function is
\begin{equation}
    \Omega_{\rm UV} = \frac{[23]\delta(\la 1 3 \ra) \delta(\la 3 5 \ra) \delta([24])\delta(P)\delta({\cal Q}) \delta([56]\Tilde{\eta}_4+[45]\Tilde{\eta}_6)}{[34]\la 1 2 \ra \la 2 5 \ra [25][56]^2} \label{hextree}
\end{equation}
It is very interesting that the kinematics in (\ref{eq:hexagon2}) is already quite constrained (\ref{NMHV6con}), but it is only after imposing one more condition and approaching the UV pole that it completely factorizes into $\lambda$, $\widetilde{\lambda}$ collinearities (\ref{UV6pt}).

We now evaluate the on-shell function for this diagram. The topology allows us to fix the GL(1)s in a symmetric way and choose the edge variables as
\begin{equation} \label{eq:hexagon4}
\raisebox{-21mm}{\includegraphics[trim={0cm 0cm 0cm 0cm},clip,scale=1.15]{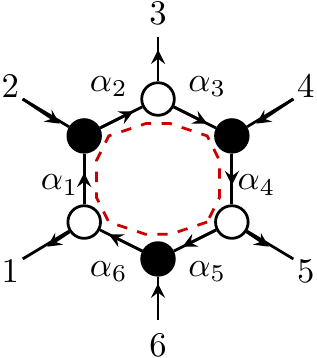}}
\begin {aligned}\qquad\qquad
\alpha_1&=\frac{\la 23 \ra}{\la 13 \ra}, \quad \alpha_2=\frac{\la 13 \ra}{\la 12 \ra},\\
\alpha_3&=\frac{\la 45 \ra}{\la 35 \ra}, \quad \alpha_4=\frac{\la 35 \ra}{\la 34 \ra}, \\
\alpha_5&=\frac{\la 16 \ra}{\la 15 \ra}, \quad \alpha_6=\frac{\la 15 \ra}{\la 56 \ra}.
\end{aligned}
\end{equation}
The Jacobian of the diagram is given by
\begin{equation}
    {\cal J} = 1-\alpha_1\alpha_2\alpha_3\alpha_4\alpha_5\alpha_6 = \frac{\la12\ra\la34\ra\la56\ra - \la23\ra\la45\ra\la16\ra}{\la12\ra\la34\ra\la56\ra} = \frac{\la13\ra\la4|3{+}5|1]}{\la12\ra\la34\ra[12]} \label{Jac6}
\end{equation}
where in the last equality we used the special kinematics (\ref{UV6pt}). The on-shell functions can be then written in many equivalent ways thanks to (\ref{UV6pt}), one representation is 
\begin{align}
    \Omega &= \underbrace{\frac{\la 4| 3{+} 5|1] \la 3 4 \ra \la 1 5 \ra\, \delta(\Xi)}{[56]^2[12]\la 1 2\ra \la 1 6 \ra \la 2 3 \ra \la 4 5\ra \la 3 5\ra}}_{\Omega^{\rm bare}} \times \underbrace{\frac{\la12\ra\la34\ra[12]}{\la13\ra\la4|3{+}5|1]}}_{{\cal J}^{-1}}= \frac{\la 1 5 \ra \la 3 4 \ra^2\, \delta(\Xi)}{[56]^2 \la 1 3 \ra \la 1 6 \ra \la 2 3 \ra \la 4 5 \ra \la 3 5 \ra} \label{Hex1}
\end{align}
where we denoted all delta functions as 
\begin{equation}
    \delta(\Xi) \equiv \delta^4(P)\delta^6({\cal Q})\,\delta(\la 1 | 2 {+}3 |4]) \delta(\la 3 | 4 {+} 5 |6])\delta^3([56]\Tilde{\eta}_4{+}[46]\Tilde{\eta}_5{+}[45]\Tilde{\eta}_6)
\end{equation}
Looking at the denominator of the inverse Jacobian ${\cal J}^{-1}$ we can see two terms: the first is $\la4|3{+}5|1]$ -- this is just a helicity factor which is canceled by the numerator of the bare form (and would continue to be canceled also for ${\cal N}<3$), and then the pole at infinity $\la13\ra=0$. Note that due to the special kinematics, the pole structure of $\Omega$ is not very transparent. The singularities, $\la23\ra=0$, $\la45\ra=0$, $\la16\ra=0$, are all physical poles and correspond to erasing edges $\alpha_1$, $\alpha_3$, $\alpha_5$. Pole $[56]=0$ is not present because it also implies $\la34\ra=0$ and there is a cancelation between the numerator and the denominator (it also corresponds to non-erasable edge $\alpha_4$ which is forbidden). We can make manifest an absence of this pole but at the expense of introducing some other spurious pole. This is a general feature of such expressions, because of mass dimension and little group weights any representation has some spurious terms that can be removed by trading them for another spurious object.

Let us look in more detail at the pole at infinity located at $\la13\ra=\la15\ra=\la35\ra=0$. This pole is not present in $\Omega^{\rm bare}$ as the factor $\la15\ra/\la35\ra$ is finite and no singularity is generated. This can be made manifest when choosing a particular parametrization. The Jacobian ${\cal J}^{-1}$ does introduce this pole and we can show after some work that the residue on the UV pole is equal to 
\begin{equation}
    \Omega_{\rm UV} = \frac{[23]\delta(\la 1 3 \ra) \delta(\la 3 5 \ra) \delta([24])\delta^4(P)\delta^6({\cal Q}) \delta([56]\Tilde{\eta}_4+[45]\Tilde{\eta}_6)}{[34]\la 1 2 \ra \la 2 5 \ra [25][56]^2} \label{UVhex}
\end{equation}
where the delta functions indicate the simple collinearity conditions. The expression (\ref{UVhex}) is equal to (\ref{hextree}) which proves our conjecture. Note that in (\ref{UVhex}) the kinematics is incredibly constrained (three extra conditions) and there are many equivalent ways how to write it. For arbitrary ${\cal N}$ the formula (\ref{Hex1}) is generalized to
\begin{align}
    \Omega &= \frac{\la 4| 3{+} 5|1]^{4{-}{\cal N}} \la 3 4 \ra \la 1 5 \ra\, \delta(\Xi)}{[56]^2[12]\la 1 2\ra \la 1 6 \ra \la 2 3 \ra \la 4 5\ra \la 3 5\ra}\times \left(\frac{\la12\ra\la34\ra[12]}{\la13\ra\la4|3{+}5|1]}\right)^{4{-}{\cal N}}  \label{Hex2}
\end{align}
and while the helicity factor $\la 4| 3{+} 5|1]$ is canceled, we do generate a higher pole at infinity for $\la13\ra=0$. We will discuss it in more details in the next section.

The next case is an arbitrary NMHV $n$-gon, with three black and $n{-}3$ white vertices. We can use the merge-expand moves to redraw this diagram as a core hexagon with three extra white blobs,
\begin{equation}
 \raisebox{-27mm}{\includegraphics[trim={0cm 0cm 0cm 0cm},clip,scale=1.05]{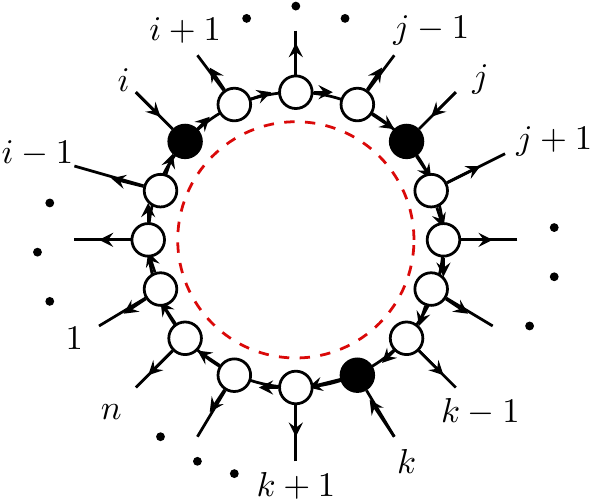}}
 \,\,\,\, = 
 \raisebox{-20mm}{\includegraphics[trim={0cm 0cm 0cm 0cm},clip,scale=1.05]{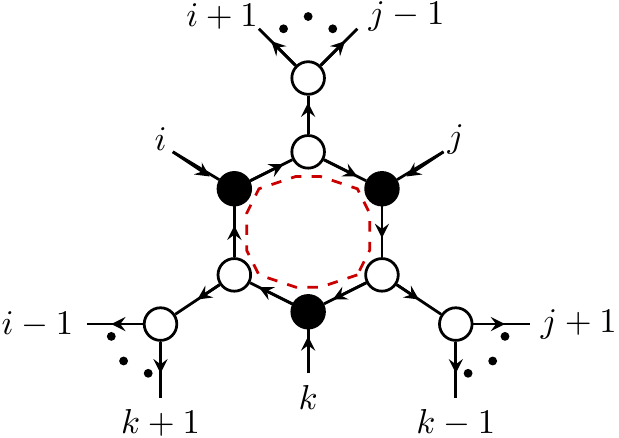}}
\end{equation}
All the $\lambda$ spinors in each group are proportional, 
\begin{equation}
    \lambda_{i{+}1}\sim\lambda_{i{+}2}\sim\dots\sim\lambda_{j{-}1}\,\,\,\mbox{etc.}
\end{equation}
and in addition we have two independent mixed conditions, for example,
\begin{equation}
    \la i{-}1|P_i|j] = \la j{-}1|P_j|k] = 0
\end{equation}
where $P_i = p_{i}+p_{i{+}1}+\dots+p_{j{-}1}$ and $P_j=p_j+p_{j{+}1}+\dots+p_{k{-}1}$. The pole at infinity is accessed by a single condition when $\la pq\ra=0$ where $p,q$ are from two different white sectors. This collapses the whole kinematics and all external $\lambda$s attached to white vertices are proportional, and all external $\widetilde{\lambda}$s attached to black vertices are proportional. As a result, we get an on-shell diagram
\begin{equation} \label{eq:generalfacorization} 
\raisebox{-15mm}{\includegraphics[trim={0cm 0cm 0cm 0cm},clip,scale=1.1]{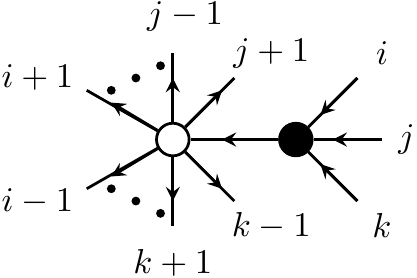}}
\end{equation}
Note that all such tree-level on-shell diagrams are \emph{non-planar} in a sense of the canonical ordering $1,2,{\dots},n$ we started with. The same procedure generalizes to higher $k$ degrees. The ``core'' N$^2$MHV one-loop on-shell diagram is an octagon,
\begin{equation} \label{eq:NNMHVoctagon}
\raisebox{-23mm}{\includegraphics[trim={0cm 0cm 0cm 0cm},clip,scale=1.05]{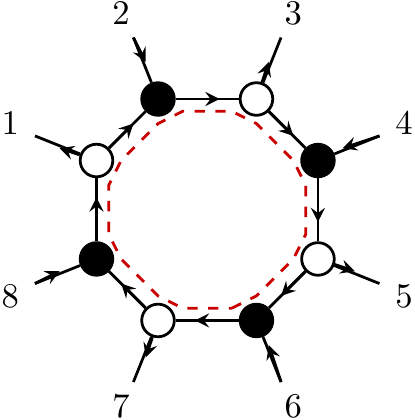}}
\end{equation}
Here we have four independent kinematical conditions imposed on the external kinematics, for example
\begin{equation}
    \la 1|2{+}3|4] = \la 3|1{+}2|8] = \la 7|8{+}1|2] = \la 5|3{+}4|2]=0 \label{8ptkin}
\end{equation}
There are no collinearity conditions between $\lambda$s and $\widetilde{\lambda}$s. The four independent constraints (\ref{8ptkin}) are quadratic in nature and there is no simplification. The solution for the internal on-shell loop momenta is
\begin{equation} \label{eq:Octagon}
\raisebox{-23mm}{\includegraphics[trim={0cm 0cm 0cm 0cm},clip,scale=1.1]{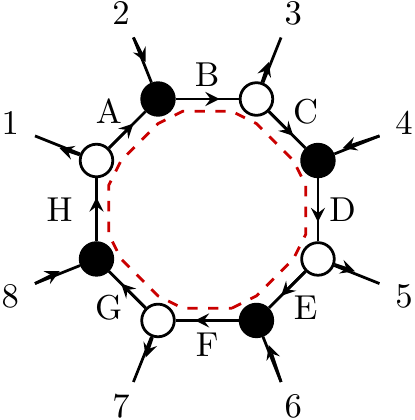}}
\begin {aligned}\qquad\qquad
\ell_A &=\frac{\la 23 \ra}{\la 13 \ra}\lambda_1 \Tilde{\lambda}_2, \,\,\, \ell_B =\frac{\la 1 2 \ra}{\la 13 \ra} \lambda_3 \Tilde{\lambda}_2,\\
\ell_C&=\frac{\la 45 \ra}{\la 35 \ra} \lambda_3 \Tilde{\lambda}_4, \,\,\, \ell_D =\frac{\la 3 4 \ra}{\la 3 5 \ra}\lambda_5 \Tilde{\lambda}_4,\\
\ell_E&=\frac{\la 6 7 \ra}{\la 5 7 \ra} \lambda_5 \Tilde{\lambda}_6, \,\,\, \ell_F=\frac{\la 5 6 \ra}{\la 5 7 \ra}\lambda_7 \Tilde{\lambda}_6,\\
\ell_G&=\frac{\la 1 8 \ra}{\la 1 7 \ra} \lambda_7 \Tilde{\lambda}_8, \,\,\, \ell_H=\frac{\la 7 8 \ra}{\la 1 7 \ra}\lambda_1 \Tilde{\lambda}_8.
\end{aligned}
\end{equation}
We see that the pole at infinity is accessed by imposing a simple condition $\la13\ra=0$ which again collapses the whole kinematics to two sectors,
\begin{equation}
\lambda_1\sim\lambda_3\sim\lambda_5\sim\lambda_7\quad\mbox{and}\quad \widetilde{\lambda}_2\sim\widetilde{\lambda}_4\sim\widetilde{\lambda}_6\sim\widetilde{\lambda}_8 \label{Octpole}
\end{equation}
and the corresponding on-shell diagram is 
\begin{equation} \label{eq:octafac}
\raisebox{-10mm}{\includegraphics[trim={0cm 0cm 0cm 0cm},clip,scale=1.15]{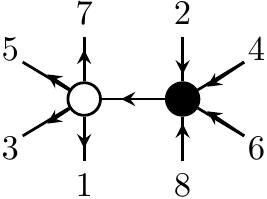}}
\end{equation}
We can see that the diagram again does not respect the canonical ordering of the original octagon (\ref{eq:NNMHVoctagon}). The on-shell function for the octagon is 
\begin{equation}
    \Omega = \underbrace{\frac{[18] [14] [23] \la 56 \ra^2\la 34 \ra  \la 7 8  \ra \, \la 4 | 6 {+} 7|8] [17]\delta(\Xi)}{[56]^2[78]^3[12]^3[34] \la 2 3 \ra^3 \la 4 5 \ra \la 6 7 \ra \la 8 1 \ra}}_{\Omega^{\rm bare}} \times \underbrace{\frac{\la 12 \ra \la 3 4 \ra [87][12]}{\la 13 \ra \la 4 | 6 {+} 7|8][17]}}_{{\cal J}^{-1}}
\end{equation}
where the delta functions are
\begin{align}
   \delta(\Xi) &=  \delta^4(P)\delta^6({\cal Q})\,\delta(\Tilde{\eta}_4[56]+\Tilde{\eta}_5 [46]+\Tilde{\eta}_6[45]) \delta(\Tilde{\eta}_6[78]+\Tilde{\eta}_7 [68]+\Tilde{\eta}_8[67])\nonumber\\
   &\hspace{1cm} \times \delta(\la 3|4{+}5|6]) \delta(\la 7|5{+}6|4]) \delta(\la 5|6{+}7|8]) \delta(\la 1|7{+}8|6])
\end{align}
The kinematics is incredibly constrained and there are many equivalent ways how to write $\Omega$. Also many of the poles are equivalent, as it is evident from the figure, e.g. setting $\la12\ra=0$ (erasing edge B) also sends $[34]=0$ etc. The pole at infinity is again not present in the bare form -- two factors of $\la35\ra\la57\ra$ are canceled by $\la17\ra^2$ in the numerator, and the form is regular for (\ref{Octpole}). The UV pole is produced by the inverse Jacobian and the residue on the pole (\ref{Octpole}) gives
\begin{equation}
    \Omega_{\rm UV} = \frac{\la 3 4 \ra\,\delta(\Xi_a)}{\la 1 2 \ra \la 2 3 \ra \la 4 7 \ra [14]^4[16][18]}
\end{equation}
where the delta functions are now denoted
\begin{align}
   \delta(\Xi_a) &=  \delta^4(P)\delta^6({\cal Q})\,\delta(\Tilde{\eta}_4[18]{+}\Tilde{\eta}_8 [14]) \delta(\Tilde{\eta}_4[16]{+}\Tilde{\eta}_6 [14]) \delta(\la 1 3 \ra) \delta(\la 3 5 \ra) \delta(\la 5 7\ra) \delta([46])\delta([28]) \nonumber\\
\end{align}
which reproduces the on-shell function for the tree on-shell diagram (\ref{eq:octafac}). Now we are ready to generalize to arbitrary one-loop diagram. It is enough to solve the problem only for a ``core diagram'' which for N$^{k{-}2}$MHV is a $2k$-gon with $k$ black and $k$ white alternating vertices. 
\begin{equation} \label{eq:nkmhv-ngon}
\raisebox{-23mm}{\includegraphics[trim={0cm 0cm 0cm 0cm},clip,scale=1.0]{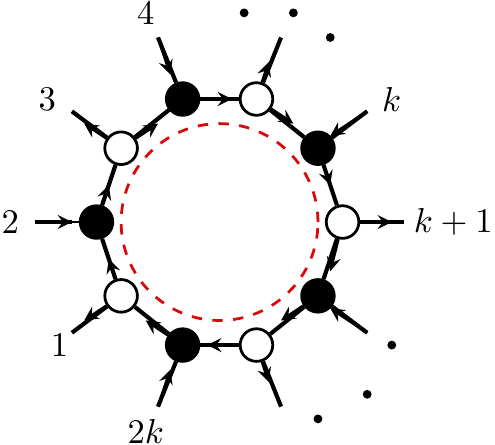}}
\end{equation}
In this diagram $2k-4$ conditions on external kinematics are imposed. We can characterize them by two cyclic classes 
\begin{equation}
    \la j|p_{j{+}1} {+} p_{j{+}2}|j{+}3] = \la j | p_{j{-}1} {+} p_{j{-}2}|j{-}3] = 0
\end{equation}
for all odd $j=1,3,\dots,2k{-}1$. These are $2k$ conditions, 4 of them are redundant. The pole at infinity is approached by imposing $\la pq\ra=0$ where $p,q$ are attached to two different white vertices (alternatively $[cd]=0$ where $c,d$ are attached to black vertices). This completely collapses the kinematics into two sectors where all $\lambda$, resp. $\widetilde{\lambda}$ are proportional,
\begin{equation}
    \lambda_1 \sim \lambda_3 \sim \lambda_5 \sim \dots \sim \lambda_{2k{-}1},\qquad 
    \widetilde{\lambda}_2 \sim \widetilde{\lambda}_4 \sim \widetilde{\lambda}_6 \sim \dots \sim \widetilde{\lambda}_{2k}
\end{equation}
and the resulting on-shell diagram is 
\begin{equation} \label{eq:nkmhv-ngonfac}
\raisebox{-14mm}{\includegraphics[trim={0cm 0cm 0cm 0cm},clip,scale=1.05]{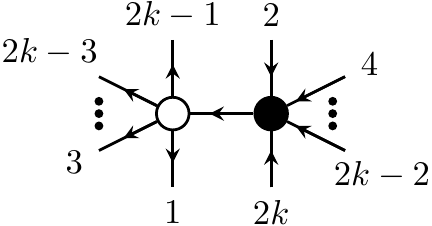}}
\end{equation}
where all legs originally connected to $\overline{\rm MHV}$ white vertices in \eqref{eq:nkmhv-ngon} are now in the white vertex and all legs originally connected to MHV black vertices are now in the black vertex. For any other $n$-gon with chains of white and black vertices we can always use merge-expand moves to identify the core polygon which contains the pole at infinity, and apply the same rule, i.e.
\begin{equation} \label{eq:ngongeneral}
\raisebox{-26mm}{\includegraphics[trim={0cm 0cm 0cm 0cm},clip,scale=1.0]{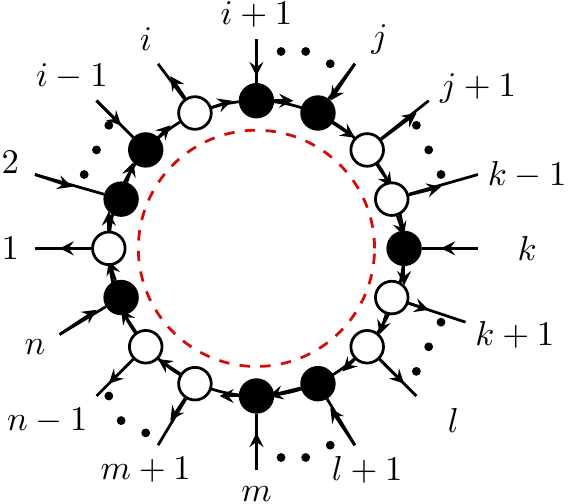}}=\raisebox{-30mm}{\includegraphics[trim={0cm 0cm 0cm 0cm},clip,scale=1.0]{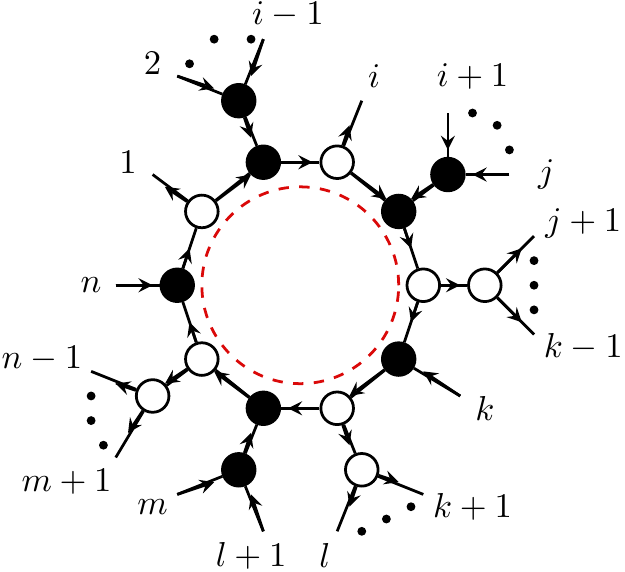}}
\end{equation}
For an arbitrary one-loop on-shell diagram, such as the one above, the residue at infinity corresponds to a tree-level graph where all external legs initially attached to white vertices are now attached together in the same white vertex, and the same applies to black vertices,
\begin{equation}\label{eq:ngongeneralfac}
 \quad \quad  \raisebox{-18mm}{\includegraphics[trim={0cm 0cm 0cm 0cm},clip,scale=1]{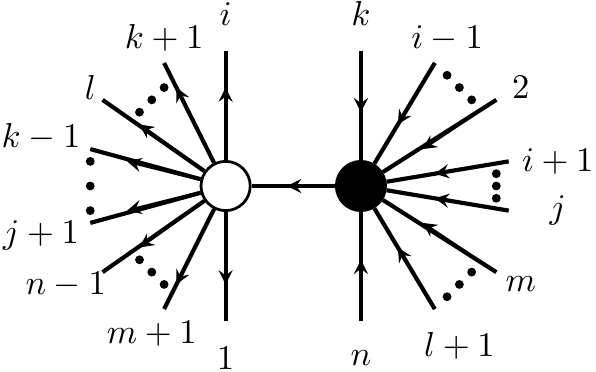}},
\end{equation}
and the external momenta are then divided into two sets: in the first set (attached to white vertex) all $\lambda$'s are proportional, in the other set (attached to black vertex) all $\widetilde{\lambda}$'s are proportional. Note that the cyclic ordering of the original graph is not preserved: the legs connected to black or white vertices are not ordered in any particular way, but the overall canonical ordering is broken. Hence we refer to this operation as a \emph{non-planar twist}. As a result, the residue on the UV pole of a planar on-shell diagram is a lower-loop non-planar on-shell diagram.

\section{All planar diagrams}

In the simple examples we discussed earlier, the UV pole was always localized in a one-loop subgraph. We will show that this is a general property of planar on-shell diagrams, and using the rules of the previous section we can then calculate poles at infinity for an arbitrary higher loop planar graph.

\subsection{Localization of UV poles in planar diagrams\label{sec:planar-proof}}

Let us consider a core $n$-gon one-loop subdiagram somewhere in the middle of a larger on-shell diagram. As discussed we can always put any one-loop subdiagram in this form by merge-expand moves. Because the whole on-shell diagram is planar this $n$-gon is connected to any other one-loop subdiagram through a pair of adjacent legs. If the legs are not adjacent then the diagram is not planar and our proof does not apply.  
\begin{equation}\label{eq:doublebox1}
 \raisebox{-22mm}{\includegraphics[trim={0cm 0cm 0cm 0cm},clip,scale=1.0]{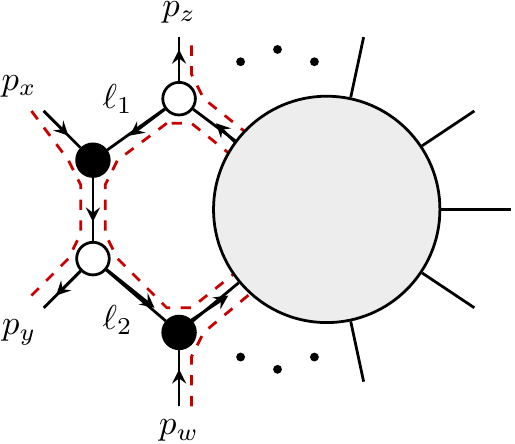}}
\end{equation}
where the central $n$-gon has a pole at infinity for $\la zy\ra = 0$ which is equivalent to $[xw] = 0$. This $n$-gon is glued into the rest of the diagram, so all momenta $p_x$, $p_y$, $p_z$, $p_w$ are in general internal momenta. We want to prove that the UV pole does not propagate beyond this one-loop $n$-gon. In other words, external momenta $p_x$, $p_y$, $p_z$, $p_w$ (from the point of view of the $n$-gon, otherwise these are internal momenta of the whole diagram) do not blow up for $\la zy\ra = [xw]=0$. We label
\begin{equation}
    p_x = \lambda_x\widetilde{\lambda}_x,\quad  p_y = \lambda_y\widetilde{\lambda}_y,\quad  p_z = \lambda_z\widetilde{\lambda}_z,\quad  p_w = \lambda_w\widetilde{\lambda}_w \label{proof1}
\end{equation}
We focus on $p_x$, $p_y$ but an analogous argument applies to other loops and we can also prove that $p_z$, $p_w$ and other momenta attached to our $n$-gon. Using the kinematical properties of three point vertices we can write internal momenta $\ell_1$, $\ell_2$ in the following way,
\begin{equation}
    \ell_1 = \frac{\alpha}{\la zy\ra}\lambda_z\widetilde{\lambda}_x,\qquad \ell_2 = \frac{\beta}{\la zy\ra}\lambda_y \widetilde{\lambda}_w \label{proof2}
\end{equation}
where $\alpha,\beta$ are coefficients. We see that both $\ell_1$, $\ell_2$ blow up for $\la zy\ra=0$ as expected. Note that $\alpha,\beta$ and spinors $\lambda_y$, $\lambda_z$, $\widetilde{\lambda}_x$, $\widetilde{\lambda}_w$ do not have a pole in $\la zy\ra=0$. We need to prove that $\lambda_x$, $\widetilde{\lambda}_y$ also do not have this pole, hence $p_x$, $p_y$ are regular for $\la zy\ra=0$. Using the momentum conservation
\begin{equation}
    \ell_1+p_x = \ell_2+p_y \label{proof3}
\end{equation}
and the expressions (\ref{proof1}), (\ref{proof2}) we contract (\ref{proof3}) with $\widetilde{\lambda}_x$, resp. $\widetilde{\lambda}_w$ and get
\begin{equation}
    [xy] = -\frac{\beta[xw]}{\la zy\ra},\qquad \lambda_y [yw] = [xw]\lambda_x + \frac{\alpha[xw]}{\la zy\ra}\lambda_z
\end{equation}
The ratio $[xw]/\la zy\ra$ is regular for $\la zy\ra=0$ and has no pole. This means that both $[xy]$ and $[yw]$ are free of the UV pole, and hence also $\widetilde{\lambda}_y$ and $p_y$ are regular for $\la zy\ra=0$. Very similarly, we can contract (\ref{proof3}) with $\lambda_y$, resp. $\lambda_z$ and conclude that $\la xz\ra$ and $\la xy\ra$ have no pole at infinity, hence also $\lambda_x$ and $p_x$ are free of this pole. This concludes the proof. We can use the same logic and prove that the UV pole $\la zy\ra=0$ does not propagate to any other loop attached to our $n$-gon, hence this pole is only localized in this one-loop subgraph.

\subsection{Three-loop example}

We will demonstrate the procedure of calculating poles at infinity on several higher loop examples. First, we take a three-loop six-point NMHV leading singularity on-shell diagram and solve for the edge variables,
\begin{equation} \label{eq:HexaDoubleBox-with-edge}
\raisebox{-23mm}{\includegraphics[trim={0cm 0cm 0cm 0cm},clip,scale=1.2]{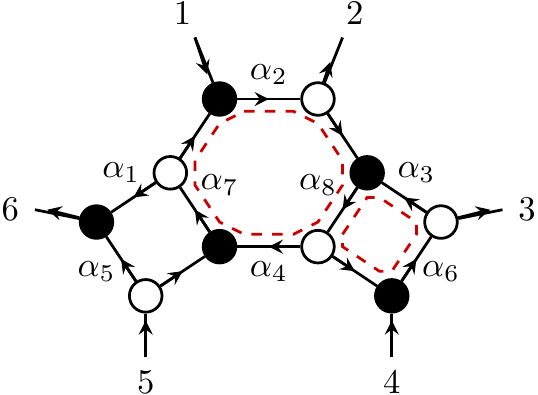}}
\begin {aligned}\,\,
\alpha_1 &=- \frac{\langle 2 | 1{+}5 | 6 ]
}{\la 2 | 5{+}6 | 1 ]},\,\,\alpha_2 =- \frac{\langle 2 | 5{+}6 | 1 ]}{s_{156}},\\
\alpha_3 &=\frac{\langle 2 | 5{+}6| 1 ]}{\langle 3 | 5{+}6 | 1 ]},\quad \,\alpha_5 =- \frac{\langle 2 | 1{+}5 | 6 ]}{\langle 2 | 1{+}6 | 5 ]},\\
\alpha_4 &= \frac{\langle 2 | 
1{+}5 | 6 ]\langle 3 | 5{+}6  | 1 ]}{\la 34 \ra [56]\langle 2| 5{+}6  | 1 ]},\\
\alpha_6 &= \frac{\langle 3 | 5{+}6  | 1 ]}{\langle 4 | 5{+}6  | 1 ]}\,,  \,\,
\alpha_7 = \frac{[56]}{[16]},\,\,\alpha_8 =- \frac{ \la 34 \ra}{\la 23 \ra}.
\end{aligned}
\end{equation}
In ${\cal N}{=}4$ SYM the on-shell function associated with this diagram is a famous $R$-invariant \cite{Drummond:2008vq,Drummond:2008cr} which appears in the context of tree-level recursion relations \cite{Drummond:2008cr}, leading singularities of loop integrands and Grassmannians \cite{Mason:2009qx,Arkani-Hamed:2009nll,Arkani-Hamed:2010zjl,Arkani-Hamed:2010pyv}, and coefficients of polylogs in the hexagon bootstrap \cite{Dixon:2011nj,Dixon:2015iva,Dixon:2016apl,Caron-Huot:2019vjl},
\begin{equation}
\Omega = \int \frac{d\alpha_1{\dots}d\alpha_8\,\delta(C\cdot Z)}{\alpha_1\alpha_2\dots\alpha_8} = {\frac{\delta^8({\cal Q})\delta^4(P)\delta([56]\widetilde\eta_1{+}[61]\widetilde\eta_5{+}[15]\widetilde\eta_6)}{s_{156}\la 23 \ra \la 34 \ra
        \la 2 | 3{+}4 |5] \la 4 | 5{+}6 |1]
 [16][56]}} \equiv R[4,5,6,1,2]
\end{equation}
Note that this expression does not depend on any helicity assignments because of the maximal supersymmetry. In the ${\cal N}{=}3$ SYM theory with the helicity assignment (\ref{eq:HexaDoubleBox-with-edge}) we get
\begin{equation} 
    \Omega = \int \underbrace{\frac{d\alpha_1\,d\alpha_2\dots d\alpha_8\,\,\delta(C\cdot Z)}{\alpha_1\alpha_2\alpha_3\alpha_4\alpha_5\alpha_6\alpha_7\alpha_8}}_{\Omega^{\rm bare}}\times \underbrace{\frac{1}{1-\alpha_2\alpha_4\alpha_7\alpha_8-\alpha_3\alpha_6\alpha_8}}_{\mathcal{J}^{-1}}.
\end{equation}
which evaluates to
\begin{equation}
\begin{aligned}
        \Omega &= \underbrace{\frac{\la 4| 1{+}5|6]\delta^6({\cal Q})\delta^4(P)\delta([56]\widetilde\eta_1+[61]\widetilde\eta_5+[15]\widetilde\eta_6)}{s_{156}\la 23 \ra \la 34 \ra
        \la 2 | 3{+}4 |5] \la 4 | 5{+}6 |1]
 [16][56]}}_{\Omega^{\rm bare}}\times \underbrace{\frac{s_{156}\la 23 \ra
 \la 4 | 5{+}6 |1]
 [16]}{
 \la 2 | 5{+}6 |1] \la 3 | 5{+}6 |1]
 \la 4 | 1{+}5 |6]}}_{\mathcal{J}^{-1}}\\
         &\hspace{2cm} = \frac{\delta^6({\cal Q})\delta^4(P)\delta([56]\widetilde\eta_1{+}[61]\widetilde\eta_5{+}[15]\widetilde\eta_6)}{\la 34 \ra
         \la 2 | 3{+}4 |1]\la 3 | 5{+}6 |1]
         \la 2 | 1{+}6 |5]
    [56]} \label{res6a}
\end{aligned}
\end{equation}
The inverse Jacobian adds two new poles $\la2|5{+}6|1]$ and $\la3|5{+}6|1]$ (the other term $\la4|1{+}5|6]$ is just a helicity factor). These are UV poles and originate from closed cycles in (\ref{eq:HexaDoubleBox-with-edge}). In order to see which UV pole corresponds to a hexagon and which one corresponds to a box, we solve for the internal momenta (where we denoted $Q_{156} = p_1{+}p_5{+}p_6$ in all formulas),
\begin{equation} \label{eq:HexaDoubleBox}
\raisebox{-23mm}{\includegraphics[trim={0cm 0cm 0cm 0cm},clip,scale=1.1]{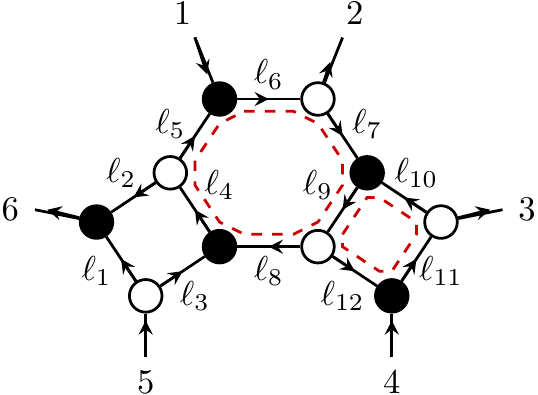}}
\end{equation}
\begin{equation}
    \begin{aligned}
\ell_1=\frac{ \langle 2 | Q_{156} | 5] }{\langle 2 | Q_{156} | 6] }\widetilde \lambda_6\lambda_5,~~~~
\ell_2=\frac{\widetilde \lambda_6 \left(\lambda_2\cdot Q_{34}\right)\cdot Q_{56}}{\langle 2 | Q_{156} | 6] },~~~~\ell_3=\frac{[56] \lambda_5 \left(\lambda_2\cdot Q_{34}\right)}{\langle 2 | Q_{156} | 6] }\\
\ell_4=\frac{[16]\left(\lambda_2\cdot Q_{34}\right)\cdot Q_{56}\left(\lambda_2\cdot Q_{34}\right)}{\langle 2 | Q_{156} | 6] \langle 2 | Q_{156} | 1] },~~~\ell_5=\frac{\widetilde\lambda_1\left(\lambda_2\cdot Q_{34}\right)\cdot Q_{56}}{\langle 2 | Q_{156} | 1] },~~~\ell_6=\frac{s_{156}}{\langle 2 | Q_{156} | 1] }\widetilde \lambda_1\lambda_2\\
\ell_7=\frac{\lambda_2(\widetilde\lambda_1\cdot Q_{56})\cdot Q_{34}}{\langle 2 | Q_{156} | 1]},~~~\ell_8=\frac{(\lambda_2\cdot Q_{34})(\widetilde \lambda_1\cdot Q_{56})}{\langle 2 | Q_{156} | 1]},~~\ell_9=\frac{\langle 23 \rangle(\widetilde \lambda_1\cdot Q_{56})\cdot Q_{34}(\widetilde \lambda_1\cdot Q_{56})}{\langle 3|Q_{156} | 1]\langle 2 | Q_{156} | 1]}\\\ell_{10}=\frac{\lambda_{3} (\widetilde \lambda_1\cdot Q_{56})\cdot Q_{34}}{\langle 3 | Q_{156} | 1]},~~~
\ell_{11}=\frac{\langle 4 | Q_{156}| 1]}{\langle 3 | Q_{156} | 1]}\widetilde \lambda_4\lambda_3,~~~~\ell_{12}=\frac{\la 3 4 \ra \widetilde \lambda_4 (\widetilde\lambda_1\cdot Q_{56})}{\langle 3 | Q_{156} | 1]} \nonumber
    \end{aligned}
\end{equation}
Looking at the denominator of solved on-shell loop momenta we see that $\la2|Q_{156}|6]=0$ is the pole at infinity in the left box diagram, the pole at infinity in the hexagon is $\la 2|Q_{156}|1]=0$, and the pole at infinity in the right box is $\la3|Q_{156}|1]=0$. The last two poles are accessible for this orientation as evident from the diagram (presence of both closed cycles) and the poles in the inverse Jacobian ${\cal J}^{-1}$, while the the pole $\la2|1{+}5|6]=0$ can not be reached in this orientation. The inverse Jacobian also removes some of the irremovable edges. For example we can not remove $\alpha_2$ edge (by sending $\alpha_2\rightarrow\infty$) forcing $s_{156}=0$ as it would lead to an illegal diagram,
\begin{equation} 
\raisebox{-19mm}{\includegraphics[trim={0cm 0cm 0cm 0cm},clip,scale=1.1]{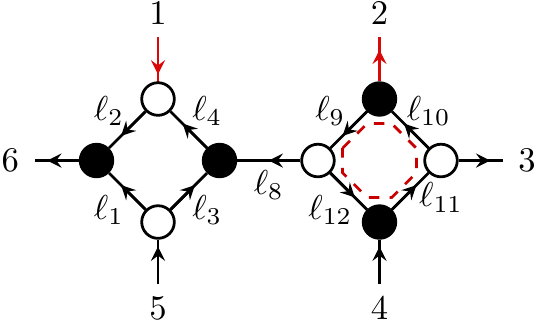}}~~.
\end{equation}
where we violated the rules about incoming/outgoing legs in black/white vertex. 

Going back to the diagram \eqref{eq:HexaDoubleBox} we concentrate on the pole at infinity in the hexagon subdiagram. There are two kinematical constraints imposed on the external legs of this hexagon $p_1,p_2,\ell_2,\ell_3,\ell_{10},\ell_{12}$, while in the context of the whole diagram the last four momenta are internal and the actual external momenta $p_1,{\dots},p_6$ are unconstrained. Using the rules developed in the last section, the residue on the UV pole turns the hexagon into a tree subdiagram,
\begin{equation} 
\raisebox{-20.5mm}
{\includegraphics[trim={0cm 0cm 0cm 0cm},clip,scale=1.1]{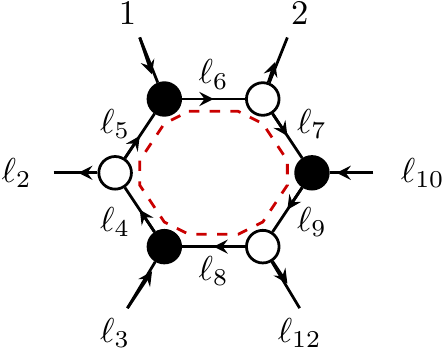}}\qquad
\longrightarrow\qquad
\raisebox{-11mm}
{\includegraphics[trim={0cm 0cm 0cm 0cm},clip,scale=1.1]{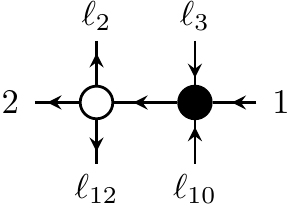}}
\label{eq:hex} 
\end{equation}
Gluing back into the rest of the diagram we get,
\begin{equation} \label{eq:doublepentagon-w-edge1}
\raisebox{-19mm}{\includegraphics[trim={0cm 0cm 0cm 0cm},clip,scale=1.1]{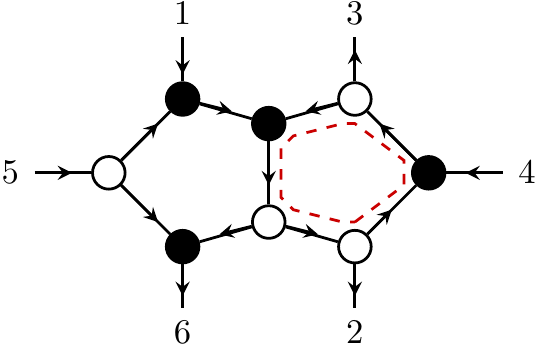}},
\end{equation}
where we used one particular way to expand the four-point black and white vertices. We can now treat it as a new on-shell diagram and assign edge variables, construct the $C$ matrix and solve for them using delta functions 
\begin{equation} \label{eq:doublepentagon-w-edge2}
\raisebox{-19mm}{\includegraphics[trim={0cm 0cm 0cm 0cm},clip,scale=1.1]{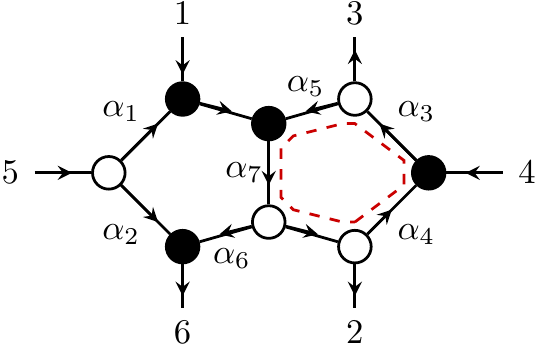}}
\begin {aligned}~~~~~~~
\alpha_1 &= \frac{[56]}{[16]},\,\,\alpha_2 = \frac{[15]}{[16]},
\alpha_3 =\frac{\langle 23 \rangle}{\langle 42 \rangle},\\ \alpha_4&= \frac{\langle 43 \rangle}{\langle 23 \rangle},\,\,
\alpha_5 = \frac{\langle 2 |1{+}5| 6 ] }{\langle 23\rangle [16]}, \\
\alpha_6 &= \frac{\langle 3 |5{+}6| 1 ] }{\langle 23\rangle [16]},\,\,\alpha_7 = \frac{\langle 23\rangle [16]}{\langle 3 |1{+}5| 6 ]}.
\end{aligned}
\end{equation}
Note that in this diagram there is one extra kinematical condition imposed: $\la2|5{+}6|1]=0$ which was exactly the hexagon UV pole in (\ref{eq:HexaDoubleBox}). The edge variables in (\ref{eq:doublepentagon-w-edge2}) are not related in any obvious way with the edge variables in (\ref{eq:doublepentagon-w-edge1}). Calculating the on-shell function for (\ref{eq:doublepentagon-w-edge2}) we get
\begin{equation}
\begin{aligned}
        \Omega &= \int \frac{d\alpha_1\dots d\alpha_7\,\delta(C\cdot Z)}{\alpha_1\alpha_2\alpha_3\alpha_4\alpha_5\alpha_6\alpha_7} \times \frac{1}{1-\alpha_3\alpha_4\alpha_5\alpha_7} \\
        &= \frac{\langle 4|1{+}5| 6]\,\delta^4(P)\delta^6({\cal Q})\delta([56]\widetilde\eta_1{+}[61]\widetilde\eta_5{+}[15]\widetilde\eta_6)\delta(\langle 2 |5{+}6| 1 ])}{s_{156}\langle 24\rangle \langle 34\rangle  \langle 3 |1{+}5| 6 ] [56][15]}\times \frac{\langle 24\rangle \langle 3 |1{+}5| 6 ]}{\langle 23\rangle \langle 4 |1{+}5| 6 ]}\\
        &=\frac{\delta^4(P)\delta^6({\cal Q})\delta([56]\widetilde\eta_1{+}[61]\widetilde\eta_5{+}[15]\widetilde\eta_6)\,\delta(\langle 2 |5{+}6| 1 ])}{s_{156}\langle 23\rangle \langle 34\rangle[56][15]}
\end{aligned}
\end{equation}
which is the same as the residue of (\ref{res6a}) on $\la2|5{+}6|1]=0$ showing explicitly our procedure in action. Note that since we are on special kinematics, $s_{156}\la 2 q\ra [1p]=\la 2|1{+}5{+}6|p]\la q|1{+}5{+}6|1]$ which is needed to show the equality.

\subsection{Non-planarity from the UV pole}

The pole at infinity does not preserve the planarity of the original on-shell diagram. We could already see it in our previous example where the original on-shell diagram (\ref{eq:HexaDoubleBox-with-edge}) had a canonical ordering of external legs, while the on-shell diagrams we obtained as a result of the UV pole calculation had external legs ordered differently $123456\rightarrow134265$. In fact, this was just a special case of a more general phenomenon that the residue on a UV pole of a planar on-shell diagram is a \emph{non-planar on-shell diagram}. Let us consider the following example, 
\begin{equation} \label{g36}
\raisebox{-29mm}{\includegraphics[trim={0cm 0cm 0cm 0cm},clip,scale=1.15]{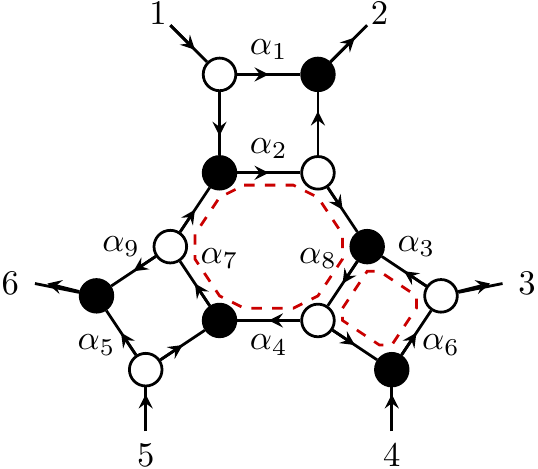}}
\end{equation}
which is an on-shell diagram for a top cell of $G_+(3,6)$. This is a unique planar 9-dimensional (has 9 independent edge variables $\alpha_j$) on-shell diagram and all other cells in $G_+(3,6)$ are associated with descendents of this diagram where one or more edges are removed. See \cite{Arkani-Hamed:2012zlh} for more details on the stratification. We can also obtain it by adding a BCFW bridge on external legs $1,2$ of the previous on-shell diagram (\ref{eq:HexaDoubleBox-with-edge}) -- which was an eight-dimensional leading singularity diagram (with no free parameters and generic kinematics). Hence in our diagram there is one unfixed parameter $\alpha_1$. The solution for the edge variables is 
\begin{equation}
	\begin{aligned}\,\,
\alpha_2 &=- \frac{\langle \widehat{2} | 5 {+} 6 | \widehat{1} ]}{s_{\widehat{1}56}},\quad
\alpha_3 =\frac{\langle \widehat{2} | 5 {+} 6 | \widehat{1} ]}{\langle 3 | 5 {+} 6 | \widehat{1} ]},\quad \alpha_4 = -\frac{\la 34 \ra [56]\langle \widehat{2} | 5 {+} 6 | \widehat{1} ]}{\langle \widehat{2} | 3 {+} 4 | 6 ]\langle 3 | 5 {+} 6 | \widehat{1} ]}\\ \alpha_5 &=- \frac{\langle \widehat{2} | 3 {+} 4 | 5 ]}{\langle \widehat{2} | 3 {+} 4 | 6 ]},~~
\alpha_6 = \frac{\langle 4 | 5 {+} 6 | \widehat{1} ]}{\langle 3 | 5 {+} 6 | \widehat{1} ]},~~
\alpha_7 = 
\frac{[\widehat{1}6]}{[56]},\,\,\alpha_8 =-  \frac{\la \widehat{2}3 \ra}{\la 34 \ra},~~ \alpha_9 =\frac{\langle \widehat{2} | 5 {+} 6 | \widehat{1} ]}{\langle \widehat{2} | 3 {+} 4 | 6 ]}.
\label{eq:alpha-sols}
\end{aligned}
\end{equation}
where we denoted $\widehat{\widetilde{\lambda}}_1=\widetilde{\lambda}_1+\alpha_1\widetilde{\lambda}_2$ and $\widehat{\lambda}_2=\lambda_2-\alpha_1\lambda_1$. 
The on-shell function is
\begin{equation}
    \Omega = \int \frac{d \alpha_1\dots d \alpha_9\, \delta(C \cdot \mathcal{Z})}{\alpha_1 \alpha_2 \alpha_3 \alpha_4 \alpha_5 \alpha_6 \alpha_7 \alpha_8 \alpha_9} \times \frac{1}{1-\alpha_2 \alpha_4 \alpha_7 \alpha_8-\alpha_3 \alpha_6 \alpha_8}
\end{equation}
Plugging in the edge variables we get,
\begin{align}
        \Omega &= \int \underbrace{\frac{d\alpha_1\,\langle 4 | \widehat{1} {+} 5 | 6]\delta(\Xi)  }{\alpha_1 s_{\widehat{1}56}\la \widehat{2}3 \ra \la 34 \ra \langle \widehat{2} | 3 {+} 4 | 5] \langle 4 | 5 {+} 6 | \widehat{1}] [\widehat{1}6][56]}}_{\Omega^{\rm bare}}
 \underbrace{\frac{s_{\widehat{1}56}\la \widehat{2}3 \ra \langle 4 | 5 {+} 6 | \widehat{1}][\widehat{1}6]}{\langle \widehat{2} | 5 {+} 6 | \widehat{1}]\langle 3 |  5 {+} 6 | \widehat{1} ](\langle 4 | \widehat{1} {+} 5 | 6])}}_{\mathcal{J}^{-1}}\\
         &= \int\frac{d\alpha_1\,\delta(\Xi) }{\alpha_1\la 34 \ra \langle \widehat{2} | 5 {+} 6 | \widehat{1} ]\langle 3 | 5 {+} 6 | \widehat{1}] \langle \widehat{2} | 3 {+} 4 | 5 ][56]}, \label{res6b}
\end{align}
where the delta function
\begin{equation}
    \delta(\Xi) \equiv \delta^6({\cal Q})\,\delta^4(P)\,\delta([56]\widetilde\eta_1{+}\alpha_1[56]\widetilde\eta_2{+}[6\widehat{1}]\widetilde\eta_5{+}[\widehat{1}5]\widetilde\eta_6).
\end{equation}
We can recognize the usual BCFW shift for $\widetilde{\eta}_1$ in the super delta function. Note that the $\alpha_1=0$ pole erases a corresponding edge and we reproduce the diagram already encountered (\ref{eq:HexaDoubleBox-with-edge}). In ${\cal N}{=}4$ SYM theory there are six erasable edges which do not impose contraints on external kinematics (poles in $\alpha_1$ in the $\Omega^{\rm bare}$). They correspond to six co-dimension-1 (i.e. eight-dimensional) cells of $G_+(3,6)$ and their on-shell functions are the six $R$-invariants. For our orientation (\ref{g36}) in ${\cal N}{<}3$ SYM we have three IR poles and two UV poles from the inverse Jacobian. The UV pole in the box is accessed by $\la3|5{+}6|\widehat{1}]=0$, the residue is 
\begin{equation}\label{gr36a}
\begin{aligned}
        \text{Res}_{\alpha_1=-\frac{\langle 3 | 5 {+} 6 | 1 ]}{\langle 3 | 5 {+} 6 | 2 ]}}\left[\Omega\right]
         &= \frac{\delta^6(\mathcal{Q})\delta^4(P)\delta([24]\widetilde\eta_1+[14]\widetilde\eta_2+[12]\widetilde\eta_4)}{\la 5 6 \ra [12] \langle 3 | 5 {+} 6 | 1 ] \langle 3 | 1 {+} 2 | 4 ] \langle 6 | 1 {+} 2 |4]}
\end{aligned}
\end{equation}
Based on our diagrammatic rule this UV pole corresponds to an on-shell diagram,
\begin{equation} \label{g36res}
\raisebox{-29mm}{\includegraphics[trim={0cm 0cm 0cm 0cm},clip,scale=1.1]{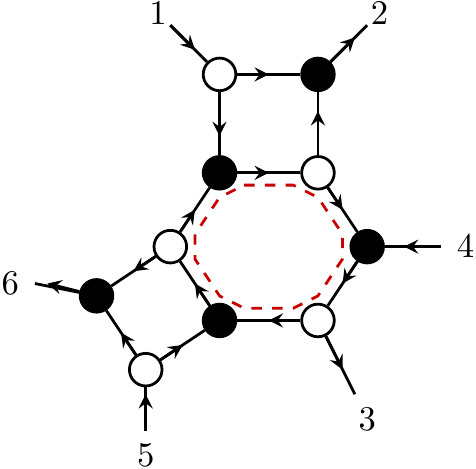}}
\end{equation}
Calculating the on-shell function we get
\begin{equation}
\begin{aligned}
        \Omega_{\rm UV} &= \underbrace{\frac{\la 5| 1{+}4|2]\,\delta^6({\cal Q})\delta^4(P)\delta([24]\widetilde\eta_1{+}[14]\widetilde\eta_2{+}[12]\widetilde\eta_4)}{s_{124}\la 35 \ra \la 56 \ra
        \la 3 | 2{+}4 |1] \la 6 | 1{+}2 |4]
 [12][24]}}_{\Omega^{\rm bare}}\times \underbrace{\frac{s_{124}\la 35 \ra
 [24]}{
 \la 3 | 1{+}2 |4] \la 5 | 1{+}4 |2]}}_{\mathcal{J}^{-1}}\\
         &\hspace{2cm} =\frac{\delta^6(\mathcal{Q})\delta^4(P)\delta([24]\widetilde\eta_1{+}[14]\widetilde\eta_2{+}[12]\widetilde\eta_4)}{\la 5 6 \ra [12] \langle 3 | 5 {+} 6 | 1 ] \langle 3 | 1 {+} 2 | 4 ] \langle 6 | 1 {+} 2 |4]}\label{reshexagon}
\end{aligned}
\end{equation}
which agrees with (\ref{gr36a}). The UV pole in the hexagon subdiagram corresponds to $\la \widehat{2} | 5 {+} 6 | \widehat{1} ]=0$ which is a quadratic condition in $\alpha_1$, 
\begin{equation}
    \langle 2 | 5 {+} 6 | 1]-\alpha_1\langle 1 | 5 {+} 6 |  1]+\alpha_1\langle 2 | 5 {+} 6 | 2]+\alpha_1^2\langle  1 | 5 {+} 6 |  2]=0
\end{equation}
and leads to two solutions,
\begin{equation}
    \begin{aligned}
\alpha_1^\pm=\frac{(\langle  2 | 5 {+} 6 |  2]-\langle  1 | 5 {+} 6 |  1])\pm \sqrt{(\langle  1 | 5 {+} 6 |  1]+\langle  2 | 5 {+} 6 |  2])^2-4s_{12}s_{56}}}{2\la 1 |5 +6 |2]}.
    \end{aligned}
\end{equation}
Plugging back into the form for the top cell \eqref{res6b} using $\langle \widehat{2} | 5 {+} 6 | \widehat{1} ]=\langle 1 | 5 {+} 6 |2 ](\alpha_1-\alpha_1^+)(\alpha_1-\alpha_1^-)$, we get an on-shell function which contains square-roots of kinematical variables. 
\begin{equation}
   \hspace{-0.5cm} \Omega = \int\frac{d\alpha_1\, \delta^6({\cal Q})\delta^4(P)\,\delta([56]\widetilde\eta_1{+}\alpha_1[56]\widetilde\eta_2{+}[6\widehat{1}]\widetilde\eta_5{+}[\widehat{1}5]\widetilde\eta_6)}{\alpha_1\la 34 \ra \langle 1 | 5 {+} 6 | 2 ]
    (\alpha_1{-}\alpha_1^+)(\alpha_1{-}\alpha_1^-)
    (\langle 3 | 5 {+} 6 | 1 ]{+}\alpha_1\langle 3 | 5 {+} 6 | 2 ]) (\langle 2 | 3 {+} 4 | 5 ]{-}\alpha_1 \langle 1 | 3 {+} 4 | 5 ])[56]}.\label{res6d}
\end{equation}
The quadratic (or higher order) equation for a parameter we localize and the presence of roots is not something unusual. For example, it also happens for the four-mass-box leading singularity, and in general we get very complicated structures. The UV pole in (\ref{eq:HexaDoubleBox-with-edge}) can be again calculated diagrammatically using our operation: replacing the hexagon with the tree diagram (\ref{eq:hexafac3}). But now, something new happens: as a result, we get a \emph{non-planar on-shell diagram}
\begin{equation} \label{eq:nonplanar}
\raisebox{-19mm}{\includegraphics[trim={0cm 0cm 0cm 0cm},clip,scale=1.1]{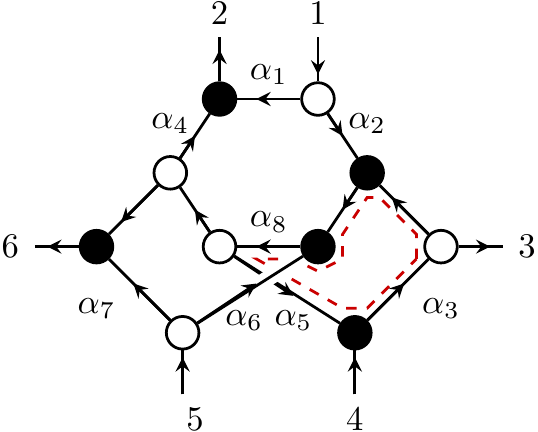}}
\end{equation}
The expression for the on-shell diagram using edge variables is the same as in the planar case and we can evaluate
\begin{equation}
\Omega_{\rm UV} =
\pm\frac{\delta^6({\cal Q})\delta^4(P)\,\delta([56]\widetilde\eta_1{+}\alpha_1^{\pm}[56]\widetilde\eta_2{+}[6\widehat{1}]\widetilde\eta_5{+}[\widehat{1}5]\widetilde\eta_6)}{\alpha_1^{\pm}\langle 34\rangle \la 1 | 5 {+}6 | 2] (\alpha_1^- {-}\alpha_1^+ )
s_{\widehat{1}56}\langle \widehat{2}3\rangle    [\widehat{1}5][56]} 
\label{nonplanar-shell}
\end{equation}
where $\widehat{1}$ and $\widehat{2}$ are defined by the same convention as \eqref{eq:alpha-sols} and $\alpha_1^{\pm}$ is:
\begin{equation}
    \begin{aligned}
\alpha_1^\pm=\frac{(\langle  2 | 5 {+} 6 |  2]-\langle  1 | 5 {+} 6 |  1])\pm \sqrt{(\langle  1 | 5 {+} 6 |  1]+\langle  2 | 5 {+} 6 |  2])^2-4s_{12}s_{56}}}{2\la 1 |5 +6 |2]}.
    \end{aligned}
\end{equation}
which agrees with the residue on either the plus or minus pole of (\ref{res6d}). This six-point NMHV non-planar diagram corresponds to one of the eight dimensional cells in $G(3,6)$ (not the positive part). All these cells for $G(3,6)$ were classified in \cite{Bourjaily:2016mnp}, and the subset of them relevant for BCFW recursion relations (for higher $k$ and $n$) in \cite{Paranjape:2022ymg}.

\subsection{UV poles and modified GRTs}

One of the most important features of on-shell functions $\Omega$ for leading singularity on-shell diagrams is that they satisfy Global Residue Theorems (GRTs). The leading singularity on-shell diagram has a dimensionality $2n{-}4$ and the on-shell function has neither free parameters, nor are there any constraints imposed on external kinematics (other than momentum conservation). For the on-shell diagram of dimensionality $2n{-}3$, the on-shell function (after solving for edge variables) has one free parameter $z$, i.e. $\Omega(z)$. We can write Cauchy's integral theorem for this function,
\begin{equation}
    \oint dz\,\Omega(z) = 0 \label{GRT1}.
\end{equation}
If the on-shell diagram has no pole at infinity, then all poles in $z$ come from erasable edges and we can write
\begin{equation}
   {\rm GRT}\,\,= \sum_j \Omega_j = 0 \quad \mbox{where} \quad \Omega_j = {\rm Res}_{z=z_j}\Omega(z)
   \label{GRT2}
\end{equation}
Let us look at the most famous example which is our diagram (\ref{g36}). In ${\cal N}{=}4$ SYM theory the on-shell form is given by $\Omega=\Omega^{\rm bare}$,
\begin{equation}
    \Omega = \int \frac{d\alpha_1\dots d\alpha_9\,\delta(C\cdot Z)}{\alpha_1\alpha_2\dots\alpha_9} = \int \frac{d\alpha_1\,\delta(\Xi)  }{\alpha_1 s_{\widehat{1}56}\la \widehat{2}3 \ra \la 34 \ra \langle \widehat{2} | 3 {+} 4 | 5] \langle 4 | 5 {+} 6 | \widehat{1}] [\widehat{1}6][56]}.
\end{equation}
There are six poles in this on-shell function corresponding to erasing edges $\alpha_1$, $\alpha_2$, $\alpha_5$, $\alpha_6$, $\alpha_7$, $\alpha_8$. Going to each of these six boundaries can be illustrated as follows,
\begin{equation} \label{eq:Top-cell}
\raisebox{-19mm}{\includegraphics[trim={0cm 0cm 0cm 0cm},clip,scale=1.1]{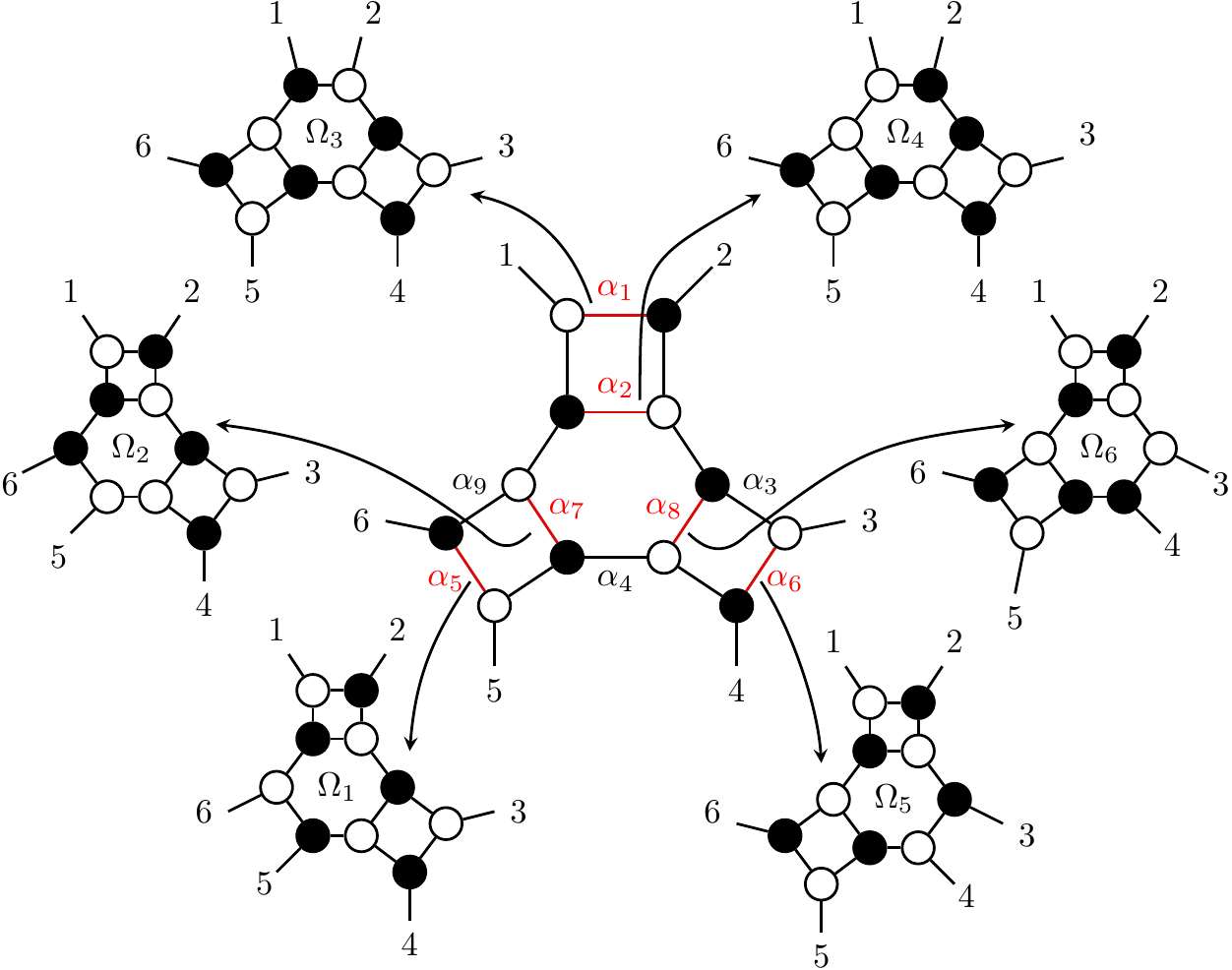}}
\end{equation}
where each arrow starts at the edge that is being erased, and all erasable edges are shown in red. This leads to the GRT in the form
\begin{equation}
    \sum_{i=1}^6 \Omega_i = 0,\quad\mbox{with}\quad   
    \Omega_1 = \frac{\delta^4(P)\delta^8({\cal Q})\,\delta([34]\widetilde{\eta}_5 {+} [45]\widetilde{\eta}_3 {+} [53]\widetilde{\eta}_4)}{s_{345}[34][45]\la16\ra\la12\ra\la 2|3{+}4|5]\la 6|1{+}2|3]} \label{GRT3}
\end{equation}
where $\Omega_1=R[23456]$ is a famous $R$-invariant, and all other $\Omega_j$ are related by cyclic shifts. The BCFW recursion relations for two different shifts represent the six-point NMHV tree-level as
\begin{equation}
    {\cal A}^{\rm tree}_{6,3} = \Omega_1 + \Omega_3 + \Omega_5 = - \Omega_2 - \Omega_4 - \Omega_6 
\end{equation}
where the equality of two representations is guaranteed by the GRT (\ref{GRT3}). All these relations were crucial in the early discovery of momentum twistor polytopes \cite{Hodges:2009hk,Arkani-Hamed:2010wgm}, on-shell diagrams and positive Grassmannians \cite{Arkani-Hamed:2009ljj,Arkani-Hamed:2009nll,Mason:2009qx,Arkani-Hamed:2012zlh} and later the Amplituhedron construction \cite{Arkani-Hamed:2013jha,Arkani-Hamed:2017vfh}.

For ${\cal N}<4$ SYM the Global Residue Theorems change dramatically. The Jacobian removes some of the poles in $\Omega(z)$ but also adds poles at infinity. The on-shell functions associated with IR poles also change as they are corrected by the Jacobian. We work in the context of ${\cal N}=3$ SYM theory and consider the same on-shell diagram for the top cell $G_+(3,6)$. At first, we choose the orientation with one pole at infinity,
\begin{equation} \label{eq:HexaDoubleBox-with-edge2}
\raisebox{-23mm}{\includegraphics[trim={0cm 0cm 0cm 0cm},clip,scale=1.1]{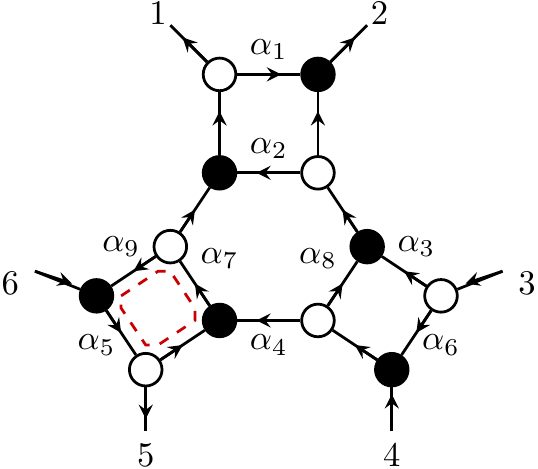}}	
\end{equation}
Solving for edge variables and plugging in the formula we get for the on-shell function,
\begin{equation}
   \begin{aligned}
         \Omega &= \int \underbrace{\frac{d\alpha_1\, \la 34\ra [\widehat{1}5] \delta(\Xi)  }{\alpha_1 s_{\widehat{1}56}\la \widehat{2}3 \ra \la 34 \ra \la \widehat{2} | 3{+}4|5] \la 4 | 5{+}6|\widehat{1}] [\widehat{1}6][56]}}_{\Omega^{\rm bare}}
 \underbrace{\frac{\la \widehat{2} |3{+}4|5][\widehat{1}6]}{\la \widehat{2} |3{+}4|6][\widehat{1}5]}}_{\mathcal{J}^{-1}}\\ 
 &= \int \frac{ d\alpha_1\, \delta(\Xi)  }{\alpha_1 s_{\widehat{1}56}\la \widehat{2}3 \ra\la 4|5{+}6|\widehat{1}] \la \widehat{2} |3{+}4|6][56]},
   \end{aligned}
\end{equation}
where, again, $ \delta(\Xi) \equiv \delta^6(\mathcal{Q})\delta^4(P)\delta([56]\widetilde\eta_1{+}\alpha_1 [56]\widetilde\eta_2{+}[6\widehat{1}]\widetilde\eta_5{+}[\widehat{1}5]\widetilde\eta_6).$
Taking the residue on the UV pole $\alpha_1 = \frac{\la 2 |3{+}4|6]}{\la 1 |3{+}4|6]}$ one obtains
\begin{equation}\label{eq:nmhvbox}
   \begin{aligned}
 \Omega_{\text{UV}} &= \frac{ \delta^6({\cal Q}) \delta^4(P) \delta([46]\widetilde\eta_3{+}[36]\widetilde\eta_4{+}[34]\widetilde\eta_6)  }{\la 15 \ra \la 1 2 \ra \la 2|3{+}4|6] \la 5|2{+}4|3][34][46]},
   \end{aligned}
\end{equation}
which matches the form obtained by collapsing the box into a tree as per the prescription. We can use this to define a modified global residue theorem (GRT):
\begin{equation}
   \begin{aligned}
\text{GRT} = \widetilde{\Omega}_{3}+\widetilde{\Omega}_{4}+\widetilde{\Omega}_{5}+\widetilde{\Omega}_{6}+\Omega_{\text{UV}} = 0
   \end{aligned}
\end{equation}
with the IR poles given by
\begin{align}
\widetilde{\Omega}_3 &=  \frac{ \delta^6({\cal Q})\delta^4(P)\delta([56]\widetilde\eta_1{+}[16]\widetilde\eta_5{+}[15]\widetilde\eta_6) }{ s_{156}\la 2 3 \ra \la 4|5{+}6|1] \la 2 |3{+}4|6][56]}, \quad
\widetilde{\Omega}_4 = \frac{\delta^6({\cal Q})\delta^4(P)\delta([23]\widetilde\eta_2{+}[24]\widetilde\eta_3{+}[23]\widetilde\eta_4)}{s_{156}  \la 1 | 5{+}6 |4] [23][34] \la 1 5 \ra \la 5 6 \ra},\nonumber \\ 
\widetilde{\Omega}_5 &=  \frac{\delta^6({\cal Q})\delta^4(P)\delta([23]\widetilde\eta_1{+}[13]\widetilde\eta_2{+}[12]\widetilde\eta_3)}{s_{123} \la 5 6 \ra \la 5 | 1{+}2 |3] \la 4 |5{+}6|2] [13][12]}, \quad
\widetilde{\Omega}_6 = \frac{ \delta^6({\cal Q})\delta^4(P)\delta([56]\widetilde\eta_4{+}[46]\widetilde\eta_5{+}[45]\widetilde\eta_6) }{ s_{456}\la 2 3 \ra  \la 1 2 \ra \la 1|5{+}6|4][56][46] }.
\end{align}
We can illustrate the GRT as before,
\begin{equation} \label{eq:Top-cell2}
\raisebox{-55mm}{\includegraphics[trim={0cm 0cm 0cm 0cm},clip,scale=1.05]{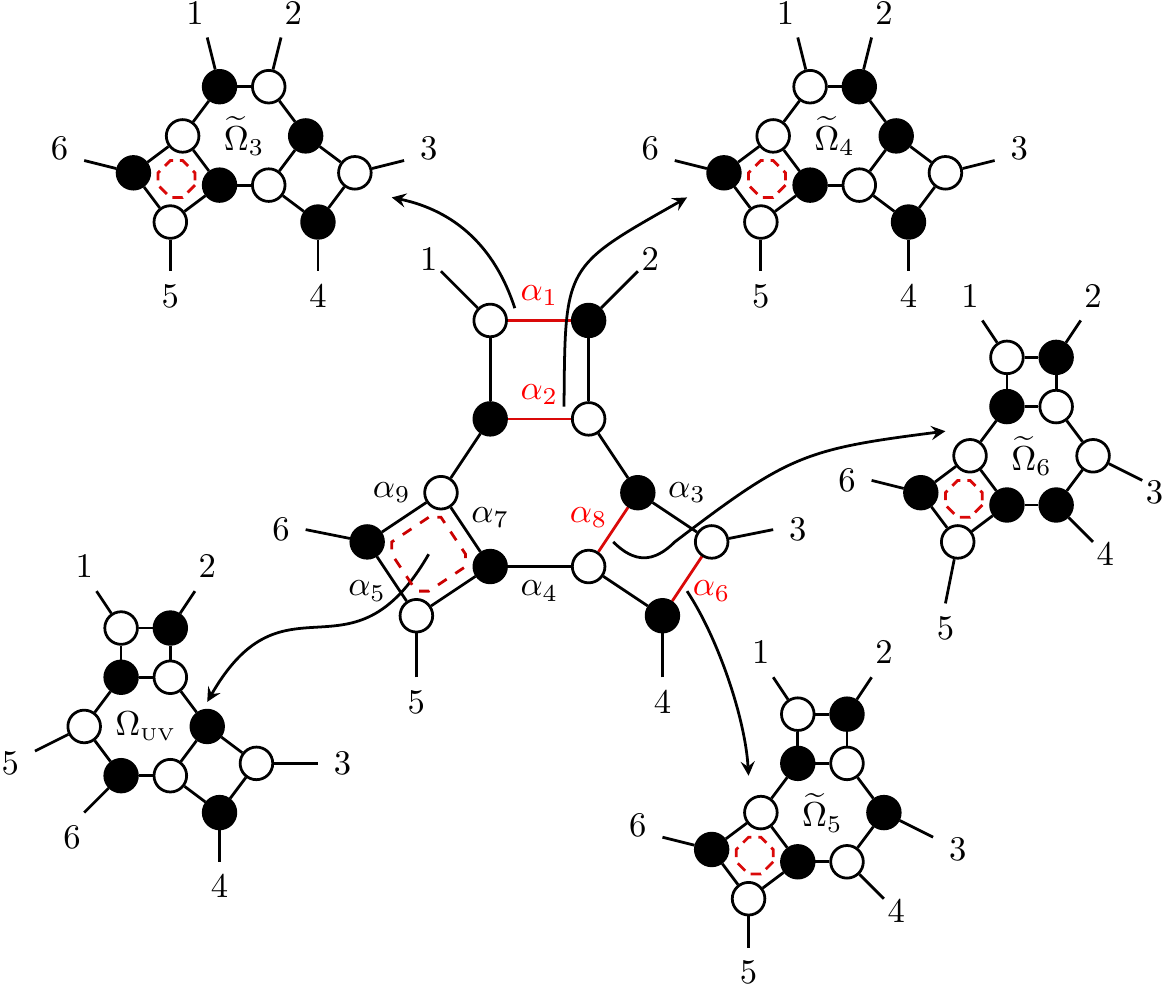}}
\end{equation}
where each arrow starts at the edge that is being erased, and all erasable edges are shown in red, while the accesible UV pole is shown using the dashed internal cycle. Notice, how $\alpha_5$ and $\alpha_7$ now no longer are removable compared to the normal GRT in \eqref{eq:Top-cell}. Also note that the on-shell functions $\widetilde{\Omega}$ contain Jacobians from the internal cycle in the box. This is a new type of GRT that relates the on-shell function for the UV pole to IR poles. Note that in this case all diagrams are planar, but the on-shell diagram for the UV pole does not respect the original canonical ordering with $5\leftrightarrow6$. Now we consider a more complicated case of the same diagram with the orientation discussed in the previous subsection, 
\begin{equation} 
\raisebox{-29mm}{\includegraphics[trim={0cm 0cm 0cm 0cm},clip,scale=1.05]{Figures/TopCell3-6-w-edge.pdf}}\quad,
\end{equation}
where the pole at infinity in the hexagon subdiagram is also accessible. In that case the Jacobian removes the IR poles for edges $\alpha_2$, $\alpha_6$, $\alpha_7$, $\alpha_8$ leaving only two IR poles $\alpha_5=\la \hat{2}|3{+}4|5]=0$ and $\alpha_1=0$. On the other hand we add 3 UV poles -- a pair of poles from the hexagon loop for $\la \hat{2}|5{+}6|\hat{1}]=0$ and one UV pole from the right box subdiagram $\la3|5{+}6|\hat{1}]=0$ as before. The global residue theorem in this case can be written as 
\begin{equation}
    {\rm GRT}\,\, = \widetilde{\Omega}_1 + \widetilde{\Omega}_3 + \Omega_{{\rm UV}_1}^++ \Omega_{{\rm UV}_1}^- + \Omega_{{\rm UV}_2} = 0\label{GRT4}
\end{equation}
where 
\begin{align}
&\widetilde{\Omega}_1 = \frac{\delta^4(P) \delta^6({\cal Q}) \delta([45]\widetilde\eta_3 + [35]\widetilde\eta_4 + [34]\widetilde\eta_5) }{[34]\langle 1 2 \rangle \langle 6 | 3{+} 5 | 4 ]\langle 2 | 3 {+} 4 | 5 ] \langle 6 | 3 {+} 4 | 5 ] },\quad
 \widetilde{\Omega}_3 = \frac{\delta^6({\cal Q})\delta^4(P)\delta([56]\widetilde\eta_1{+}[61]\widetilde\eta_5{+}[15]\widetilde\eta_6)}{\la 34 \ra
         \la 2 | 5{+}6 |1]\la 3 | 5{+}6 |1]
         \la 2 | 1{+}6 |5]
    [56]} \nonumber \\
&\hspace{2cm}\Omega^{\pm}_{\text{UV}_1} = 
\pm\frac{\delta^6({\cal Q})\delta^4(P)\,\delta([56]\widetilde\eta_1+\alpha_1^{\pm}[56]\widetilde\eta_2+[6\widehat{1}]\widetilde\eta_5{+}[\widehat{1}5]\widetilde\eta_6)}{ \la 1 | 5 {+}6 | 2] (\alpha_1^- -\alpha_1^+ )
s_{\widehat{1}56}\langle \widehat{2}3\rangle \langle 34\rangle   [56][\widehat{1}5]\alpha_1^{\pm}} \nonumber\\
&\hspace{2cm}\Omega_{{\rm UV}_2} = \frac{\delta^6(\mathcal{Q})\delta^4(P)\delta([24]\widetilde\eta_1+[14]\widetilde\eta_2+[12]\widetilde\eta_4)}{\la 5 6 \ra [12] \langle 3 | 5 {+} 6 | 1 ] \langle 3 | 1 {+} 2 | 4 ] \langle 6 | 1 {+} 2 |4]} ,\label{res6e}
\end{align}
with
\begin{equation}
    \begin{aligned}
\alpha_1^\pm=\frac{(\langle  2 | 5 {+} 6 |  2]+\langle  1 | 5 {+} 6 |  1])\pm \sqrt{(\langle  1 | 5 {+} 6 |  1]+\langle  2 | 5 {+} 6 |  2])^2-4s_{12}s_{56}}}{2\la 1 |5 +6 |2]}.
    \end{aligned}
\end{equation}
This is a new type of residue theorem for on-shell functions for two IR poles (modified $R$-invariants) and two UV poles -- one is a planar $R$-invariant for a different ordering and the other is a pair of non-planar on-shell functions. Note that while each on-shell function $\Omega^{\pm}_{\text{UV}_1}$ contains square-roots, they do cancel in the sum as evident from the GRT (\ref{GRT4}). Once again we can illustrate the GRT in this case,
\begin{equation} \label{eq:Top-cell3}
\raisebox{-45mm}{\includegraphics[trim={0cm 0cm 0cm 0cm},clip,scale=1.1]{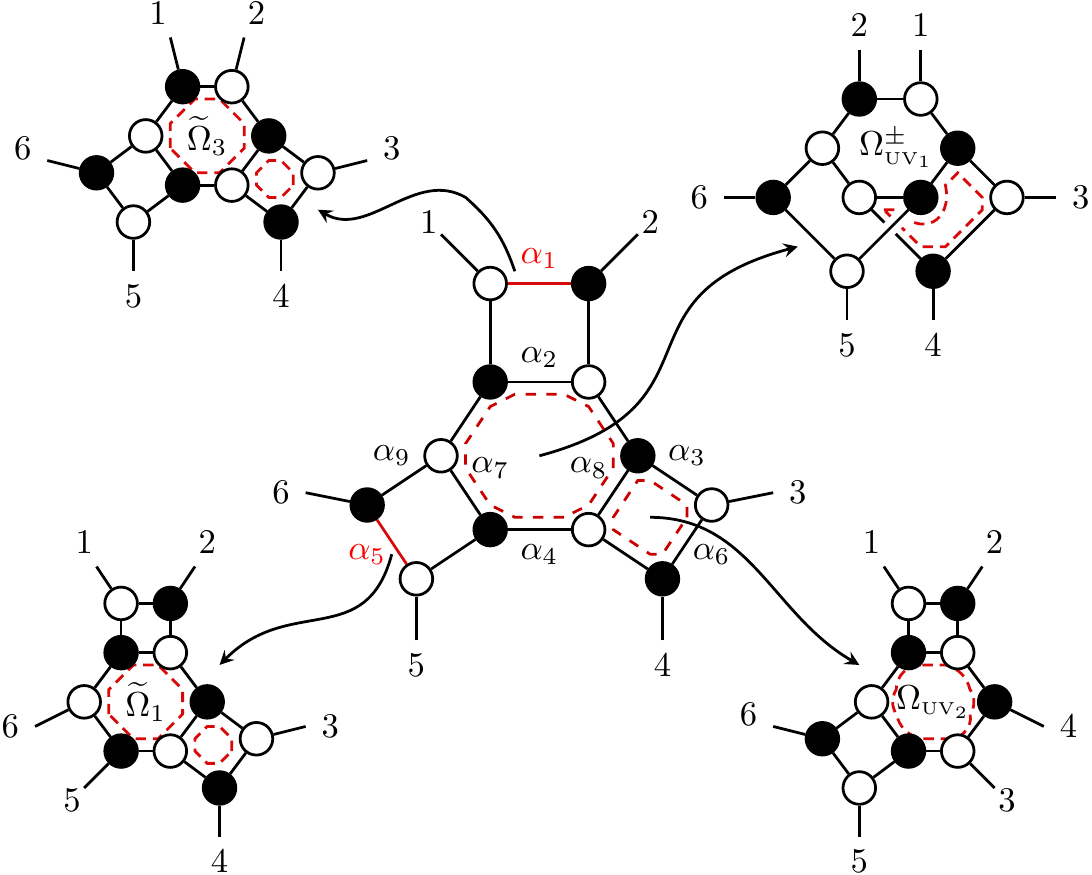}}
\end{equation}
from which it is clear that the only two removable edges left are $\alpha_5$ and $\alpha_1$. 

In ${\cal N}{=}4$ SYM theory we do not have any poles at infinity whatsoever, so we can never generate non-planar diagrams from planar diagrams. However, even in ${\cal N}=4$ SYM theory we can just directly consider non-planar on-shell diagrams, but with no orientations and internal cycles. The on-shell functions in ${\cal N}=4$ SYM for the same diagrams are then
\begin{align}
\raisebox{-22mm}{\includegraphics[trim={0cm 0cm 0cm 0cm},clip,scale=1.05]{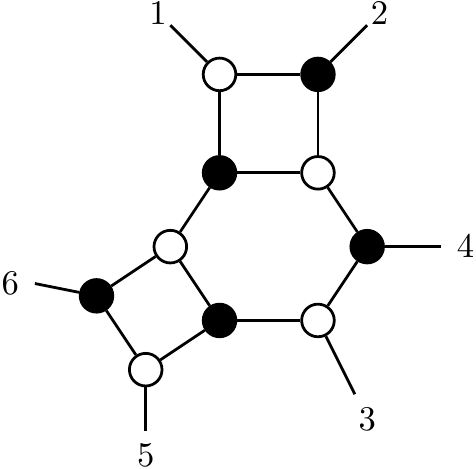}} &= \frac{\delta^4(P)\delta^8(\mathcal{Q})\delta([24]\widetilde\eta_1+[14]\widetilde\eta_2+[12]\widetilde\eta_4)}{s_{124}\la 35 \ra \la 5 6 \ra  \langle 3 | 5 {+} 6 | 1 ]  \langle 6 | 1 {+} 2 |4][12][24]}\\
 \raisebox{-20mm}{\includegraphics[trim={0cm 0cm 0cm 0cm},clip,scale=1.05]{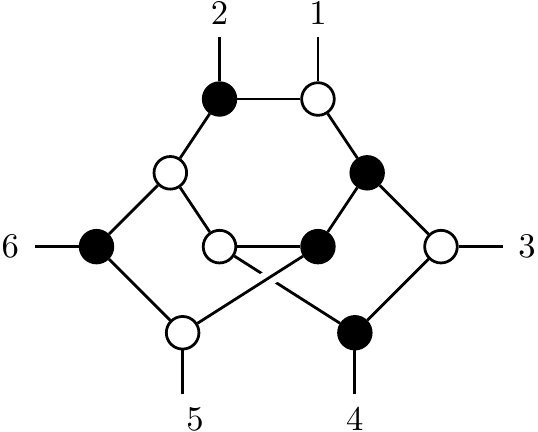}} &= \pm \frac{\delta^4(P)\delta^8({\cal Q})\,\delta([56]\widetilde\eta_1{+}\alpha_1^{\pm}[56]\widetilde\eta_2{+}[6\widehat{1}]\widetilde\eta_5{+}[\widehat{1}5]\widetilde\eta_6)}{ (\alpha_1^- {-}\alpha_1^+ )s_{\widehat{1}56}\langle \widehat{2}4\rangle \langle 34\rangle  \la 1 | 5 {+}6 | 2] \langle 3 |\widehat{1}{+}5| 6 ] [56][\widehat{1}5]\alpha_1^\pm}
\end{align}
where the first on-shell function is just a regular R-invariant with $5\leftrightarrow6$ from canonical ordering, as also evident from the on-shell diagram. The on-shell function for the second, genuine non-planar, on-shell diagram is of a new type and is not related to an R-invariant as also evident from the presence of square-roots. It would be interesting to identify this form with the items in the list of appendix B of \cite{Bourjaily:2016mnp}. 

Having thoroughly gone through the procedure of taking UV poles of general $n$-gons and how to incorporate this into larger on-shell diagrams in $\mathcal{N}{=}3$ we are ready to tackle the higher order polls at infinity that arise in $\mathcal{N}{<}3$ in the next section.
%
\section{Higher order poles at infinity}

So far we have studied the minimal version of the problem of UV poles in the context of ${\cal N}=3$ SYM theory where all poles at infinity are simple poles. We want to generalize it to gauge theories with lower supersymmetries, including pure Yang-Mills theory. As explained in previous sections, there are two parts of the story: (i) locations of poles at infinity and the kinematics of on-shell diagrams which are produced on these poles; and (ii) the actual residues at infinity, i.e. on-shell functions associated with on-shell diagrams. The first part of the story is general and not specific to any theory, hence our discussion is valid for all relevant QFTs (with fundamental 3-point vertices) including gravity. The second part, the on-shell function, is specific to a particular theory and here, indeed, we solved only the case of ${\cal N}=3$ SYM and need to find a generalization to other theories. 

\subsection{UV pole as a residue}

Let us consider ${\cal N}=2$ SYM theory, the case of general ${\cal N}$ is just a trivial generalization. The main challenge is to figure out how to deal with higher poles at infinity which come from the inverse Jacobian ${\cal J}^{-1}$. For ${\cal N}=2$ these are double poles. First, we redo the box on-shell diagram (\ref{four4N3}) and get 
\begin{equation}
        \raisebox{-15mm}{\includegraphics[trim={0cm 0cm 0cm 0cm},clip,scale=1.2]{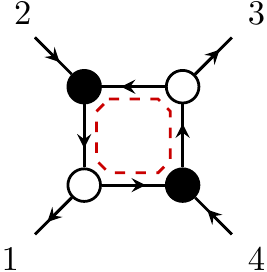}}
\begin{aligned}
    = \frac{\delta^4(P)\delta^4({\cal Q})\la 24 \ra^2}{\la 12 \ra \la 23 \ra \la 34 \ra \la 14 \ra} \times \left( \frac{\la 23 \ra \la 14 \ra }{\la 13 \ra \la 24 \ra}\right)^2 = \frac{\la23\ra\la14\ra\,\delta^4(P)\delta^4({\cal Q})}{\la 12 \ra \la 34 \ra \la 13\ra^2}
\end{aligned} \label{four5}
\end{equation}
where the supermomentum delta function has the same form, but now in $\eta_i^I$, $I=1,2$. Our goal is to calculate the residue on the UV pole $\la13\ra=0$. Just based on the little group weights and the mass dimension, the resulting expression must be 
\begin{equation}
    \underset{\la13\ra=0}{\rm Res}\,\, \frac{\la23\ra\la14\ra\,\delta^4(P)\delta^4({\cal Q})}{\la 12 \ra \la 34 \ra \la 13\ra^2} \sim \frac{\la24\ra\,\delta^4(P)\delta^4({\cal Q})\delta(\la13\ra)}{\la12\ra\la34\ra}
\end{equation}
Note that on the support of $\la13\ra=0$ we have a relation $\la12\ra\la34\ra=\la23\ra\la14\ra$. We want to derive this result by taking the residue of (\ref{four5}). 

The first issue is to find a good parametrization that allows us to approach the pole in a nice way. One approach is to write $\la13\ra=\lambda_1^{(1)}\lambda_3^{(2)}{-}\lambda_1^{(2)}\lambda_3^{(1)}$, fix three of the components and then solve for the last one, e.g. $\lambda_3^{(2)}$, making sure that $\la13\ra=0$.
However, we need to have in mind that through momentum conservation this parameter also appears in other momenta, which makes this approach less clean. Therefore, it is simplest to shift the external kinematics, e.g. using a BCFW shift $\lambda_3\rightarrow \lambda_3 + z \lambda_4$, $\widetilde{\lambda}_4 \rightarrow \widetilde{\lambda}_4 - z \widetilde{\lambda}_3$, we get $\Omega(z)$ but with an extra $\delta(z)$ ensuring we preserve the same kinematics. Then we take a residue on $\la13\ra \rightarrow \la 13\ra {+} z \la14\ra=0$ using the formula (\ref{higher}), 
\begin{equation} 
    \Omega_{\rm UV} =\underset{z=-\frac{\la13\ra}{\la14\ra}}{\rm Res} 
    \frac{\la14\ra(\la23\ra + z\la24\ra)\,\delta^4(P)\delta^4({\cal Q})\delta(z)}{\la12\ra\la34\ra(\la13\ra+z\la14\ra)^2} = \frac{\la24\ra\,\delta^4(P)\delta^4({\cal Q})\delta(\la13\ra)}{\la12\ra\la34\ra}
    \label{UVN2b}
\end{equation}
where we used $\delta\left(-\frac{\la13\ra}{\la14\ra}\right) = - \frac{\delta(\la13\ra)}{\la14\ra}$ and for the calculation of the residue on the higher pole we used the definition
\begin{equation} \label{higher}
\underset{x=x_0}{\rm Res}\,f(x) = \frac{1}{(m-1)!}\lim_{x \to x_0} \frac{{\rm d}^{m-1}}{{\rm d}x^{m-1}}\left(\left(x-x_0\right)^mf(x)\right)
\end{equation}
Same result is obtained using the shift $\lambda_3\rightarrow \lambda_3 + z \lambda_2$, $\widetilde{\lambda}_2 \rightarrow \widetilde{\lambda}_2 - z \widetilde{\lambda}_3$,
\begin{equation}
    \Omega_{\rm UV}=\underset{z=-\frac{\la13\ra}{\la12\ra}}{\rm Res} 
    \frac{\la14\ra\la23\ra\,\delta^4(P)\delta^4({\cal Q})\delta(z)}{\la12\ra(\la34\ra+z\la24\ra)(\la13\ra+z\la12\ra)^2}= \frac{\la24\ra\,\delta^4(P)\delta^4({\cal Q})\delta(\la13\ra)}{\la23\ra\la14\ra}
    \label{UVN2c}
\end{equation}
which is the same as \eqref{UVN2b} using $\la12\ra\la34\ra=\la14\ra\la23\ra$ on the support of $\delta(\la 13 \ra)$. We can show that the result is independent of the shift. For general ${\cal N}<4$ we can repeat the same exercise and get,
\begin{equation}
\Omega_{\rm UV} = \frac{\la24\ra^{3-{\cal N}}\,\delta^4(P)\delta^{2{\cal N}}({\cal Q})\delta(\la13\ra)}{\la12\ra\la34\ra}
\end{equation}
Note that m-shell function for the on-shell diagram (\ref{resinf2}) with additional helicity factor for ${\cal N}<3$ supersymmetry. Therefore, our diagrammatic rule works perfectly here for any ${\cal N}$ for this case. Let us look at the double box on-shell diagram in ${\cal N}=2$ SYM,
\begin{equation}
    \raisebox{-15mm}{\includegraphics[trim={0cm 0cm 0cm 0cm},clip,scale=1.15]{Figures/DoubleBoxCycle1.pdf}}
    \begin{aligned}
    &=\int dz \frac{\la 2 4 \ra^2 \delta^4(P)\delta^4({\cal Q})}{z(\la 2 3 \ra + z\la 24 \ra)\la 1 4 \ra \la 3 4 \ra \la 1 2 \ra} \times \left(\frac{\la 1 4 \ra ( \la 2 3 \ra + z\la 2 4 \ra)}{\la 2 4 \ra (\la 1 3 \ra + z\la 1 4 \ra)}\right)^2\\
    &\hspace{2cm} = \int dz\frac{\la14\ra(\la23\ra+z\la24\ra)\,\delta^4(P)\delta^4({\cal Q})}{z\la12\ra\la34\ra(\la13\ra+z\la14\ra)^2}
\end{aligned} \label{DoubleBoxCycle1}
\end{equation}
This on-shell diagram has only two poles: IR pole for $z=0$ (deleting the far right edge) and the double UV pole which blows up the left loop. The residue on the IR pole is
\begin{equation}
    \Omega_{\rm IR} = \underset{z=0}{\rm Res}\,\, \Omega(z) = \frac{\delta^4(P)\delta^4({\cal Q})\,\la14\ra\la23\ra}{\la12\ra\la34\ra\la13\ra^2}
\end{equation}
which is a box with an internal loop (\ref{four5}) as expected. The residue on the UV pole is
\begin{equation}
    \Omega_{\rm UV} \,\, =\underset{z=-\frac{\la13\ra}{\la14\ra}}{\rm Res} \Omega(z) = -\frac{\delta^4(P)\delta^4({\cal Q})\,\la14\ra\la23\ra}{\la12\ra\la34\ra\la13\ra^2} \label{UVN2a}
\end{equation}
and the sum $\Omega_{\rm IR}+\Omega_{\rm UV} = 0$ as dictated by the GRT for $\Omega$. Note that the on-shell diagram associated with the UV pole (based on our rule) should be just a box without an internal loop and an ordering $(1342)$. However, the on-shell function obtained in (\ref{UVN2a}) is not an on-shell function for this diagram,
\begin{equation}
    \raisebox{-13mm}{\includegraphics[trim={0cm 0cm 0cm 0cm},clip,scale=1.1]{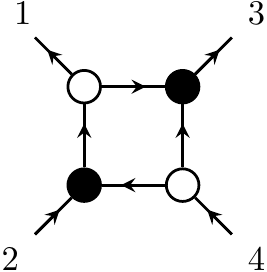}}
\end{equation}
But rather, is its given by
\begin{equation}
    \begin{aligned}
     \Omega_{\rm UV} = \underbrace{\frac{\la24\ra^2\,\delta^4(P)\delta^4({\cal Q})}{\la13\ra\la34\ra\la42\ra\la21\ra}}_{\Omega_D} \times \underbrace{\frac{\la14\ra\la23\ra}{\la13\ra\la24\ra}}_{\cal K}
\end{aligned}
\end{equation}
where $\Omega_D$ is the on-shell function associated with the diagram and ${\cal K}$ is some prefactor. This formula generalizes to arbitrary ${\cal N}$,
\begin{equation}
    \Omega_{\rm UV} = \frac{\delta^4(P)\delta^{2{\cal N}}({\cal Q})\,\la24\ra^{4-{\cal N}}}{\la13\ra\la34\ra\la42\ra\la21\ra} \times \left(\frac{\la14\ra\la23\ra}{\la13\ra\la24\ra}\right)^{3-{\cal N}} = \frac{(\la14\ra\la23\ra)^{3-{\cal N}}\delta^4(P)\delta^{2{\cal N}}({\cal Q})}{\la13\ra^{4-{\cal N}}\la34\ra\la42\ra\la21\ra}
\end{equation}
We can see that the situation is not as simple as in the ${\cal N}{=}3$ SYM case, the on-shell function is now corrected by some prefactor. This prefactor does not add any new poles, it only increases the degree of some of the poles. We will see shortly that it is generated by a derivative operator acting on this on-shell diagram.

\subsection{Infinity operator}

To start with, let us consider a generalization of the two-loop double box on-shell diagram,
\begin{equation}\label{blobdiag}
 \raisebox{-17.5mm}{\includegraphics[trim={0cm 0cm 0cm 0cm},clip,scale=1.1]{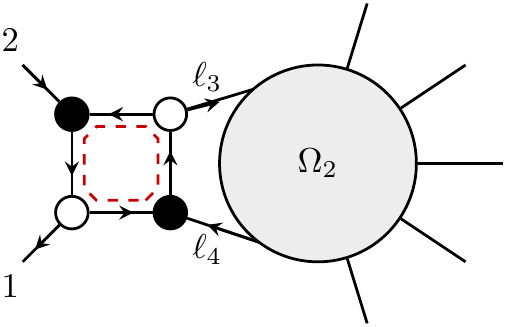}}
\end{equation}
where the one-loop box on-shell diagram with an internal loop is glued to an arbitrary other on-shell diagram through two on-shell internal legs $\ell_3$, $\ell_4$. The on-shell function for the diagram for ${\cal N}=2$ SYM can be then written as 
\begin{equation} 
    \Omega = \int d^2\widetilde{\eta}_{\ell_3} d^2\widetilde{\eta}_{\ell_4}
    \int \frac{d^2\lambda_{\ell_3}\,d^2\widetilde{\lambda}_{\ell_3}}{{\rm GL(1)}}
    \int \frac{d^2\lambda_{\ell_4}\,d^2\widetilde{\lambda}_{\ell_4}}{{\rm GL(1)}}
   \,\Big\{ \Omega_1 \times \Omega_2\Big\}\label{OmegaDef}
\end{equation}   
The on-shell functions for left and right parts of the diagram are:
\begin{equation}
    \Omega_1 =  \frac{\la 2\ell_4\ra^2\,\delta^4(P_1)\delta^4({\cal Q}_1)}{\la12\ra\la2\ell_3\ra\la \ell_3\ell_4\ra\la \ell_41\ra}\left(\frac{\la1\ell_4\ra\la2\ell_3\ra}{\la2\ell_4\ra\la1\ell_3\ra}\right)^2 = \frac{\la2\ell_3\ra\la1\ell_4\ra\,\delta^4(P_1)\delta^4({\cal Q}_1)}{\la12\ra\la\ell_3\ell_4\ra\la1\ell_3\ra^2}
\end{equation}
\begin{equation}
\Omega_2 = \delta^4(P_2)\delta^4({\cal Q}_2)\, \omega_2
\end{equation}
where $P_1$, $P_2$ are momentum conservation and $Q_1$, $Q_2$ are supermomentum conservation for each part of the diagram. $\omega_2$ for the right part of the diagram is arbitrary, and while it can contain additional delta functions or extra parameters, it does not depend on legs $1,2$ (only through $\ell_3$, $\ell_4$). Now we want to calculate the residue on the pole at infinity $\la1\ell_3\ra=0$ of the box diagram. In order to do that we have to take $\lambda_{\ell_3}$ to be collinear with $\lambda_1$. However, due to momentum conservation any changes in $\lambda_{\ell_3}$ also propagate into other momenta, and it is relevant when we calculate the derivative of the function in (\ref{higher}). We can remove this dependence by integrating over $\ell_4$, then $\ell_3$ becomes independent,
\begin{equation}
    \Omega = \int d^2\widetilde{\eta}_{\ell_3} 
    \int \frac{d^2\lambda_{\ell_3}\,d^2\widetilde{\lambda}_{\ell_3}}{{\rm GL(1)}}
    \Bigg\{\frac{\la12\ra}{\la1\ell_3\ra^2}\times
   \underbrace{ \int d^2\widetilde{\eta}_{\ell_4}\int \frac{d^2\lambda_{\ell_4}\,d^2\widetilde{\lambda}_{\ell_4}}{{\rm GL(1)}}\frac{\la2\ell_3\ra\la1\ell_4\ra\,\delta^4(P_1)\delta^4({\cal Q}_1)}{\la12\ra^2\la\ell_3\ell_4\ra}\times \Omega_2}_{\omega}\Bigg\} \label{Glue5}
\end{equation}
where we defined an on-shell function $\omega$ which automatically implements the momentum conservation between $\ell_3$, $\ell_4$ and external legs, and the momentum $\ell_3$ is now independent and not subject to any constraints. In this form we can take a residue on the pole $\la1\ell_3\ra=0$. In order to approach the pole we parametrize
\begin{equation}
\lambda_{\ell_3} = \alpha \lambda_1 + \beta \lambda_2 \label{param}
\end{equation}
and rewrite 
\begin{equation}
    \Omega = \int d^2\widetilde{\eta}_{\ell_3}
    \int \frac{d\alpha\,d\beta\,d^2\widetilde{\lambda}_{\ell_3}}{{\rm GL(1)}}
    \Bigg\{\frac{1}{\beta^2}\times\omega\Bigg\} \label{Glue6}
\end{equation}
The UV pole is then given by the residue on $\beta=0$,
\begin{equation}
\Omega_{\rm UV} = \int d^2\widetilde{\eta}_{\ell_3} 
    \int \frac{d\alpha\,d^2\widetilde{\lambda}_{\ell_3}}{{\rm GL(1)}} \Bigg\{\frac{d\omega}{d\beta}\Big|_{\beta=0}\Bigg\}
\end{equation}
We can also undo the parametrization (\ref{param}) and rewrite the result in a more invariant way, 
\begin{equation}
    \Omega_{\rm UV} = \int d^2\widetilde{\eta}_{\ell_3} 
    \int \frac{d^2\lambda_{\ell_3}\,d^2\widetilde{\lambda}_{\ell_3}}{{\rm GL(1)}}
    \Bigg\{
    \delta(\la1\ell_3\ra)\left\langle \lambda_2 \frac{d\omega}{d\lambda_{\ell_3}}\right\rangle
    \Bigg\} \label{Glue7}
\end{equation}
In the angle bracket we take the derivative of $\omega$ with respect to $\lambda_{\ell_3}$ and then contract with $\lambda_2$. This is the on-shell function associated with the pole at infinity in (\ref{blobdiag}). Now we need to compare (\ref{Glue7}) with the on-shell diagram for the collapsed box diagram,
\begin{equation}\label{blobdiag2}
 \raisebox{-17.5mm}{\includegraphics[trim={0cm 0cm 0cm 0cm},clip,scale=1.1]{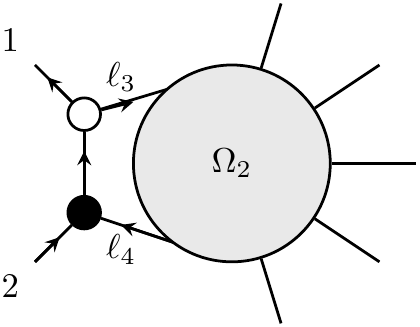}}
\end{equation}
The on-shell function for this diagram is 
\begin{equation}
    \Omega^{\rm bare}_{\rm UV} = \int d^2\widetilde{\eta}_{\ell_3} 
    \int \frac{d^2\lambda_{\ell_3}\,d^2\widetilde{\lambda}_{\ell_3}}{{\rm GL(1)}} \Bigg\{\delta(\la1\ell_3\ra)
    \underbrace{\int d^2\widetilde{\eta}_{\ell_4}\int \frac{d^2\lambda_{\ell_4}\,d^2\widetilde{\lambda}_{\ell_4}}{{\rm GL(1)}}
   \,\frac{\la2\ell_4\ra\, \delta^4(P_1)\delta^4({\cal Q}_1)}{\la12\ra\la\ell_3\ell_4\ra}\times \Omega_2}_{\omega^{\rm bare}}\Big\}\label{Glue8}
\end{equation}   
where we again eliminated $\ell_4$. Comparing (\ref{Glue7}) and (\ref{Glue8}) we find the relation 
\begin{equation}
 \left\langle \lambda_2 \frac{d\omega}{d\lambda_{\ell_3}}\right\rangle = \frac{\la2\ell_3\ra[2\ell_3]}{\la12\ra[1\ell_3]}\left\langle \lambda_2\frac{d\omega^{\rm bare}}{d\lambda_{\ell_3}}\right\rangle \label{deriv}
\end{equation}
Plugging back we get
\begin{equation}
    \Omega_{\rm UV} = \int d^2\widetilde{\eta}_{\ell_3}
    \int \frac{d^2\lambda_{\ell_3}\,d^2\widetilde{\lambda}_{\ell_3}}{{\rm GL(1)}} \Bigg\{\delta(\la1\ell_3\ra) \frac{\la2\ell_3\ra[2\ell_3]}{\la12\ra[1\ell_3]}\left\langle \lambda_2\frac{d\omega^{\rm bare}}{d\lambda_{\ell_3}}\right\rangle\Bigg\}
     \label{Glue9}
\end{equation}   
We can then interpret the result in the following way: the pole at infinity of the original on-shell diagram (\ref{blobdiag}) for ${\cal N}=2$ SYM is equal to the ``infinity operator''
\begin{equation}
 {\cal O} = \frac{\la2\ell_3\ra[2\ell_3]}{\la12\ra[1\ell_3]}\left\langle\lambda_2\,\frac{d}{d\lambda_{\ell_3}}\right\rangle \label{operator}
\end{equation}
acting on the bare on-shell function
\begin{equation}\label{blobdiag4}\Omega_{\rm UV}= \frac{\la2\ell_3\ra[2\ell_3]}{\la12\ra[1\ell_3]}\left\langle\lambda_2\,\frac{d}{d\lambda_{\ell_3}}\right\rangle \quad \otimes\hspace*{-0.5cm}
 \raisebox{-19mm}{\includegraphics[trim={0cm 0cm 0cm 0cm},clip,scale=1.1]{Figures/4ptboxblob4.pdf}}
\end{equation}
or, more schematically $\Omega_{\rm UV} = {\cal O}\otimes \Omega^{\rm bare}_{\rm UV}$ in the sense of the formulas (\ref{Glue8}) and (\ref{deriv}). Note that for ${\cal N}=3$ SYM the operator ${\cal O}=\mathbb{I}$ and the residue of $\Omega$ on the pole at infinity is equal to $\Omega_{\rm UV} = \Omega^{\rm bare}_{\rm UV}$. This relation is very easy to generalize for arbitrary ${\cal N}<4$. In that case the operator ${\cal O}$ is given by 
\begin{equation}
 {\cal O}^{\cal N} = \frac{1}{(3{-}{\cal N})!}\left(\frac{\la2\ell_3\ra[2\ell_3]}{\la12\ra[1\ell_3]}\left\langle\lambda_2\,\frac{d}{d\lambda_{\ell_3}}\right\rangle\right)^{3{-}{\cal N}}
\end{equation}
Note that the derivative commutes with the prefactor. As an example of our procedure, we look at the double box on-shell diagram studied in the previous section. We interpret the diagram in the context of (\ref{blobdiag}) as a one-loop box with legs $1,2,\ell_3,\ell_4$ glued to the tree-level diagram with legs $3,4,\ell_3,\ell_4$,
\begin{equation}
    \raisebox{-15mm}{\includegraphics[trim={0cm 0cm 0cm 0cm},clip,scale=1.2]{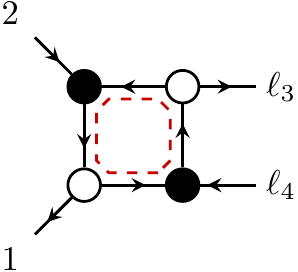}}
    \quad\times\quad     \raisebox{-13.5mm}{\includegraphics[trim={0cm 0cm 0cm 0cm},clip,scale=1.2]{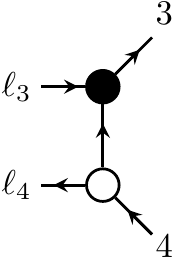}} \label{doublebox3}
\end{equation}
The on-shell function $\Omega_2$ is now
\begin{equation}
  \Omega_2 = \frac{\la\ell_34\ra^2\,\delta^4(P_2)\delta^4({\cal Q}_2)\delta(\la4\ell_4\ra)}{\la\ell_3 3\ra\la34\ra\la\ell_3\ell_4\ra} = \frac{\la\ell_34\ra[3\ell_4]\,\delta^4(P_2)\delta^4({\cal Q}_2)\delta([3\ell_3])}{\la\ell_3 3\ra\la34\ra\la\ell_3\ell_4\ra} \label{Omega2}
\end{equation}
where we rewrote the delta function on the support of the momentum conservation,
\begin{equation}
    \delta(\la4\ell_4\ra) = [\ell_43]\,\delta(\la4|\ell_4|3]) = -[\ell_43]\,\delta(\la4|\ell_3|3]) = \frac{[3\ell_4]}{ \la\ell_34\ra}\,\delta([3\ell_3])
\end{equation}
where the function $\omega$, further integrated over $\widetilde{\eta}_{\ell_3}$ and $\widetilde{\lambda}_{\ell_3}$ for simplicity (this does not interfere with $\lambda_{\ell_3}$ derivative) is equal to
\begin{align}
 \int d^2\widetilde{\eta}_{\ell_3}\,\int \frac{d^2\widetilde{\lambda}_{\ell_3}}{{\rm GL(1)}}\,\omega = \nonumber \\
 &\hspace{-3.3cm}= \int d^2\widetilde{\eta}_{\ell_3}\,d^2\widetilde{\eta}_{\ell_4}\int \frac{d^2\widetilde{\lambda}_{\ell_3}}{{\rm GL(1)}}\int \frac{d^2\lambda_{\ell_4}\,d^2\widetilde{\lambda}_{\ell_4}}{{\rm GL(1)}}
   \,\frac{\la1\ell_4\ra[\ell_43]\la2\ell_3\ra\la4\ell_3\ra\, \delta^4(P_1)\delta^4({\cal Q}_1)\delta^4(P_2)\delta^4({\cal Q}_2)\delta([3\ell_3])}{\la12\ra^2\la34\ra\la3\ell_3\ra\la\ell_3\ell_4\ra^2} \nonumber\\
   & \hspace{-3.3cm} =\int \frac{d^2\widetilde{\lambda}_{\ell_3}}{{\rm GL(1)}}\frac{(\la1\ell_3\ra[\ell_33]+\la14\ra[34])\la2\ell_3\ra\la4\ell_3\ra\, \delta^4(P)\delta^4({\cal Q})\delta((p_3{+}p_4{-}\ell_3)^2)\delta([3\ell_3])}{\la12\ra^2\la34\ra\la3\ell_3\ra} 
\end{align}
where the integration over $\widetilde{\eta}_{\ell_3}$, $\widetilde{\eta}_{\ell_4}$ reproduced an overall supermomentum conservation $\delta^4({\cal Q})$ and an extra factor $\la\ell_3\ell_4\ra^2$ while the integral over $\ell_4$ lead to an overall momentum conservation $\delta^4(P)$ and an extra delta function $\delta((p_3{+}p_4{-}\ell_3)^2)$. We can also integrate over $\widetilde{\lambda}_{\ell_3}$ which fixes $\widetilde{\lambda}_{\ell_3}=\widetilde{\lambda}_3$ and factorizes the delta function 
\begin{equation}
    \delta((p_3{+}p_4{-}\ell_3)^2) = \frac{1}{[34]}\delta(\la34\ra{-}\la 4\ell_3\ra)
\end{equation}
Plugging back we get 
\begin{equation} 
 \int d^2\widetilde{\eta}_{\ell_3}\,\int \frac{d^2\widetilde{\lambda}_{\ell_3}}{{\rm GL(1)}}\,\omega = \frac{\la 14\ra\la2\ell_3\ra\la4\ell_3\ra\, \delta^4(P)\delta^4({\cal Q})\delta(\la34\ra{-}\la 4\ell_3\ra)}{\la12\ra^2\la34\ra\la3\ell_3\ra}\label{UVpoledef}
\end{equation}
Now we calculate the derivative with respect to $\lambda_{\ell_3}$ and contract with $\lambda_2$,
\begin{equation}
\int d^2\widetilde{\eta}_{\ell_3}\,\int \frac{d^2\widetilde{\lambda}_{\ell_3}}{{\rm GL(1)}}\,\left\langle \lambda_2\frac{d\omega}{d\lambda_{\ell_3}}\right\rangle = \frac{\la14\ra\la2\ell_3\ra(\la2\ell_3\ra\la34\ra^2{+}\la23\ra\la4\ell_3\ra^2)\,\delta^4(P)\delta^4({\cal Q})\delta((\la34\ra{-}\la 4\ell_3\ra)^2)}{\la12\ra^2\la34\ra\la3\ell_3\ra^2}
\end{equation}
where all delta functions are here considered in the context of contour integrals and residues, i.e. under the integral $\delta(z)\sim \frac{1}{z}$. This is important since the derivative also acts on the delta function. Plugging back into (\ref{Glue7}) and integrating over $\lambda_{\ell_3}$ we get
\begin{equation}
\Omega_{\rm UV} = \delta^4(P)\,\delta^4({\cal Q})\,\int d^2\lambda_{\ell_3} \frac{\la14\ra\la2\ell_3\ra(\la2\ell_3\ra\la34\ra^2{+}\la23\ra\la4\ell_3\ra^2)\,\delta^4(P)\delta^4({\cal Q})\delta((\la34\ra{-}\la 4\ell_3\ra)^2)\delta(\la1\ell_3\ra)}{\la12\ra^2\la34\ra\la3\ell_3\ra^2}
\end{equation}
In order to make the integration easier, we can expand $\ell_3$ using (\ref{param}),
\begin{align}
\Omega_{\rm UV} &= \frac{\la14\ra}{\la12\ra\la34\ra}\,\delta^4(P)\,\delta^4({\cal Q})\, \oint d\alpha\,d\beta \frac{\alpha^2\la12\ra\la34\ra^2+\alpha\la23\ra(\alpha\la14\ra+\beta\la24\ra)^2}{\beta(\la34\ra {+} \alpha \la14\ra {+} \beta \la 24\ra)^2(\alpha \la 13\ra {+} \beta \la23\ra)^2}\nonumber\\
&= -\frac{\la14\ra\la23\ra\,\delta^4(P)\,\delta^4({\cal Q})}{\la12\ra\la34\ra\la13\ra^2} \label{ResA1}
\end{align}
where the integral was evaluated on the contour given by the delta functions, $\beta=0$ and $\alpha=-\la34\ra/\la14\ra$ (this was a double pole). Now we compare (\ref{ResA1}) with the infinity operator (\ref{operator}) acting on the on-shell diagram
\begin{equation}\label{Box8}
 \raisebox{-15mm}{\includegraphics[trim={0cm 0cm 0cm 0cm},clip,scale=1.1]{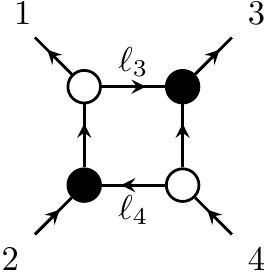}}
\end{equation}
which is obtained by collapsing the left box with the UV pole in (\ref{doublebox3}). The on-shell function $\omega^{\rm bare}$ for $\Omega_2$ for the diagram (\ref{Box8}), again integrated over $\widetilde{\eta}_{\ell_3}$ and $\widetilde{\lambda}_{\ell_3}$ is 
\begin{align}
\int d^2\widetilde{\eta}_{\ell_3}\,\int \frac{d^2\widetilde{\lambda}_{\ell_3}}{{\rm GL(1)}}\,\omega^{\rm bare} &=\\
&\hspace{-3cm} =\int d^2\widetilde{\eta}_{\ell_3}\,d^2\widetilde{\eta}_{\ell_4}\int \frac{d^2\widetilde{\lambda}_{\ell_3}}{{\rm GL(1)}}\int \frac{d^2\lambda_{\ell_4}\,d^2\widetilde{\lambda}_{\ell_4}}{{\rm GL(1)}}
   \,\frac{\la2\ell_4\ra\la4\ell_3\ra[3\ell_4]\, \delta^4(P_1)\,\delta^4({\cal Q}_1)\delta^4(P_2)\delta^4({\cal Q}_2)\delta([3\ell_3])}{\la12\ra\la34\ra\la\ell_3\ell_4\ra^2\la3\ell_3\ra}
   \nonumber\\
   &= \frac{\la24\ra\la4\ell_3\ra\, \delta^4(P)\delta^4({\cal Q})\delta(\la34\ra{-}\la4\ell_3\ra)}{\la12\ra\la34\ra\la3\ell_3\ra}
\end{align}
If we now integrated over $\lambda_{\ell_3}$ we would just recover the ordinary on-shell function for the box on-shell diagram (\ref{Box8}),
\begin{equation}
\Omega^{\rm bare} = \int d^2\lambda_{\ell_3}  \frac{\la24\ra\la4\ell_3\ra\, \delta^4(P)\delta^4({\cal Q})\delta(\la34\ra{-}\la4\ell_3\ra)\delta(\la 1\ell_3\ra)}{\la12\ra\la34\ra\la3\ell_3\ra} = \frac{\la24\ra\delta^4(P)\delta^4({\cal Q})}{\la12\ra\la13\ra\la34\ra}
\end{equation}
where the contour integral over the delta functions fixed $\lambda_{\ell_3} = \frac{\la34\ra}{\la14\ra}\lambda_1$. But instead of that we have to calculate the derivative,
\begin{align}
&\int d^2\widetilde{\eta}_{\ell_3}\,\int \frac{d^2\widetilde{\lambda}_{\ell_3}}{{\rm GL(1)}}\,\frac{\la2\ell_3\ra[2\ell_3]}{\la12\ra[1\ell_3]}\left\langle \lambda_2\frac{d\omega^{\rm bare}}{d\lambda_{\ell_3}}\right\rangle\nonumber\\
&\hspace{2cm}= 
\frac{[23]\la24\ra\la2\ell_3\ra(\la34\ra^2\la2\ell_3\ra{+}\la23\ra\la4\ell_3\ra^2)\,\delta^4(P)\delta^4({\cal Q})\delta((\la34\ra{-}\la4\ell_3\ra)^2)}{[13]\la12\ra^2\la34\ra\la3\ell_3\ra^2}
\end{align}
and then integrate over $\lambda_{\ell_3}$,
\begin{align}
    \Omega_{\rm UV} &= \int d^2\lambda_{\ell_3}\,\frac{[23]\la24\ra\la2\ell_3\ra(\la34\ra^2\la2\ell_3\ra{+}\la23\ra\la4\ell_3\ra^2)\,\delta^4(P)\delta^4({\cal Q})\delta((\la34\ra{-}\la4\ell_3\ra)^2)\delta(\la1\ell_3\ra)}{[13]\la12\ra^2\la34\ra\la3\ell_3\ra^2} \nonumber\\
    &\hspace{2cm}= -\frac{\la14\ra\la23\ra\,\delta^4(P)\,\delta^4({\cal Q})}{\la12\ra\la34\ra\la13\ra^2}
     \label{ResA2}
\end{align}   
with a perfect agreement with (\ref{ResA1}). In pure Yang-Mills theory (${\cal N}{=}0$) we act with the operator ${\cal O}$ three times, and get 
\begin{align}
    \Omega_{\rm UV} &= \frac{1}{6} \int d^2\widetilde{\eta}_{\ell_3}\,\int \frac{d^2\lambda_{\ell_3}\,d^2\widetilde{\lambda}_{\ell_3}}{{\rm GL(1)}}\,\left(\frac{\la2\ell_3\ra[2\ell_3]}{\la12\ra[1\ell_3]}\right)^3\left\langle \lambda_2\frac{d}{d\lambda_{\ell_3}}\left\langle \lambda_2 \frac{d}{d\lambda_{\ell_3}}\left\langle \lambda_2\frac{d\omega^{\rm bare}}{d\lambda_{\ell_3}}\right\rangle\right\rangle\right\rangle\nonumber\\ 
    &\hspace{2cm}= -\frac{\la14\ra^3\la23\ra^3\,\delta^4(P)}{\la12\ra\la34\ra\la13\ra^4}
     \label{ResA3}
\end{align}   
where $\omega^{\rm bare}$ is the same as before up to an extra $\la24\ra^2$ helicity factor and supermomentum delta function. Finally, let us summarize the procedure of the calculation of the UV pole:
\begin{itemize}
\item Calculate the bare on-shell function $\omega^{\rm bare}$ of a lower-loop on-shell diagram obtained by diagrammatic rules: a collapse of the loop and a non-planar twist. This includes an integration over the internal leg $\ell_4$ which eliminates the dependencies from the momentum conservation.
\item Take the appropriate number of derivatives with respect to $\lambda_{\ell_3}$ and plug into (\ref{ResA3}). This is the value of the original on-shell diagram on the UV pole.
\end{itemize}

\subsection{General on-shell diagram}

We are now ready to solve a more general problem where the box diagram with the UV pole is inside an arbitrary on-shell diagram,
\begin{equation}\label{blobdiag3}
 \raisebox{-24mm}{\includegraphics[trim={0cm 0cm 0cm 0cm},clip,scale=1.1]{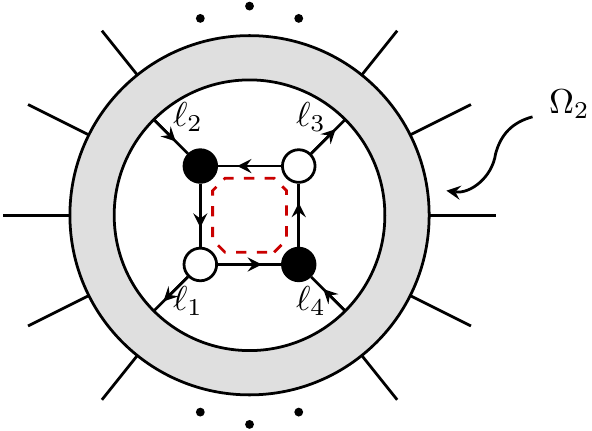}}
\end{equation}
The on-shell function for the diagram is
\begin{equation}
    \Omega = \int d^2\widetilde{\eta}_{\ell_1} \dots d^2\widetilde{\eta}_{\ell_4}
    \int \frac{d^2\lambda_{\ell_1}\,d^2\widetilde{\lambda}_{\ell_1}}{{\rm GL(1)}}\dots
    \int \frac{d^2\lambda_{\ell_4}\,d^2\widetilde{\lambda}_{\ell_4}}{{\rm GL(1)}}
   \,\Big\{ \Omega_1 \times \Omega_2\Big\}
\end{equation}   
where on-shell functions for the internal box and the rest of the diagram are 
\begin{equation}
\Omega_1 = \frac{\la \ell_2\ell_4\ra^2 \delta^4(P_1)\,\delta^4({\cal Q}_1)}{\la\ell_1\ell_2\ra\la\ell_2\ell_3\ra\la\ell_3\ell_4\ra\la\ell_4\ell_1\ra} \left(\frac{\la\ell_1\ell_4\ra\la\ell_2\ell_3\ra}{\la\ell_2\ell_4\ra\la\ell_1\ell_3\ra}\right)^2,\quad \Omega_2 = \delta^4(P_2)\,\delta^4({\cal Q}_2)\,\omega_2
\end{equation}
We rewrite this expression as 
\begin{equation}
\Omega = \int d^2\widetilde{\eta}_{\ell_1} \dots d^2\widetilde{\eta}_{\ell_3}
    \int \frac{d^2\lambda_{\ell_1}\,d^2\widetilde{\lambda}_{\ell_1}}{{\rm GL(1)}}\dots \int \frac{d^2\lambda_{\ell_3}\,d^2\widetilde{\lambda}_{\ell_3}}{{\rm GL(1)}} \Bigg\{ \frac{\la \ell_1\ell_2\ra}{\la \ell_1\ell_3\ra^2} \times \omega \Bigg\}
\end{equation}
where the function $\omega$ is defined as 
\begin{equation}
\omega = 
    \int d^2\widetilde{\eta}_{\ell_4} \int \frac{d^2\lambda_{\ell_4}\,d^2\widetilde{\lambda}_{\ell_4}}{{\rm GL(1)}}
   \,\Big\{\frac{\la \ell_2\ell_3\ra\la\ell_1\ell_4\ra\,\delta^4(P_1)\delta^4({\cal Q}_1)} {\la\ell_1\ell_2\ra^2\la\ell_3\ell_4\ra} \times \Omega_2\Big\}
\end{equation}
This is a straight generalization of the case discussed in the last subsection. We included the integration over internal particle $\ell_4$ in the definition of $\omega$ in order to eliminate the momentum conservation on four internal legs, such that we can then take a derivative of the expression without any ambiguities. But this is just one choice, we can obviously choose another internal leg to be eliminated in this way. The residue on the pole at infinity $\la\ell_1\ell_3\ra=0$ is then given by 
\begin{equation}\label{blobdiag5}\Omega_{\rm UV}= {\cal O}\otimes\Omega^{\rm bare}_{\rm UV} \qquad\mbox{where} \qquad \Omega^{\rm bare}_{\rm UV} = 
\hspace{0.1cm} \raisebox{-23mm}{\includegraphics[trim={0cm 0cm 0cm 0cm},clip,scale=1.1]{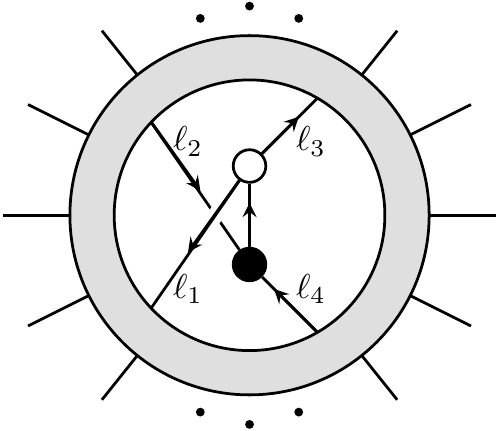}}
\end{equation}
where the operator ${\cal O}$ is given by
\begin{equation}
{\cal O} = \left(\frac{\la\ell_2\ell_3\ra[\ell_2\ell_3]}{\la\ell_1\ell_2\ra[\ell_1\ell_3]}\right)\left\langle\lambda_{\ell_2}\,\frac{d}{d\lambda_{\ell_3}}\right\rangle 
\end{equation}
and the action on the (bare) on-shell function in (\ref{blobdiag4}) is given by the analogous formula to (\ref{UVpoledef}),
\begin{equation}
    \Omega_{\rm UV} = \int d^2\widetilde{\eta}_{\ell_1} \dots d^2\widetilde{\eta}_{\ell_3}
    \int \frac{d^2\lambda_{\ell_1}\,d^2\widetilde{\lambda}_{\ell_1}}{{\rm GL(1)}}\dots \int \frac{d^2\lambda_{\ell_3}\,d^2\widetilde{\lambda}_{\ell_3}}{{\rm GL(1)}}  \Bigg\{\delta(\la\ell_1\ell_3\ra) \frac{\la\ell_2\ell_3\ra[\ell_2\ell_3]}{\la\ell_1\ell_2\ra[\ell_1\ell_3]}\left\langle \lambda_{\ell_2}\frac{d\omega^{\rm bare}}{d\lambda_{\ell_3}}\right\rangle\Bigg\}
     \label{UVpoledef2}
\end{equation}   
where the bare on-shell function $\omega^{\rm bare}$ for (\ref{blobdiag5}) is
\begin{equation}
\omega^{\rm bare} = \int d^2\widetilde{\eta}_{\ell_4} \int \frac{d^2\lambda_{\ell_4}\,d^2\widetilde{\lambda}_{\ell_4}}{{\rm GL(1)}}
   \Bigg\{\frac{\la\ell_2\ell_4\ra\, \delta^4(P_1)\delta^4({\cal Q}_1)}{\la\ell_1\ell_2\ra\la\ell_3\ell_4\ra}\times \Omega_2\Bigg\}
\end{equation}
which is just a straightforward generalization of our discussion before. Note that the only limitation is that we required that the leg $\ell_4$ is an internal leg. In case it is an external leg, and we do not integrate over it (hence we do not resolve the momentum conservation), we just have to choose another leg to integrate over first. The generalization to an arbitrary number of supersymmetries is again straightforward, we just act with $3{-}{\cal N}$ copies of ${\cal O}$, i.e.
\begin{equation}
\Omega_{\rm UV} = {\cal O}^{3{-}{\cal N}}\otimes \Omega^{\rm bare}_{\rm UV}
\end{equation}
We are now ready to generalize this procedure to higher $n$-gons. Let us now consider the NMHV hexagon subdiagram, 
\begin{equation}\label{blobdiag6}
 \raisebox{-26mm}{\includegraphics[trim={0cm 0cm 0cm 0cm},clip,scale=1.05]{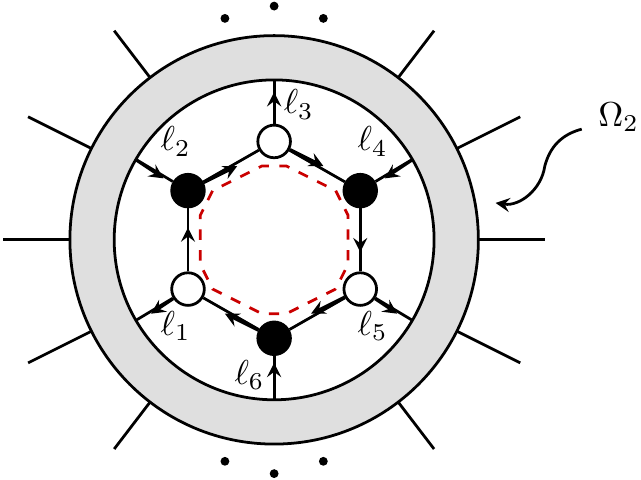}}.
\end{equation}
The on-shell function for ${\cal N}=2$ SYM is
\begin{equation}
    \Omega = \int d^2\widetilde{\eta}_{\ell_1} \dots d^2\widetilde{\eta}_{\ell_6}
    \int \frac{d^2\lambda_{\ell_1}\,d^2\widetilde{\lambda}_{\ell_1}}{{\rm GL(1)}}\dots
    \int \frac{d^2\lambda_{\ell_6}\,d^2\widetilde{\lambda}_{\ell_6}}{{\rm GL(1)}}
   \,\Big\{ \Omega_1 \times \Omega_2\Big\},\label{Hex03}
\end{equation}   
where the on-shell function for the hexagon can be written as 
\begin{align}
    \Omega_1 &= \frac{\la\ell_4| \ell_3{+} \ell_5|\ell_1]^{2} \la \ell_3 \ell_4 \ra \la \ell_1 \ell_5 \ra\, \delta(\Xi)}{[\ell_5\ell_6]^2[\ell_1\ell_2]\la \ell_1 \ell_2\ra \la \ell_1 \ell_6 \ra \la \ell_2 \ell_3 \ra \la \ell_4 \ell_5\ra \la \ell_3 \ell_5\ra}\times \left(\frac{\la\ell_1\ell_2\ra\la\ell_3\ell_4\ra[\ell_1\ell_2]}{\la\ell_1\ell_3\ra\la\ell_4|\ell_3{+}\ell_5|\ell_1]}\right)^{2}\nonumber\\
    &\hspace{3cm}= \frac{\la \ell_3 \ell_4 \ra^3 \la \ell_1 \ell_5 \ra\la\ell_1\ell_2\ra[\ell_1\ell_2]\, \delta(\Xi)}{[\ell_5\ell_6]^2\la \ell_1 \ell_6 \ra \la \ell_2 \ell_3 \ra \la \ell_4 \ell_5\ra \la \ell_3 \ell_5\ra\la\ell_1\ell_3\ra^2}   \label{Hex4}
\end{align}
where the delta functions are 
\begin{equation}
\delta(\Xi_1)\equiv \delta^4(P_1)\delta^4({\cal Q}_1)\,\delta(\la\ell_1|\ell_2{+}\ell_3|\ell_4])\delta(\la\ell_3|\ell_4{+}\ell_5|\ell_6])\delta([\ell_5\ell_6]\widetilde{\eta}_{\ell_4}{+}[\ell_6\ell_4]\widetilde{\eta}_{\ell_5}{+}[\ell_4\ell_5]\widetilde{\eta}_{\ell_6})
\end{equation}
The residue on the pole at infinity $\la\ell_1\ell_3\ra=0$ is then given by
\begin{equation}
    \Omega_{\rm UV} = \int d^2\widetilde{\eta}_{\ell_1} \dots d^2\widetilde{\eta}_{\ell_5}
    \int \frac{d^2\lambda_{\ell_1}\,d^2\widetilde{\lambda}_{\ell_1}}{{\rm GL(1)}}\dots \int \frac{d^2\lambda_{\ell_5}\,d^2\widetilde{\lambda}_{\ell_5}}{{\rm GL(1)}}  \Bigg\{\delta(\la\ell_1\ell_3\ra) \left(\frac{\la\ell_2\ell_3\ra[\ell_2\ell_3]}{\la\ell_1\ell_2\ra[\ell_1\ell_3]}\right)\left\langle \lambda_{\ell_2}\frac{d\omega^{\rm bare}}{d\lambda_{\ell_3}}\right\rangle\Bigg\}
     \label{UVpoledef3}
\end{equation}   
where $\omega^{\rm bare}$ is now the on-shell function associated with the graph
\begin{equation}\label{blobdiag7}
 \raisebox{-25mm}{\includegraphics[trim={0cm 0cm 0cm 0cm},clip,scale=1.15]{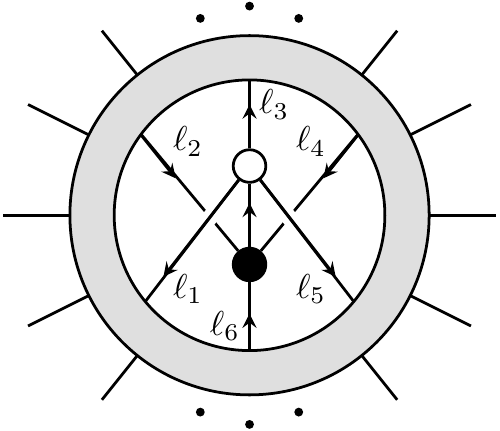}}
\end{equation}
before we integrate over momenta $\ell_1,{\dots},\ell_5$ (in fact, we can integrate over all momenta except $\lambda_{\ell_3}$ in which we take the derivative). This straightforwardly generalizes to any one-loop $n$-gon, and also to any number of supersymmetries -- we again act $(3{-}{\cal N})$ times with the operator ${\cal O}$. 

It is worth noting that in our procedure of approaching and calculating the higher pole at infinity we made a certain choice: we parametrized $\lambda_{\ell_3} = \alpha \lambda_{\ell_1} + \beta\lambda_{\ell_2}$ and calculated the pole for $\beta=0$ which then transformed into a certain form of the ${\cal O}$. Note that the final result does not depend on this choice (the calculation of the residue is independent on the parametrization). One assumption we made in our calculation was that the leg $\ell_3$ was internal (as we integrated over it and removed any dependencies between $\ell_j$ through momentum conservation). If this was an external leg we would have to approach the pole via another leg. For the general on-shell diagram with $L>1$ at least one of the legs of the $n$-gon (which UV pole we approach) is internal and we can use it in our procedure. The only exception is the one-loop $n$-gon itself as an on-shell diagram, in this case we can just attach a BCFW bridge as we did for the box diagram in the previous section. The result for the UV pole does not depend on the particular procedure used to approach that pole.

\section{Non-planar diagrams\label{sec:non-planar}}

The non-planar on-shell diagrams have been extensively studied in the past. Based on the principles of generalized unitarity they represent the cuts of non-planar loop integrands. In the dual description, the on-shell function is reproduced by the simple dlog form in ${\cal N}=4$ SYM theory while in the lower SUSY Yang-Mills theories we have to include the Jacobians \cite{Arkani-Hamed:2012zlh}, and the formula is the same as for the planar diagrams. Unlike in the planar case not much is known about the connection of non-planar on-shell diagrams to the Grassmannian geometry, permutations, stratifications, identity moves etc. Some interesting observations were made for MHV on-shell diagrams in the context of ${\cal N}=4$ SYM theory \cite{Arkani-Hamed:2014bca} where the on-shell function for each diagram was written as a special linear combination of Parke-Taylor factors. A more systematic study was performed in \cite{Franco:2015rma} and the classification of all on-shell diagrams for 6 point NMHV amplitudes was performed in \cite{Bourjaily:2016mnp}. The first attempts to associate NMHV on-shell diagrams with Grassmannian geometries were provided in \cite{Paranjape:2022ymg} in the context of non-adjacent BCFW recursion relations. 

In our discussion of UV poles in on-shell diagrams, we restricted only to planar diagrams and also our proof of the localization of UV poles in one loop subdiagrams required the planarity of the whole diagram. We will show that indeed the same argument does not hold for non-planar on-shell diagrams and the poles at infinity propagate beyond just a one-loop subgraph. We explore one particular five-point non-planar diagram,
\begin{equation}\label{eq:hexagon8}
 \raisebox{-15mm}{\includegraphics[trim={0cm 0cm 0cm 0cm},clip,scale=1.1]{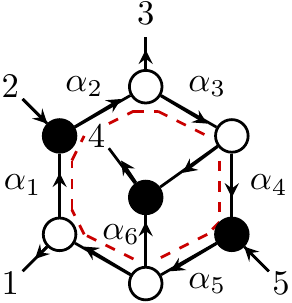}},
 \begin {aligned}\qquad\qquad
\alpha_1&=\frac{\la 32 \ra}{\la 13 \ra}, \quad \alpha_2=\frac{\la 13 \ra}{\la 12 \ra},\\
\alpha_3&=\frac{\la 14 \ra}{\la 13 \ra}, \quad \alpha_4=\frac{\la 51 \ra}{\la 14 \ra}, \\
\alpha_5&=\frac{\la 13 \ra}{\la 53 \ra}, \quad \alpha_6=\frac{\la 43 \ra}{\la 13 \ra}.
\end{aligned}
\end{equation}
The Jacobian from the cycle is
\begin{equation}
    \begin{aligned}
        \mathcal{J}=1-\alpha_1\alpha_2\alpha_3\alpha_4\alpha_5 = \frac{\la 13 \ra \la 25 \ra}{ \la 12 \ra \la 35 \ra},
    \end{aligned}
\end{equation}
where we can identify the UV pole as $\la 13\ra\to 0$, while $\la25\ra$ is the helicity factor. We could instead have picked an orientation with a different internal cycle as follows,
\begin{equation}\label{eq:hexagon9}
 \raisebox{-15mm}{\includegraphics[trim={0cm 0cm 0cm 0cm},clip,scale=1.1]{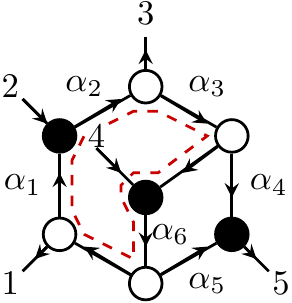}},
 \begin {aligned}\qquad\qquad
\alpha_1&=\frac{\la 32 \ra}{\la 13 \ra}, \quad \alpha_2=\frac{\la 13 \ra}{\la 12 \ra},\\
\alpha_3&=\frac{\la 14 \ra}{\la 13 \ra}, \quad \alpha_4=\frac{\la 51 \ra}{\la 14 \ra}, \\
\alpha_5&=\frac{\la 35 \ra}{\la 13 \ra}, \quad \alpha_6=\frac{\la 13 \ra}{\la 34 \ra}.
\end{aligned}
\end{equation}
Here the Jacobian from the cycle is
\begin{equation}
    \begin{aligned}
        \mathcal{J}=1-\alpha_1\alpha_2\alpha_3\alpha_6 = \frac{\la 13 \ra \la 24 \ra}{ \la 12 \ra \la 34 \ra}.
    \end{aligned}
\end{equation}
We find once again that the pole at infinity is at $\la 13\ra\to 0$, even though we are looking at different loop. We can uncover the same problem by solving for internal momenta,
\begin{equation}\label{eq:hexagon10}
 \raisebox{-15mm}{\includegraphics[trim={0cm 0cm 0cm 0cm},clip,scale=1.1]{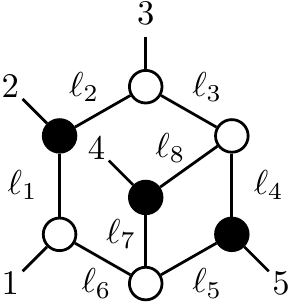}},
 \begin {aligned}\qquad
\ell_1&=\frac{\la 23 \ra}{\la 13 \ra}\lambda_{1}\widetilde{\lambda}_{2}, \quad \ell_2=\frac{\la 12 \ra}{\la 13 \ra}\lambda_{3}\widetilde{\lambda}_{2}, \quad 
\ell_3=\frac{(Q_{23}\cdot \lambda_1)}{\la 13 \ra}\lambda_{3},\\ \quad \ell_4&=\frac{\la 15 \ra}{\la 13 \ra}\lambda_{3}\widetilde{\lambda}_{5},\quad 
\ell_5=\frac{\la 35 \ra}{\la 13 \ra}\lambda_{1}\widetilde{ \lambda}_{5}, \quad \ell_6=\frac{(Q_{12}\cdot \lambda_3)}{\la 13\ra}\lambda_{1},\\ 
\quad \ell_7&=\frac{\la 34 \ra}{\la 13 \ra}\lambda_{1}\widetilde{\lambda}_{4},\quad 
\ell_8=\frac{\la 14 \ra}{\la 13 \ra}\lambda_{3}\widetilde{\lambda}_{4}.
\end{aligned}
\end{equation}
and we see that the UV pole $\la13\ra$ is present in multiple loops rather than only one-loop subgraph. There is a purely graph theoretic reason for these subtleties: we can not invariantly talk about ``fundamental'' and ``overlapping'' internal loops. In the planar case, the first type corresponds to poles at infinity, while the second type does not. Take e.g. the pentabox \eqref{eq:pentabox9} with the orientation in the box flipped. Schematically we can show the pole and cycle structure as,
\begin{equation}\label{eq:Pentabox10}
 \raisebox{-18mm}{\includegraphics[trim={0cm 0cm 0cm 0cm},clip,scale=1.2]{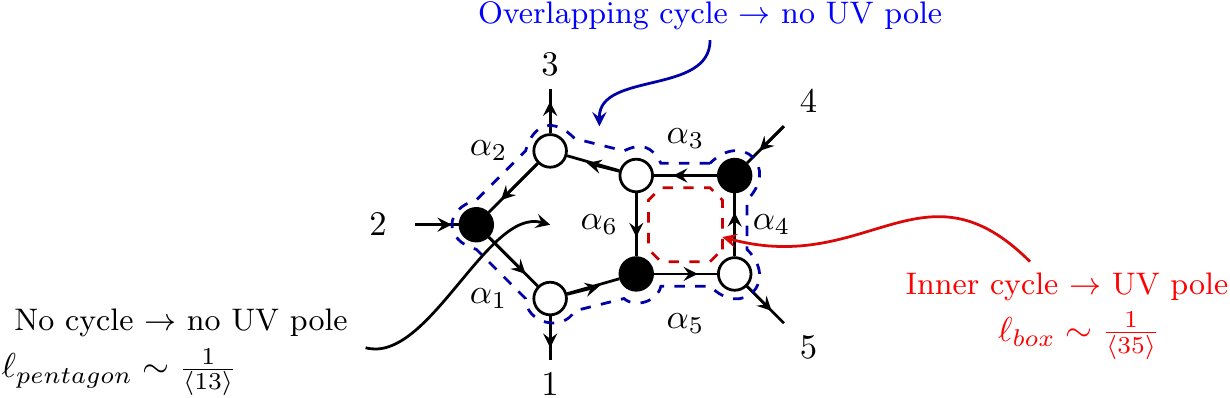}}.
\end{equation}
Even with the overlapping cycle, the Jacobian still only introduces the UV pole of the box $\la 35 \ra$.
For the non-planar graphs there is no difference between these two types of cycles, and we can always redraw a diagram in a way which turns one type into the other,
\begin{equation}\label{eq:hexagon15}
 \raisebox{-22mm}{\includegraphics[trim={0cm 0cm 0cm 0cm},clip,scale=1.1]{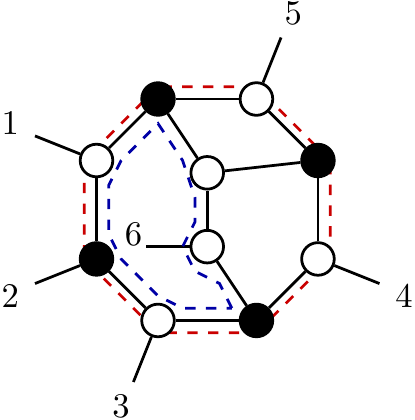}}\quad = \quad \raisebox{-22mm}{\includegraphics[trim={0cm 0cm 0cm 0cm},clip,scale=1.1]{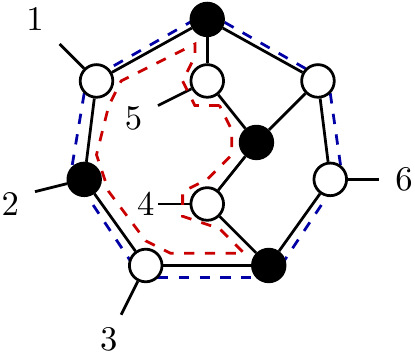}}.
\end{equation}
This makes it clear that we need to generalize our rule to capture the UV poles in non-planar diagrams where multiple loops blow up at the same time. This also concerns the gravity on-shell diagrams \cite{Herrmann:2016qea,Heslop:2016plj,Farrow:2017eol,Armstrong:2020ljm} which are intrinsically non-planar. 

\section{Conclusion and Outlook}

In this paper we studied the poles at infinity in on-shell diagrams for ${\cal N}{<}4$ SYM theory. We showed that for planar diagrams the poles at infinity are located in one-loop subgraphs and the kinematical condition on the pole reduces the graph to a simpler on-shell diagram with a \emph{non-planar twist}, hence the diagram is generally non-planar. The on-shell function on the UV pole is then given by a certain derivative operator acting on the on-shell function for a new non-planar on-shell diagram. This comes naturally from the definition of the residue on a higher pole as a derivative. We also showed that for non-planar on-shell diagrams the situation is more complicated as the poles at infinity are not localized in one-loop subgraphs and a different procedure is needed.

Our work is a direct attempt to systematically study the poles at infinity in scattering amplitudes. While the residues of tree-level amplitudes and loop integrands in the deep IR are governed by principles of unitarity and lead to factorizations and unitarity cuts, the poles in the UV are naively not constrained by similar principles. We do not face this problem directly in the context of scattering amplitudes, but we chose simpler on-shell gauge-invariant objects -- on-shell diagrams, and solve the problem here. Note that tree-level amplitudes (and in special cases also loop integrands) can be expressed as sums of on-shell diagrams, so it is suggestive that our results would be relevant to amplitudes problem as well. The two main lessons from our analysis are: (i) the kinematics of UV poles correspond to non-planar objects, in this case non-planar on-shell diagrams, (ii) on-shell functions are given by derivatives of the bare on-shell diagrams. In future work, we want to use this insight to address this problem directly in the context of tree-level amplitudes and loop integrands.

The further goal is to formulate the general prescription for the calculation of poles at infinity for arbitrary tree-level amplitudes, and loop integrands -- similar to rules of factorizations and cuts. This would have major implications on our ability to calculate amplitudes using recursion relations, and possibly search for positive geometries that could capture amplitudes in theories with non-trivial UV physics.  

\vspace{-0.3cm}

\acknowledgments

\vspace{-0.3cm}

We thank Nima Arkani-Hamed, Enrico Herrmann and Shruti Paranjape for useful discussions and comments. This work is supported by a DOE grant No. SC0009999,  GA\v{C}R 21-26574S and the funds of the University of California.


\bibliographystyle{JHEP}
\bibliography{main.bib}

\end{document}